\documentclass[fleqn,usenatbib]{mnras}

\usepackage{newtxtext,newtxmath}

\usepackage[T1]{fontenc}

\DeclareRobustCommand{\VAN}[3]{#2}
\let\VANthebibliography\thebibliography
\def\thebibliography{\DeclareRobustCommand{\VAN}[3]{##3}\VANthebibliography}

\usepackage{graphicx}	
\usepackage{amsmath}	
\usepackage{siunitx}

\newcommand{\civ}{\ion{C}{IV}}	
\newcommand{\mgii}{\ion{Mg}{II}}
\newcommand{\siv}{\ion{Si}{IV}}	
\newcommand{\hi}{\ion{H}{I}}
\newcommand{\maggiv}{MAGG IV}

\title[MAGG. V. Linking galaxies to \ion{C}{IV} and \ion{Si}{IV} absorbers]{MUSE Analysis of Gas around Galaxies (MAGG) - V: Linking ionized gas traced by \ion{C}{IV} and \ion{Si}{IV} absorbers to Ly$\alpha$ emitting galaxies at $z\approx 3.0-4.5$.}

\author[M. Galbiati et al.]{
Marta Galbiati,$^{1}$\thanks{E-mail:m.galbiati29@campus.unimib.it}
Michele Fumagalli,$^{1,2}$\thanks{E-mail:michele.fumagalli@unimib.it}
Matteo Fossati,$^{1,3}$
Emma K. Lofthouse,$^{1,3}$
Rajeshwari Dutta,$^{1,3}$
\newauthor
J. Xavier. Prochaska,$^{4,5}$
Michael T. Murphy,$^{6}$
Sebastiano Cantalupo$^{1}$
\\
$^{1}$Dipartimento di Fisica G. Occhialini, Universit\`a degli Studi di Milano-Bicocca, Piazza della Scienza 3, 20126 Milano, Italy \\
$^{2}$INAF - Osservatorio Astronomico di Trieste, via G. B. Tiepolo 11, 34143 Trieste, Italy \\
$^{3}$INAF - Osservatorio Astronomico di Brera, via Brera 28, 21021 Milano, Italy \\
$^{4}$Department of Astronomy and Astrophysics, University of California, Santa Cruz, CA 95064, USA \\
$^{5}$Kavli Institute for the Physics and Mathematics of the Universe, 5-1-5 Kashiwanoha, Kashiwa 277-8583, Japan \\
$^{6}$Centre for Astrophysics and Supercomputing, Swinburne University of Technology, Hawthorn, Victoria 3122, Australia
}

\date{}

\pubyear{\the\year}

\begin{document}
\label{firstpage}
\pagerange{\pageref{firstpage}--\pageref{lastpage}}
\maketitle

\begin{abstract}
We use 28 quasar fields with high-resolution (HIRES and UVES) spectroscopy from the MUSE Analysis of Gas Around Galaxies survey to study the connection between Ly$\alpha$ emitters (LAEs) and metal-enriched ionized gas traced by \civ\ in absorption at redshift $z\approx3-4$. In a sample of 220 \civ\ absorbers, we identify 143 LAEs connected to \civ\ gas within a line-of-sight separation $\pm500\rm\,km\,s^{-1}$, equal to a detection rate of $36\pm5$ per cent once we account for multiple LAEs connected to the same \civ\ absorber. The luminosity function of LAEs associated with \civ\ absorbers shows a $\approx 2.4$ higher normalization factor compared to the field. \civ\ with higher equivalent width and velocity width are associated with brighter LAEs or multiple galaxies, while weaker systems are less often identified near LAEs. The covering fraction in groups is up to $\approx 3$ times larger than for isolated galaxies. Compared to the correlation between optically-thick \hi\ absorbers and LAEs, \civ\ systems are twice less likely to be found near LAEs especially at lower equivalent width. Similar results are found using \siv\ as tracer of ionized gas. We propose three components to model the gas environment of LAEs: i) the circumgalactic medium of galaxies, accounting for the strongest correlations between absorption and emission; ii) overdense gas filaments connecting galaxies, driving the excess of LAEs at a few times the virial radius and the modulation of the luminosity and cross-correlation functions for strong absorbers; iii) an enriched and more diffuse medium, accounting for weaker \civ\ absorbers farther from galaxies.
\end{abstract}




\section{Introduction}

Based on our current understanding of the onset of the first episodes of star formation in halos hosting population III stars, heavy elements are produced in a primordial Universe. Due to the massive nature of these first stars, the newly-produced metals are not always retained in compact objects, but are ejected into the interstellar medium (ISM) and, owing to energetic events, also in the circumgalactic medium (CGM) and the intergalactic medium (IGM), contributing to the early enrichment of the Universe \citep[e.g.][]{Scannapieco2002,Schneider2002,Maio2011,Wise2012}. Likewise, stellar winds and supernova explosions associated with the formation of the subsequent generations of stars in galaxies are believed to be the source of a substantial fraction of the heavy elements we observe today near and outside galaxies \citep[e.g.][]{Aguirre2001,Oppenheimer2006,Shen2013}, with mechanisms like winds from active galactic nuclei, gravitational interactions or ram-pressure stripping accounting for an additional fraction of the metals found outside galaxies \citep[e.g.][]{Fossati2016,Hafen2019}. Due to the tight correspondence between the production of metals by stars and the subsequent ejection into the CGM and IGM, the study of the chemical enrichment of the more diffuse gas outside and around galaxies as a function of time is a powerful tool to complement our view of the star formation history in galaxies  \citep[e.g.][]{Bouche2006,Rafelski2014,Madau2014,Fumagalli2016}, and a fundamental step for developing complete chemical evolution models \citep[e.g.][]{Tremonti2004,Finlator2008,Welsh2019}.

The most effective way to map the chemical enrichment of the low-density IGM and CGM is, at present, the study of hydrogen and metal absorption lines imprinted on the spectra of background sources, such as quasars. Compilations of high-resolution spectra, as well as moderate-to-low resolution spectroscopy in larger surveys, paint a view of a widespread metal enrichment of the Universe at least for gas around or above the mean density \citep[e.g.][]{Schaye2003,Simcoe2004}, with only a few rare exceptions of unpolluted regions observed below $z\lesssim 4.5$ \citep[e.g.][]{Fumagalli2011,Robert2019}. 

 The analysis of the \civ\ doublet with its two strong transitions at 1548.195\AA\ and 1550.770\AA\ and a characteristic ratio of the optical depth for unsaturated lines equal to two (given by the ratio of the corresponding oscillator strengths) has been particularly effective for this type of study, enabling the identification of this ion also in moderate signal-to-noise and low resolution data. In particular, recent searches of \civ\ doublets in Sloan Digital Sky Survey quasar spectra by \citet{Cooksey2013} and in quasar spectroscopy from the Keck and Very Large Telescope (VLT) archives by \citet{Hasan2020} consistently show a steady increase of the number of \civ\ absorbers per unit redshift with time. As the observed incidence of \civ\ is proportional to the comoving number density of \civ\ bearing systems times their cross section, one can learn about the link between \civ\ and galaxies in a model-dependent fashion. For instance, assuming that \civ\ absorbers with equivalent width (EW) $>0.06$~\AA\ arise in the CGM of Lyman break galaxies (LBGs), \citet{Cooksey2013} estimated that the observed incidence can be modelled by filling the inner $\approx 50~\rm kpc$ of LBGs halos with ionized carbon with time. More recently,  \citet{Hasan2021} expanded this argument both in mass and redshift, combining \civ\ absorption statistics with dark matter halos to constrain the link between gas and galaxies. 

Leveraging the deployment of echellette spectrographs in the near infrared, \citet{Simcoe2011} have expanded previous efforts to characterise the evolution of \civ\ in quasar surveys at even higher redshift \citep[see also][]{Pettini2003,Songaila2005}, reaching the epoch of reionization at $z\approx 6$ and confirming the presence of a downturn in the mass density of \civ\ beyond $z\approx 5$. These studies in large samples are complemented by the orthogonal approach followed by
\citet{Dodorico2016} who have obtained an extremely-high $S/N$ spectrum of a single quasar, measuring the distribution of some of the weakest \civ\ lines currently detected around column density
$N_{\rm CIV}= 10^{11.4}~\rm cm^{-2}$ (see also \citealp{Ellison2000}). Their analysis reveals a factor $\approx 2$ higher incidence of \civ\ lines associated with \hi\  than what is expected in LBG halos, implying the presence of metals in IGM filaments or in the CGM of galaxies not selected as LBGs (e.g. at lower mass; see also \citealp{Pieri2006,Hasan2021}).	

While absorption line studies currently remain the most powerful approach to trace the entire distribution of the selected ions regardless of the astrophysical environment in which they reside (e.g. IGM, CGM at different halo masses), more direct links between the production sites of heavy elements and their current location requires an explicit correlation between the ions detected in absorption and the surrounding galaxy population traced in emission. Focusing on \civ\ as a tracer, \citet{Dutta2021} have completed the largest survey to date of galaxies in quasar fields between $0.5 \lesssim z \lesssim 2$, finding that \civ\ is 
relatively more extended than \ion{Mg}{II} around galaxies \citep[see also][]{Liang2014,Schroetter2021}. Further, they reported a higher incidence of \civ\ around massive and star-forming galaxies \citep[see also][]{Chen2001,Bordoloi2014,Burchett2016}, although this ion appears comparably less sensitive than \ion{Mg}{II} with respect to the properties of the galaxies and of the environment. However, in the local Universe, \citet{Burchett2016} have uncovered an environmental dependence, with an higher detection rate in lower-density regions. Moreover, \citet{Schroetter2021} further explored the connection between \civ\ absorbers and galaxies, finding a non-negligible fraction of instances where ionized gas can be associated to the IGM or undetected low-mass galaxies.

At redshift $2\lesssim z \lesssim 3$, \citet{Adelberger2005} have provided the first clear evidence of an association between strong \civ\ absorbers and massive (with halo mass $M_{\rm H}\approx 10^{12}~\rm M_\odot$) star-forming galaxies (with star formation rate, SFR, $\approx 10-50~\rm M_\odot~yr^{-1}$) selected as LBGs \citep[see also][]{Crighton2011}, with these types of halos accounting for $\approx 1/3$ of all \civ\ absorbers with EW $>0.4~$\AA. Their study has also suggested that galaxies in denser environments are more likely to show \civ\ in absorption compared to isolated galaxies. The ubiquitous presence of \civ\ near LBGs is also confirmed by later studies building on the same sample presented by \citet{Adelberger2005}, with \citet{Steidel2010} using galaxy pairs to infer that LBGs account for almost 50\% of strong (EW $>0.15~$\AA) \civ\ systems at  $z\approx 2$ (see also \citealp{Turner2014}).

These studies rely on spectroscopic follow-up of relatively-bright continuum-selected galaxies and, especially at $z>2$, focus on the massive end of the galaxy population. For a complete view of the correlation between galaxies and metals, this type of analysis needs to be extended at the low-mass end, and to encompass also passive or highly-obscured galaxies.
The serendipitous discovery of a faint Ly$\alpha$ emitting galaxy (LAE) associated with a metal-rich absorption system in an LBG survey by \citet{Crighton2015} provides a clear example of the need to extend this analysis to lower-mass galaxies without relying on continuum selection \citep[see also][]{Diaz2015}, an approach that is now possible (see e.g. \citealp{Fumagalli2016,Fumagalli2017}) thanks to large  integral field spectrographs and in particular the  Multi Unit Spectroscopic Explorer (MUSE; \citealp{Bacon2010}) at VLT.  
To this end, the recent $z\approx 3-4$ MUSE observations by \citet{Muzahid2021} have identified 96 LAEs in 8 quasar fields. By correlating these galaxies with \civ\ absorption, these authors found an elevated \civ\ optical depth near LAEs similarly to the case of LBGs. They also report evidence of an excess of absorption near LAEs in groups compared to isolated ones. Examples of similar studies extending to higher redshift can also be found in the literature \citep{Bielby2020,Diaz2021}.

In this paper, we exploit the larger MUSE Analysis of Gas around Galaxies (MAGG, \citealp{Lofthouse2019}) survey to study in detail the correlation between 292 LAEs and 220 \civ\ absorbers. To verify whether the link between LEAs and ionized gas traced by \civ\ is specific to the selected transition or general for a ionized gas phase, we also expand our study to the associations between LAEs and \siv, which is an additional doublet that arises from moderately-ionized gas and it is conveniently accessible in a comparable wavelength range to \civ. MAGG is built on a MUSE Large Programme (ID 197.A-0384; PI Fumagalli) that was explicitly designed to study the link between gas and galaxies at $z\approx3-4$ by targeting 28  fields with quasars at $ z\approx3.2-4.5$ for which archival high-resolution ($ R\gtrsim30,000$) spectroscopy is available 
(\citealt{lofthouse2022}, hereafter \maggiv; see also \citealp{Dutta2020,Fossati2021}). The MAGG selection results in a sample of quasars with magnitudes $m_{\rm r}\leq 19$ mag and at least one strong hydrogen absorption line system ($N_{\rm HI}\ge 10^{17}~\rm cm^{-2}$) at redshift $z\ge 3.05$. 

The structure of this paper is as follows: in Sect.~\ref{sec:data}-\ref{sec:analysis} we present an overview of the observations and data analysis, including a description of how the LAEs,  \civ\ and \siv\ samples have been assembled. Readers not interested in  technical details on data can skip these two sections. In Sect.~\ref{sec:result} we turn to the main question of studying in detail the correlation between LAEs and \civ\ absorbers as a function of galaxy properties and environment, 
also comparing with the results obtained by relying on \siv\ rather than \civ\ as a tracer of ionized gas. We discuss our main findings and conclusions in Sect.~\ref{sec:discussion}-\ref{sec:summary}. Throughout, we assume a flat $\Lambda$CDM cosmology with $H_0=67.7~\rm km~s^{-1}~Mpc^{-1}$ and $\Omega_{\rm m} = 0.307$ \citep{Planck2016}, magnitudes are expressed in the AB system,  distances are in physical units unless explicitly stated (for instance when computing the correlation functions), and errors are at $1\sigma$ confidence level.

\section{Observations and Data Reduction}\label{sec:data}

\subsection{MUSE observations}

Each quasar field included in the MAGG survey has been observed with MUSE between period 97 and period 103, for a total on-source observing time of $\approx\,4$ hours per field using the wide field mode with extended wavelength coverage ($465-930$~nm, $R\approx 2000-3500$). As part of this survey, we also include archival data from the GTO observations (PI Schaye; \citealp{Muzahid2021}) that matched the selection criteria of the original MAGG sample (i.e., the presence of $z>3$ \ion{H}{I} optically-thick absorption systems along the line of sight; see details in table 1 of \citealt{Lofthouse2019}).

 All the observations have been executed on clear nights at airmass $\rm \leq\,1.6$ so that the image quality is better than 0.8 arcsec full width at half maximum (FWHM). Image quality for individual fields can be found in table 1 of \citealt{Lofthouse2019}.
 In wide field mode, the MUSE field of view (FoV) is $\rm\approx1\times1\,\rm arcmin^{2}$. In our observations, the quasar typically lies at the center of the FoV, enabling deep spectroscopic surveys of regions of $\rm\approx 500\times500\,\rm kpc^{2}$ at redshift $z\approx3$.
As documented in the MUSE instrument manual, sensitivity variations across the FoV may arise from small differences in the performance of the MUSE spectrographs, and these are mitigated by including small dithers and instrument rotations of 90 degrees.

The detailed procedure of MUSE data reduction is described by \citet{Lofthouse2019}, and only the key steps are summarized here. The raw data are first processed with the ESO MUSE pipeline \citep[][versione 2 or greater]{Weilbacher2014} which applies bias and flat calibrations, reduces the sky flats, and reconstructs the cube of individual exposures following wavelength calibration. Each individual exposure is then post-processed with the {\sc CubExtractor} package ({\sc CubEx}, \citealp{Cantalupo2019}), which is one of the available tools to improve the quality of the datacubes and mitigate the known imperfections arising after the basic ESO pipeline reduction \citep[see figure 1 in][]{Lofthouse2019}. In particular, the residual differences in the relative illumination of the 24 MUSE IFUs are corrected with the {\sc cubefix} tool in {\sc CubEx} which scans the cube as a function of wavelength to re-align the relative illumination of the IFUs. Next, the {\sc cubesharp}  tool (also part of {\sc CubEx}) is used to perform local sky subtraction removing the residuals left from the ESO pipeline reduction, by taking into account spatial variations in the line spread function of the instrument. A second iteration using both these tools is then performed masking continuum-detected sources to enable an improved determination of the background. At the end of the reduction process, in order to allow a more accurate identification of possible contaminants, all the single exposures are combined in four final products: an average cube of all the exposures (mean cube), a median cube and two cubes, each containing only one half of all the exposures. These last two products are useful to confirm the detection of sources in two independent datasets.

As fully described by \citet{Lofthouse2019}, the uncertainties of each pixel are computed by propagating the detector noise through the whole reduction process. Since the individual pixels are transformed across several steps, including the non-linear interpolation leading to the final data cube, the resulting pixel variance does not accurately reproduce the effective standard deviation of each volumetric pixel. Following the procedure detailed by \citet{Fossati2019}, we derive a series of data cubes with accurate estimates of the standard deviation by bootstrapping the pixels in individual exposures in order to compute the noise in each final product.

\subsection{High-resolution quasar spectroscopy}

All the quasars included in the MAGG survey have  high-resolution ($R\gtrsim30,000$) and moderate or high signal-to-noise ($S/N\gtrsim10$ per pixel, for details see table 2 of \citealt{Lofthouse2019}) spectra obtained with the Ultraviolet and Visual Echelle Spectrograph (UVES, \citealp{Dekker2000}) at the VLT, the High Resolution Echelle Spectrometer (HIRES, \citealp{Vogt1994}) at Keck, or the Magellan Inamori Kyocera Echelle (MIKE; \citealp{Bernstein2003}) at the Magellan telescopes. Moderate dispersion spectra from X-SHOOTER \citep{Vernet2011} at the VLT and Echellette Spectrograph and Imager (ESI, \citealp{Sheinis2002}) at Keck are also used to extend the observed wavelength range. Details on the reduction procedure for each instrument and a summary of the data available for each quasar  are described by \citet{Lofthouse2019} in section 3.1 and listed in their table 2. The main observational information that are summarized therein include: the wavelength range covered by the data, the spectral resolution, and the final signal-to-noise ratio at selected wavelengths.

\section{Data Analysis}\label{sec:analysis}

\subsection{Catalogue of \civ\ absorption lines}

We visually inspect the high-resolution quasar spectra searching for \civ\ $\lambda\lambda$ 1548, 1550 resonant doublets, identified through the characteristic rest wavelength separation of 2.575 \AA, corresponding to $\rm 489\,km\,s^{-1}$, and the equivalent width ratio 2:1 for $W_{\rm r}^{1548}:W_{\rm r}^{1550}$ in the unsaturated regime. 
For the sightlines where multiple observations from different instruments are available, we choose to inspect the spectrum with the highest signal-to-noise ratio for each sightline in a specific wavelength range.

\begin{figure} 
\centering
\includegraphics[width=\columnwidth]{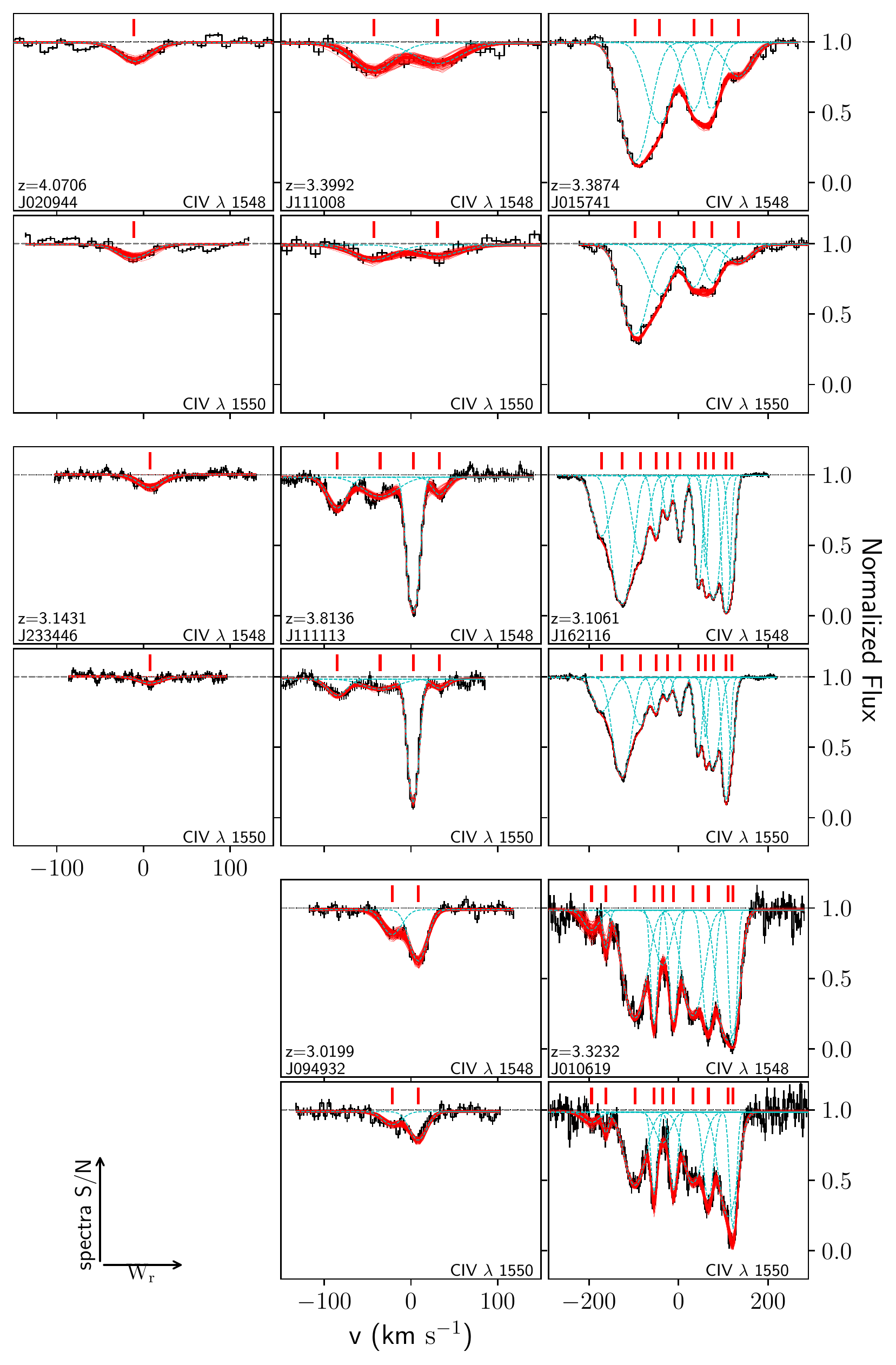}
\caption{Examples of \civ\ doublet absorption line systems detected in MAGG quasar spectra at different $S/N$ ($S/N\leq20$, $20<S/N<40$ and $S/N\geq40$) and randomly extracted from three bins of \civ\ equivalent width ($W_{\rm r}/\text{\AA}\leq0.03$, $0.03<W_{\rm r}/\text{\AA}<0.20$ and $W_{\rm r}/\text{\AA}\geq0.20$). The normalized spectrum (step black lines with $1\sigma$ error bars) is reproduced by Voigt profiles sampling the posterior distribution of the parameters obtained by MC-ALF (solid red lines). Each individual \civ\ component is marked by a vertical red tick and shown as cyan dotted line.}
\label{fig:CIVlines}
\end{figure}

\begin{figure*} 
\centering
\includegraphics[width=\textwidth]{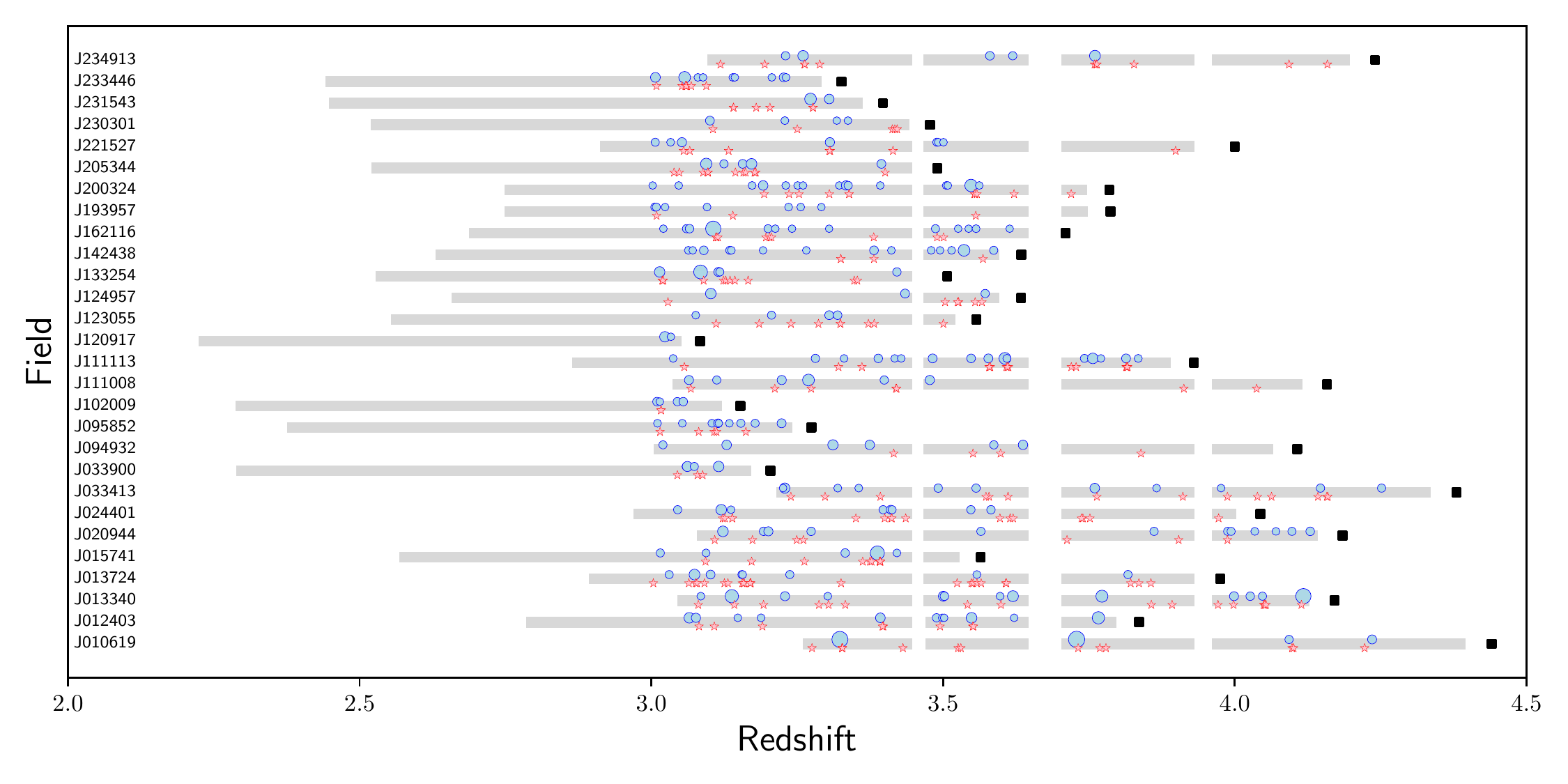}
\caption{Summary of the final MAGG-\civ\ sample. The grey bands mark the searchable \civ\ redshift path of each sightline. The gaps correspond to known telluric bands we masked to avoid false positives. The doublets at $z\geq3$ are drawn as blue dots and sized proportionally to the equivalent width of each line. LAEs detected in the MUSE cubes $z\geq3$ and lying in the \civ\ redshift path are shown as red stars. The central quasar of each field is shown as a black square. A velocity window $|\Delta v|<3000\rm\,km\,s^{-1}$ from the redshift of each quasar is also masked to avoid proximity effects.}
\label{fig:CIVzpath}
\end{figure*}

\begin{figure} 
\centering
\includegraphics[width=\columnwidth]{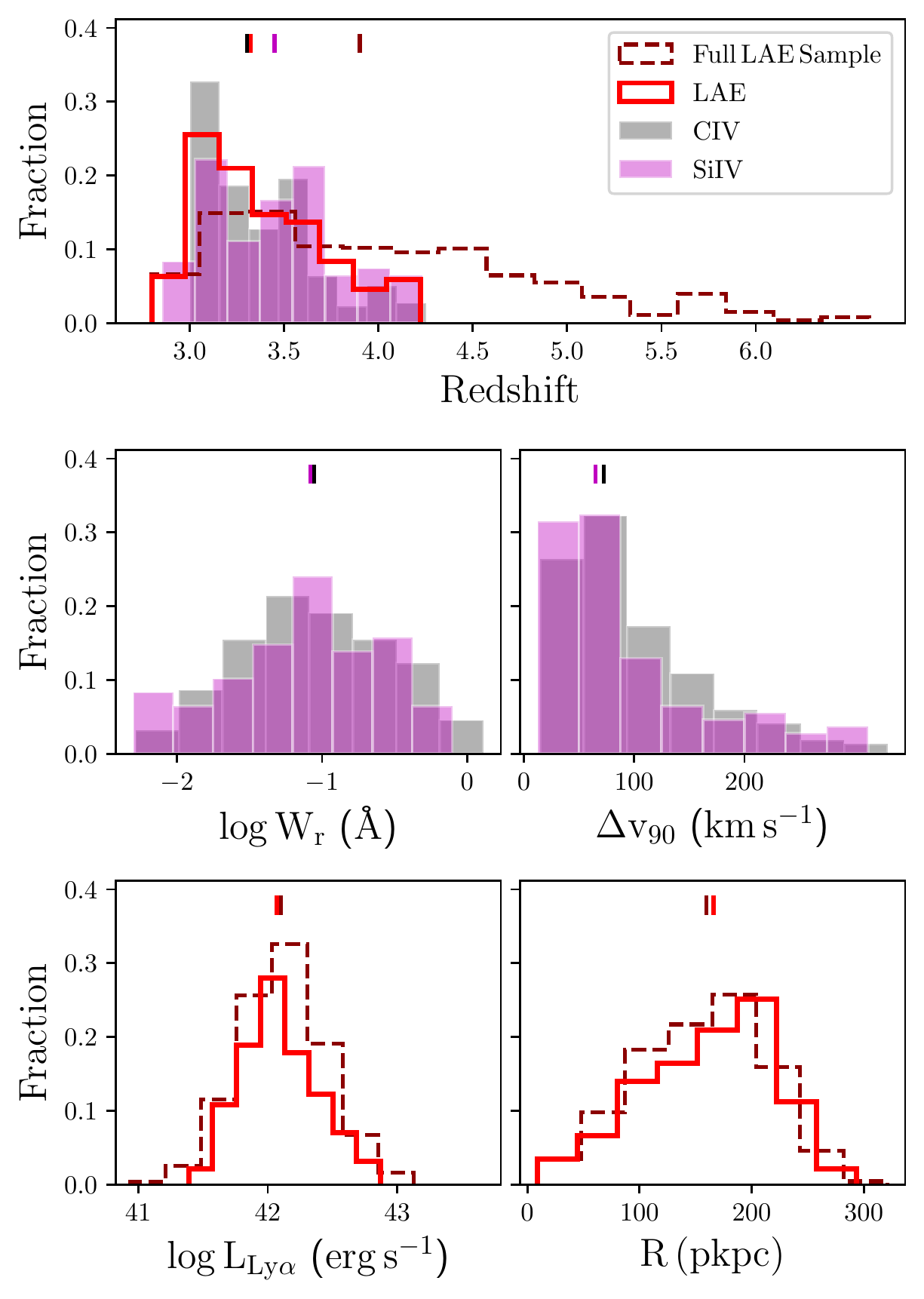}
\caption{Summary of the properties measured for the \civ\ (black) and \siv\ (magenta) doublets (redshift, rest-frame equivalent width and velocity width) and those of the emitters (redshift, $Ly\alpha$ luminosity and impact parameter) identified in the MAGG fields at $SNR>7$ (dashed red line) and in the absorbers redshift path (solid red line). Vertical ticks mark  the median of each distribution.}
\label{fig:SampleProperties}
\end{figure}

The inspection is limited to in the wavelength range lying redward the quasar $\rm Ly\alpha$ forest, so as to exclude wavelengths in which metals are only the minority of the observed transitions and thus hard to detect due to the severe blending. We also exclude small windows that overlap with known strong telluric bands at 6870-6935 \AA\ and 7595-7700 \AA\, since these contaminant lines of the Earth's atmosphere could be more easily misinterpreted as false positives. To obtain the highest possible purity of the sample, we also reject all the candidate \civ\ doublets that show nonphysical line ratios (i.e., not matching the ratio set by the oscillator strengths) in the range 7180-7295 \AA, a further window that is contaminated by the presence of additional weak telluric features. Lastly, we mask a velocity window $|\Delta v|<3000\rm\,km\,s^{-1}$ from the redshift of each quasar to avoid proximity effects.
The resulting sample includes 467 candidate absorbers in the redshift range $z\approx 2-4$. To reduce the subjectivity of human classification performed at this stage, candidates absorbers are vetted by two authors (MG and RD) independently. Since the galaxies targeted in the optical band observed by MUSE lie at $z>3$, we further restricted our sample to the systems above this redshift, for a total of 332 candidate \civ\ absorbers.

The availability of high resolution spectra, with intrinsic widths of the lines often resolved, allows us to model the absorption lines as Voigt profiles, see Figure~\ref{fig:CIVlines} for a few examples. Once the spectra are continuum-normalized (see \citealp{Lofthouse2019} for the detailed procedure), we fit each candidate doublet with Voigt profiles by running the MC-ALF code \citep{Fossati2019}. Using a Bayesian formalism in which the likelihood is sampled with a nested sampling algorithm and the best model is chosen via the Akaike Information Criterion, this code models the absorption profiles assuming an initial number of components within an interval based on visual inspection of each line (e.g, typically ranging between $\approx5-10$ \AA\ for most profiles) and then estimates the minimum number of Voigt components required to reproduce the observed flux. Each component is defined by the redshift, the Doppler parameter and the column density. We set priors restricting the line parameters between $ 1\leq b/(\mathrm{km\,s^{-1}})\leq30$ for the Doppler parameter and $12\leq \log\,(N/{\rm \,cm^{-2}})\leq 16$ for the column density. To account for possible uncertainties in continuum-fitting, we multiply the normalized continuum by a constant (allowed to vary in the range $0.98-1.02$) which is included as a free parameter. To improve the quality of the fit, we allow the spectral resolution to vary slightly around the nominal value. Specifically, the range extends over $\rm \pm1\,km\,s^{-1}$ and $\rm \pm5\,km\,s^{-1}$ for high-resolution (HIRES, UVES, MIKE) and moderate-resolution (X-SHOOTER, ESI) spectra, respectively. The model is then convolved with a Gaussian kernel to match the line spread function of the observed data. To reproduce the shape of blended absorbers, we introduce filler Voigt components that account for absorption lines physically uncorrelated with \civ\ doublets. 

Once the Voigt model is computed for all the candidate doublets, we identify the significant detections by computing the rest-frame equivalent width for all the posterior samples returned by the chains of each MC-ALF fit. Filler components are excluded. Based on this estimate, we inspect again all the candidate doublets and consider significant detections those systems with an equivalent width of the weakest line, $W_{\rm r}^{1550}$ that, once compared to its error, is above $3\sigma$. This step leaves us with 264 systems at $z>3$. All the 
selected doublets are visually inspected one last time to manually exclude heavy blended lines for which we cannot reliably reconstruct a unique model. We also exclude the weakest features that passed the automatic selection but appear to be arising from imperfect continuum normalization and for which the doublet line ratio deviates from the modelled one set by the ratio of the oscillator strengths. 
The final sample includes 220 \civ\ absorption-line systems with redshift $3.0\lesssim z\lesssim4.3$ (median $ z\approx3.31$; see Figure~\ref{fig:SampleProperties}). In assembling this sample, all the components found within a velocity separation of $\pm500\rm\,km\,s^{-1}$ from the highest column density one are considered as an individual absorption system. We show a summary of the final sample in Figure~\ref{fig:CIVzpath}, where the doublets are plotted in the CIV redshift path of each sightline and sized as a function of the rest frame equivalent width. For each \civ\ system in the catalogue we provide a measure of the rest-frame equivalent width (median $W_{\rm r}^{1548}\approx0.089$ \AA) and a measure of the kinematics traced by the line width in velocity space, defined by the interval enclosing the $\rm 90\%$ of the optical depth (median $\Delta v_{90}\approx74.1\rm \,km\,s^{-1}$). The \civ\ absorption systems included in this sample and their properties (redshift, $W_{r}$ and $\Delta v_{90}$) are listed in Table \ref{tab:FullTable_CIV} and available as online material (see .

\begin{table*}
\centering
\begin{tabular}{ccccc}
\hline
Sightline & Instrument & $\mathrm{z}$ & $\mathrm{W_{1548}\,(\text{\AA})}$ & $\mathrm{\Delta v_{90}\,(km\,s^{-1})}$ \\
\hline
J010619+004823 & MIKE & 3.3232 & $1.108 \pm 0.008$ & $247.707 \pm 4.198$ \\
J010619+004823 & MIKE & 3.7289 & $1.146 \pm 0.013$ & $285.509 \pm 14.692$ \\
J010619+004823 & MIKE & 4.0931 & $0.093 \pm 0.007$ & $37.800 \pm 8.400$ \\
J010619+004823 & ESI & 4.2354 & $0.188 \pm 0.006$ & $229.808 \pm 10.793$ \\
J012403+004432 & UVES & 3.0650 & $0.393 \pm 0.003$ & $155.025 \pm 2.501$ \\
\hline
\end{tabular}
\caption{List of the properties ($z,\,W_{1548},\,\Delta v_{90}$) measured for the \civ\ absorption line systems detected at $3\sigma$ and included in the sample studied in this work. We also provide information about the sightline and the spectrum in which each absorber is detected. Only the first five lines are shown here, the full table is available as online supplement.}
\label{tab:FullTable_CIV}
\end{table*}

\begin{figure} 
\centering
\includegraphics[width=\columnwidth]{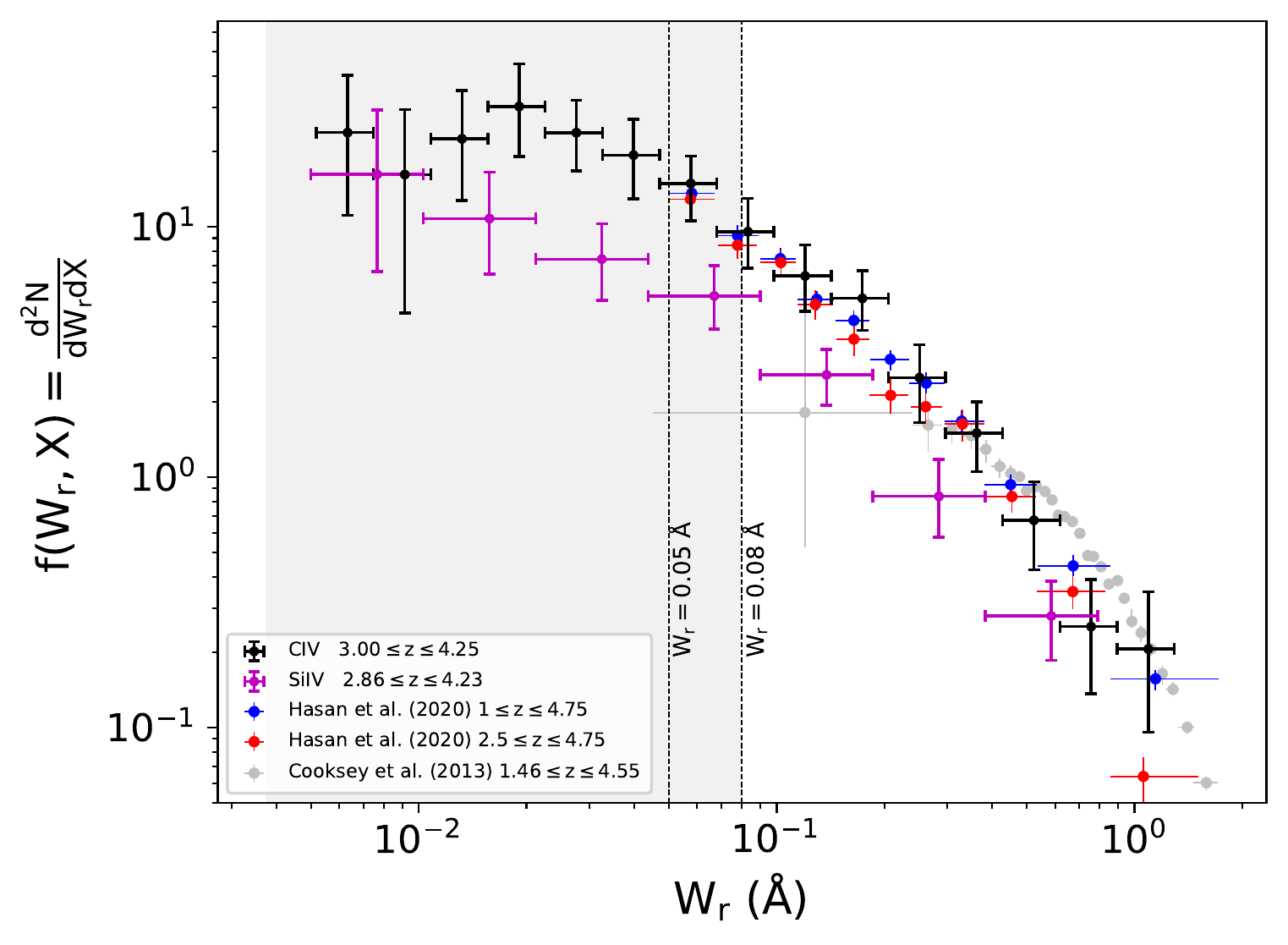}
\caption{Equivalent width frequency distribution function $f(W_{\rm r,1548},X)$ for the MAGG \civ\ sample (black dots). Vertical error bars are the $10^{\rm th}$ and $90^{\rm th}$ percentiles from bootstrap re-sampling, while horizontal error bars reproduce the width of each redshift bin. The MAGG sample is compared with the completeness-corrected results from \citet{Hasan2020} (red and blue dots) and \citet{Cooksey2013} (grey dots). The shaded region mark the equivalent width interval over which our search suffers from incompleteness. Also shown (magenta dots), the equivalent width distribution for the \siv\ line at 1393\AA.}
\label{fig:CIVfwx}
\end{figure}

To assess the completeness of this sample we compare the final observed equivalent width distribution from this search to known completeness-corrected functions from the literature.
Specifically, we compute the equivalent width frequency distribution, $f(W_{\rm r},X)$, which counts the number of \civ\ systems per unit equivalent width per unit survey co-moving path, $X$. The uncertainties on the estimate of $f(W_{\rm r},X)$ are derived as the 10$^{\rm th}$ and 90$^{\rm th}$ percentiles of the distribution obtained by bootstrapping over the sightlines with $10^{3}$ repetitions. 
To gauge the completeness of our search, in Figure~\ref{fig:CIVfwx} we compare the observed equivalent width frequency distribution obtained for the \civ\ line at 1548\AA\ in MAGG, as a function of the rest-frame equivalent width, with the completeness corrected function derived by \citet{Hasan2020} and \citet{Cooksey2013}, which are selected as significant examples in the literature in terms of statistical modelling of $f(W_{\rm r},X)$. 
In detail, \citet{Hasan2020} inspected 369 quasar sightlines at $1.1\leq z_{\rm qso}\leq5.3$ observed with Keck/HIRES and VLT/UVES, compiling a sample of 1318 absorbers in the redshift range $1.0 \leq z \leq 4.75$ that is $50\%$ complete at $W_{\rm r}\geq0.05$ \AA. \citet{Cooksey2013} searched $\approx 26000$ sightlines from the SDSS DR7 quasar catalogue at $ z_{\rm qso}\geq1.7$. The resulting sample is a collection of 16459 absorbers with $1.46 \leq z \leq 4.55$ that is $50\%$ complete at $W_{\rm r}\geq0.6$~\AA. 

According to the comparison shown in Figure~\ref{fig:CIVfwx},
our $f(W_{\rm r},X)$ is in good agreement with the observations reported in the literature. 
Overall, this good agreement validates the procedures adopted to extract our \civ\ sample. We also note that the slope of MAGG data starts decreasing around $ W_{\rm r}\approx 0.08$ \AA\ and flattens significantly at lower column densities, a feature we attribute to the progressively-increasing incompleteness of our sample. Moreover, the MAGG quasar spectroscopy and the data from \citet{Hasan2020} are comparable in quality, both being derived from echelle data at $S/N\gtrsim 10$. Therefore, if we take as reference the sensitivity function shown in figure 4 from \citet{Hasan2020} where the $50\%$ completeness limits are highlighted for different redshift ranges, we can assume that the MAGG \civ\ is approximately $50\%$ complete around an equivalent width of $W_{\rm r}^{1548}\approx 0.05$~\AA, and generally complete for $W_{\rm r}^{1548}\gtrsim 0.08$~\AA.

\subsection{Catalogue of \siv\ absorption lines}

With the goal of identifying an additional tracer of ionized gas comparable to \civ\ in the wavelength range covered by our observations, the same strategy developed to assemble the catalogue of \civ\ doublets is implemented to search and analyze \siv\ $\lambda\lambda$ 1393, 1402 absorption-line systems. A first sample of 145 \siv\ absorbers with redshift $2.9\lesssim z\lesssim 4.4$ is built by visually inspecting the quasar spectra. Running MC-ALF code with the same priors and setting defined for \civ\ doublets, we fit each of these lines and estimate the minimum number of Voigt components required to reproduce the observed flux. We exclude all the candidates that do not show a characteristic rest wavelength separation of $\approx9$ \AA\ and optical depth ratio $\tau_{1393}(\nu)/\tau_{1402}(v)\approx2$ (as set by the ratio of the oscillator strengths) between the two transitions of the doublet in the unsaturated regime. The final sample includes 108 \siv\ systems with a rest-frame equivalent width of the weakest part of the doublet, $W_{\rm r}^{1402}$, detected above $3\sigma$. 

For each absorber we provide a measure of the redshift (median $z\approx3.45$), rest-frame equivalent width (median $W^{1393}_{\rm r}\approx0.077$ \AA) and the width in velocity space (median $\Delta v_{90}\approx64.7\,\rm km\,s^{-1}$). The properties of these systems are listed in Table \ref{tab:FullTable_SiIV} and available as online material, while a summary of the distribution of these properties is shown in Figure~\ref{fig:SampleProperties}, with median values listed in Table \ref{tab:Property_Metals}. To estimate the completeness of the sample, we derive the equivalent width frequency distribution, $f(W_{\rm r},X)$, which we compare with the \civ\ one  in Figure~\ref{fig:CIVfwx}. We note that the shape of the \siv\ is similar to what we obtained for the \civ\ sample, and we conclude that the completeness estimates above can be also applied to the \siv\ catalogue, which we take to be $\approx50\%$ complete for $W_{\rm r}^{1393}\gtrsim0.05$~\AA, and complete for $W_{\rm r}^{1393}\gtrsim0.08$~\AA.

\begin{table*}
\centering
\begin{tabular}{ccccc}
\hline
Sightline & Instrument & $\mathrm{z}$ & $\mathrm{W_{1548}\,(\text{\AA})}$ & $\mathrm{\Delta v_{90}\,(km\,s^{-1})}$ \\
\hline
J010619+004823 & MIKE & 4.2348 & $0.065 \pm 0.009$ & $55.649 \pm 10.501$ \\
J012403+004432 & UVES & 3.3922 & $0.106 \pm 0.002$ & $79.997 \pm 2.500$ \\
J012403+004432 & UVES & 3.5488 & $0.321 \pm 0.002$ & $197.532 \pm 2.501$ \\
J012403+004432 & UVES & 3.6755 & $0.388 \pm 0.002$ & $115.727 \pm 1.874$ \\
J012403+004432 & UVES & 3.7661 & $0.057 \pm 0.001$ & $105.014 \pm 1.250$ \\
\hline
\end{tabular}
\caption{Same as Table \ref{tab:FullTable_CIV} for the \siv\ absorbers detected at $3\sigma$. The first 5 lines are shown, while the full table is available as online material.}
\label{tab:FullTable_SiIV}
\end{table*}

\begin{table}
\centering
\begin{tabular}{cccc}
\hline
Property & $z$ & $W_{\rm r}$ (\AA) & $\Delta v_{90}\rm\,(km\,s^{-1})$ \\
\hline
\civ & 3.31 & 0.089 & 74.1 \\
\siv & 3.45 & 0.077 & 64.7 \\
\hline
\end{tabular}
\caption{Summary of the median properties of \civ\ and \siv\ absorbers, where $W_{\rm r}$ refers to the strongest part of each doublet.}
\label{tab:Property_Metals}
\end{table}

\subsection{Identifying galaxies in MUSE data}

To link the ionized gas detected in absorption via \civ\ with the surrounding galaxy population, we proceed to compile catalogues of galaxies detected in the MUSE data. The first step is to identify continuum-bright galaxies. \cite{Lofthouse2019} and \maggiv\ provide extensive details on how continuum sources are handled. Briefly,  we first run {\sc sextractor} \citep{Bertin1996} on the reconstructed white-light image obtained by collapsing the cube along the wavelength axis. 
We then extract 1D spectra from the cubes following the 2D segmentation maps generated by {\sc sextractor} and use the M. Fossati branch\footnote{matteofox.github.io/Marz} of the {\sc marz} tool \citep{Hinton2016} to measure spectroscopic redshifts for the extracted sources.  
At the end of this procedure, we identify over 1200 sources with redshift, corresponding to $90\%$ completeness down to $\approx 24.85$~mag. 

With the purpose of extending the study of the CGM at high redshift to lower-mass star-forming galaxies, we next search the MUSE field of view for LAEs, here defined as any galaxy emitting bright $\rm Ly\alpha$ emission lines on a typically faint or non-detected continuum. Following the procedure described by \citet{Lofthouse2019} and \maggiv, we first run the three-dimensional automatic extraction performed by {\sc CubEx}
to identify connected groups of voxels (three-dimensional pixels) that lie above a defined $S/N=3$ on the cubes with subtracted continuum sources, including both galaxies and the quasar point spread function (PSF). For the subtraction of the PSF, we follow the non-parametric method detailed in section 3.1 of \citet{Arrigoni2019}. 

The presence of residuals from the continuum of stars and  galaxies is mitigated by masking the spatial position of all the continuum-detected sources with known redshift (see \citealp{Borisova2016}). 
This process produces a set of cubes free of any continuum contamination which can be used to build a sample of compact LAEs. 

More specifically, to be identified by the {\sc CubEx} algorithm, each line emitter candidate requires: i) a minimum volume, imposed by a number of voxels $\geq27$ above the threshold $S/N$; ii) a lower limit of the width in the wavelength direction of at least at least 3 wavelength channels; iii) an upper limit in the wavelength direction of 20 channels to exclude any possible contamination from residuals arising in continuum sources. Since these selection criteria do not remove all the contaminants, such as residual cosmic rays, visual inspection of each extracted LAE is required to weed out the remaining contaminants. Compact LAEs are also distinguished from residuals of sky lines, significant noise fluctuations at the edge of the field or lower redshift emission lines (typically, \civ\ $\lambda\lambda$ 1548, 1550; \ion{C}{III}] $\lambda\lambda$ 1907, 1909; [\ion{O}{II}] $\lambda\lambda$ 3727, 3729; [\ion{O}{III}] $\lambda$ 5008; \ion{H}{$\beta$} $\lambda$ 4862). In order to assure the best quality sample, a first classification is performed by two authors (MG and EL) and confirmed by other two authors (MF and MFo), independently.

\begin{figure} 
\centering
\includegraphics[width=\columnwidth]{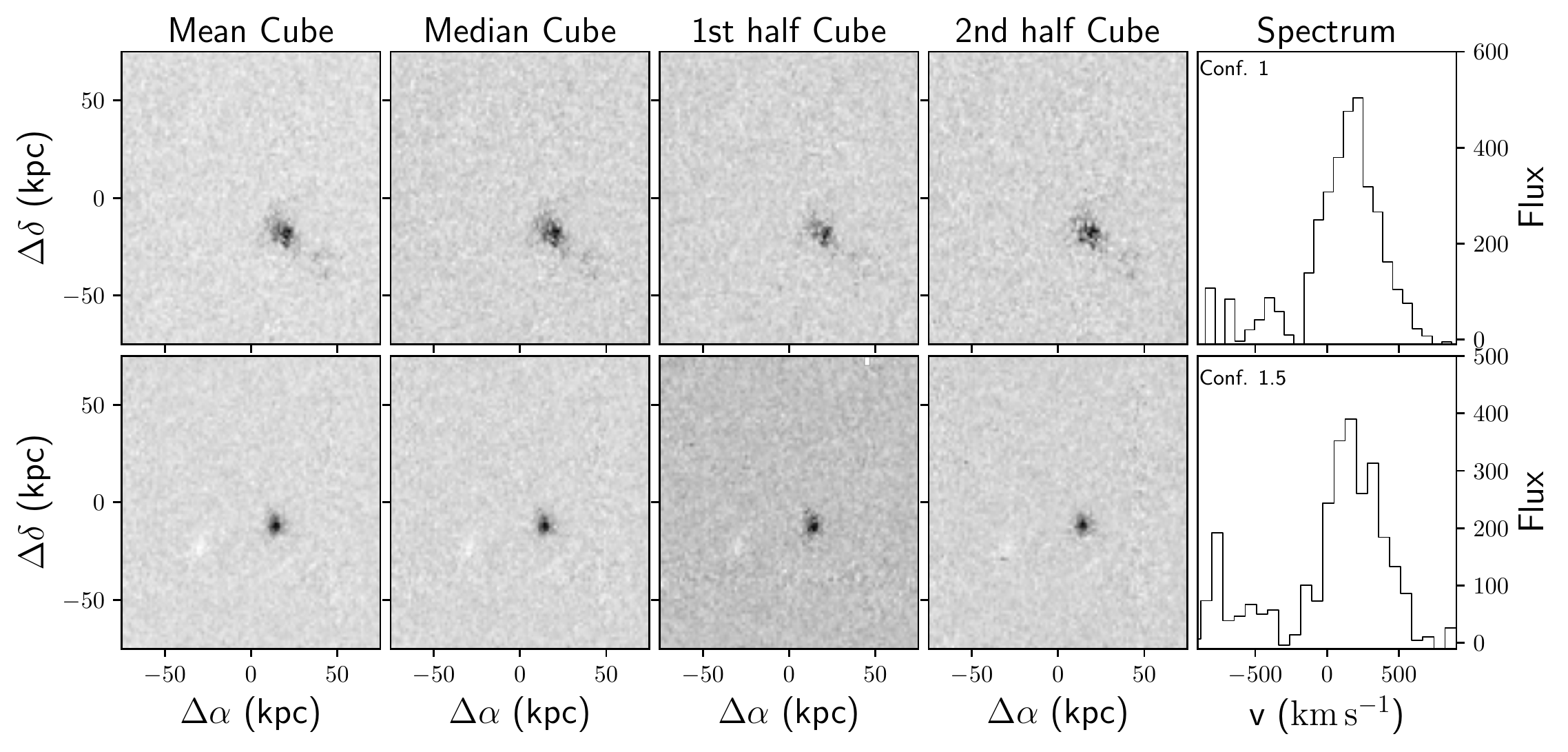}
\caption{Examples of high-confidence LAEs identified in the fields. Columns from left to right show the images extracted by {\sc CubEx} in the mean, median and the half-exposure cubes. The last column shows the 1D spectra limited to the wavelength channels centered on the redshift of the $\rm Ly\alpha$. The flux is measured in units of $\rm 10^{-20}\,erg\,s^{-1}\,cm^{-2}$ \AA$^{-1}$.}
\label{fig:LAEexample}
\end{figure}

The visual inspection of the 28 MAGG fields results in a sample of 994 LAEs at redshift $2.81\lesssim z\lesssim6.60$ and detected at integrated $S/N\geq7$. All the identified emitters are divided into three confidence levels, which updates the original classification by \citealp{Lofthouse2019} based on the integrated signal-to-noise ratio (ISN) and confidence of the classification: i) \textit{Confidence 1} class contains emitters with $ISN\geq7$ which are confidently detected and unambiguously recognized as LAE due to e.g. asymmetric line profiles, or presence of additional absorption lines other than Ly$\alpha$;
the \textit{confidence 1.5} sub-class contains instead confidently-detected sources at $ISN\geq7$ that are recognized as LAEs but for which there is less confidence in ruling out lower-redshift sources (e.g. with double peaked profiles possibly mimicking [OII] emitters but not recognized as such due to the lack of other oxygen lines or with atypical line separations/ratios); 
ii) \textit{Confidence 2} emitters have $ISN<7$ but are deemed to be LAEs, with a sub-class, marked as \textit{confidence 2.5}, containing any system that is noisy and shows anomalies due to noise in its segmentation map and half-exposure cubes; 
iii) \textit{Confidence 3} emitters are recognized as LAEs but lie at the edge of the FoV in noisy regions or only partially overlap with the detector, making any measurement of their properties (e.g. integrated flux, centroid) unreliable. Figure~\ref{fig:LAEexample} provides an example of sources classified under class 1 and 1.5, which are used in this work. Based on this classification, we assembly a sample of LAEs detected at higher integrated S/N ($ISN\geq7$) compared to the \maggiv\ sample ($ISN\geq5$) by excluding all the emitters with confidence $\rm\geq2$ from the following analysis. The final sample includes 921 high-confidence LAEs.

\begin{table*}
\centering
\begin{tabular}{cccccccc}
\hline
Field & $\mathrm{RA\,(deg)}$ & $\mathrm{DEC\,(deg)}$ & $\mathrm{z}$ & $\mathrm{\log [L_{Ly\alpha}/(erg\,s^{-1})]}$ & $\mathrm{R\,(pkpc)}$ & SNR & Confidence \\
\hline
J010619+004823 & 16.5788 & 0.8118 & 2.9263 & $41.620 \pm 0.083$ & 157 & 8.3 & 1.0 \\
J010619+004823 & 16.5834 & 0.8042 & 3.1082 & $42.289 \pm 0.023$ & 112 & 20.4 & 1.0 \\
J010619+004823 & 16.5860 & 0.8001 & 3.1213 & $41.627 \pm 0.099$ & 242 & 8.5 & 1.0 \\
J010619+004823 & 16.5834 & 0.8075 & 3.2081 & $42.033 \pm 0.036$ & 95 & 13.9 & 1.0 \\
J010619+004823 & 16.5782 & 0.8011 & 3.2112 & $42.169 \pm 0.044$ & 159 & 15.0 & 1.0 \\
\hline
\end{tabular}
\caption{List of the LAEs detected in MAGG at $SNR\geq7$. Information of the sighline, coordinates, SNR and confidence from our classification are included as well as the properties we measured (redshift $z$, luminosity $\log[L_{\rm Ly\alpha}/(\rm erg\,s^{-1})]$, projected distance to the quasar $R$). Typical errors for redshifts are of the order of $(6-7)\times 10^{-4}$. Only the first five lines are shown here: the full table is provided as online material}.
\label{tab:FullTable_LAE}
\end{table*}

For each LAE included in the final MAGG sample, we derive a measure of the redshift, $\rm Ly\alpha$ luminosity and projected distance from the line-of-sight (i.e., the impact parameter). For this purpose, from the 3D segmentation cubes produced by {\sc CubEx}, the 1D spectra is extracted along the full wavelength range and used to derive an estimate of the redshifts. $\rm Ly\alpha$ photons are known to be subject to radiative transfer effects which affect both their spatial and frequency distribution, resulting in typical asymmetric or double-peaked emission lines. Therefore, we apply the following convention \citep[see also][\maggiv]{Verhamme2018,Muzahid2020}: in case of a double-peaked line, we take the redshift of the red peak; if a single peak is observed instead, we assume that this is the red peak since the blue one is more easily absorbed. 

To derive the $\rm Ly\alpha$ luminosity we adopt instead a curve of growth (CoG) analysis. This method better adapts to the typical $\rm Ly\alpha$ emission profile that is spatially extended and characterized by two distinct components: a bright core and a faint diffuse halo \citep{Wisotzki2016, Leclercq2017}. The latter is easily excluded in the flux measured from the segmentation cube at moderate $S/N$. Here, we follow the method described by \citet{Fossati2019} (see also \citealp{Marino2018}). For each emitter we build a pseudo narrow-band image by summing up the spectral channels within $\pm 14$ \AA\ from the source redshift and masking the neighbours. A series of circular apertures with increasing radii, over which we produce the CoG flux, is centered on the {\sc CubEx} coordinates of each source. As a result, we take as reliable CoG flux the estimate obtained at the last radius where the total flux increases by more than $2.5\%$ than the previous step. All the diagnostic plots are visually inspected to check that no contamination other than the LAEs has been included in the fluxes. 
Fluxes are then corrected for Milky Way extinction using the re-calibrated extinction map \citep{Schlafly2011} and assuming the Milky Way extinction law by \citet{Fitzpatrick1999}.
Finally, we compute the luminosity distance at the redshift of each source and convert the total flux into a $\rm Ly\alpha$ luminosity. A complete list of the $\rm Ly\alpha$ emitters identified in the MAGG survey at $SNR>7$, including a measure of the redshift, $Ly\alpha$ luminosity and impact parameter, is provided in Table \ref{tab:FullTable_LAE} and available as online material.

In the whole MAGG sample, we identify 292 LAEs lying in the \civ\ redshift path. This subset also includes Ly$\alpha$ emission from 23 galaxies detected in continuum and forms the candidates for associations with the absorbers. Although we provide a complete survey of UV-selected galaxies, including continuum-detected LBGs and LAEs, our sample significantly favors line emitters. Thus, additional $z>3$ sources, e.g. LBGs with absorption features but not Ly$\alpha$ emission (1 detected in \civ\ redshift path), passive or heavily obscured galaxies, are not well represented in this work. In Figure~\ref{fig:SampleProperties} and Table~\ref{tab:Property_LAEs} we summarize the properties measured for LAEs (redshift and $\rm Ly\alpha$ luminosity) in the full MAGG sample and in the sub-set lying in the \civ\ redshift path. The latter are observed in a redshift range $2.80\lesssim z\lesssim4.23$, have a median $ z\approx3.24$ ($ z\approx3.91$ for the full MAGG sample) and emit a luminosity $ 41.39\lesssim \log\,(L_{\rm Ly\alpha}/\rm erg\,s^{-1})\lesssim42.87$, with a median $ \log\,(L_{\rm Ly\alpha}/\rm erg\,s^{-1})\approx42.07$ (median $ \log\,(L_{\rm Ly\alpha}/\rm erg\,s^{-1})\approx42.10$ for the full MAGG sample). The emitters are detected at a projected distance from the line-of sight $ 20\lesssim R/\rm kpc\lesssim 295$, with a median $ R\approx166\,\rm kpc$ ($ R\approx160\,\rm kpc$ for the full MAGG sample).

\begin{table}
\centering
\begin{tabular}{cccc}
\hline
Property & $z$ & $\log[L_{\rm Ly\alpha}/(\rm erg\,s^{-1})]$ & $R\rm\,(kpc)$ \\
\hline
Full Sample     & 3.91 & 42.10 & 160 \\
In \civ\ z-path & 3.24 & 42.07 & 166 \\
\hline
\end{tabular}
\caption{Summary of the median properties of LAEs reported for the full sample and for the emitters detected in the \civ\ redshift path.}
\label{tab:Property_LAEs}
\end{table}

\subsection{Identifying galaxy groups}\label{sec:groups}

According to the hierarchical scenario of structure formation, galaxies assemble in groups or clusters, with only a fraction evolving in isolated environments. In dense galaxy environments, interactions between galaxies and with the surrounding medium are observed to affect the properties of different gas phases of the CGM, both at low and high redshift (see e.g., \citealp{Nielsen2018, Fossati2019, Dutta2020, Muzahid2021}). Therefore, a better understanding of the gas-galaxy connection and co-evolution requires galaxy groups to be identified in the MAGG survey. For this purpose, without including any constraint on the halo mass, we recognize a galaxy to be part of a group if it is not isolated within a line-of-sight separation $|\Delta v|\leq500\,\rm km\,s^{-1}$ in the MUSE FoV. By linking galaxies with this criterion, we identify 53 groups in the CIV absorbers redshift path shown in Figure~\ref{fig:CIVzpath} (152 in the full MAGG sample) so that $\approx47\%$ of the galaxies are part of a group ($\approx41\%$ for the full MAGG sample). The majority of these identified groups ($\approx57\%$) are composed of two galaxies ($\approx61\%$ for the full MAGG sample), but  groups of three to six members (seven for the full MAGG sample) are also found. 

\section{Results}\label{sec:result}

\subsection{Measuring correlations between \civ\ absorbers and LAEs}

\subsubsection{Connecting \civ\ absorbers with galaxies}

\begin{figure} 
\centering
\includegraphics[width=\columnwidth]{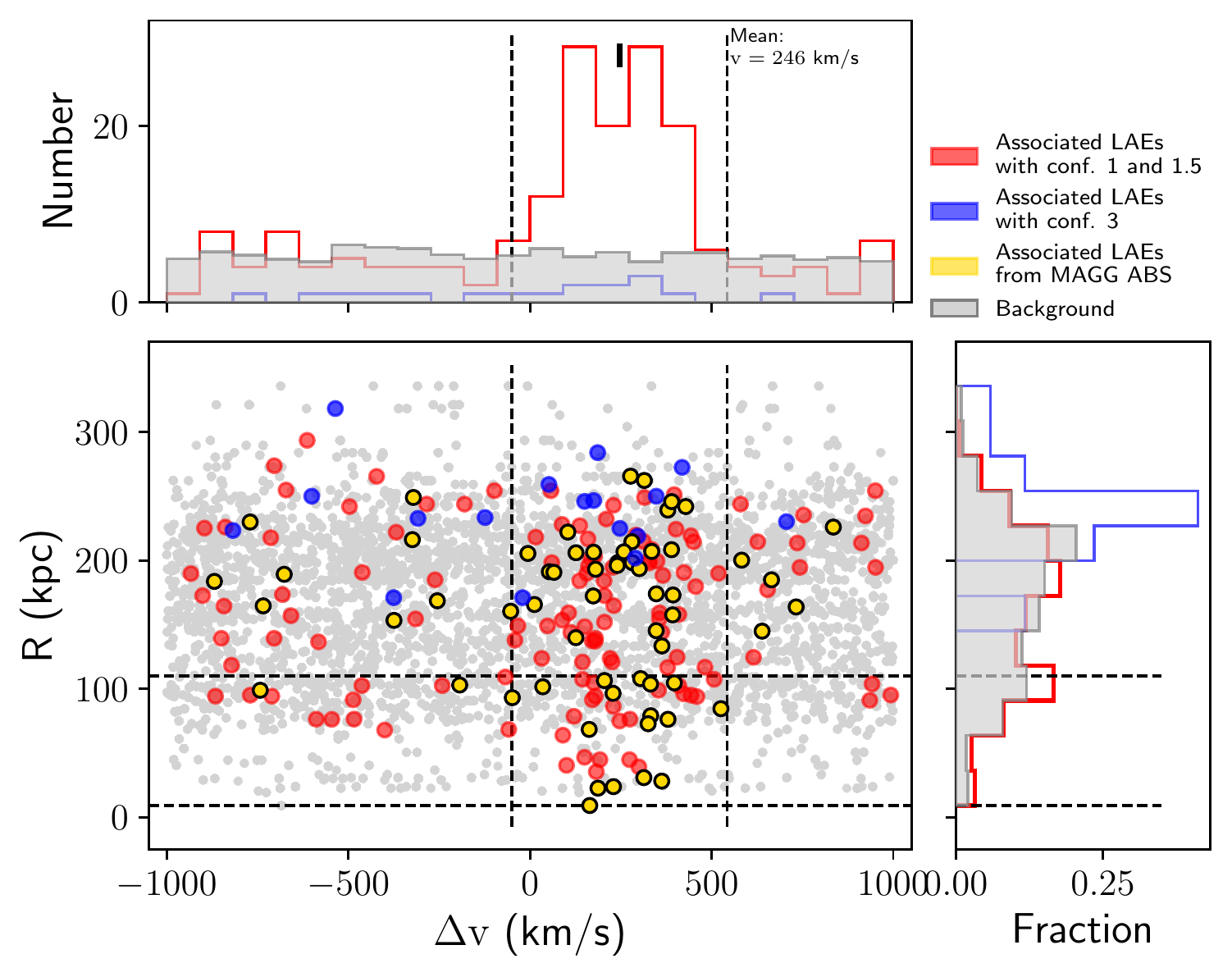}
\caption{Distributions of LAEs near \civ\ absorbers as a function of line-of-sight velocity separation and projected distance. The high-confidence LAEs are marked in red, while the galaxies at the edge of the field are in blue. In gold, we mark the galaxies connected to \civ\ absorbers that are also associated to high-column hydrogen absorbers from the \maggiv\ paper. A random
 population of LAEs, which is representative of background and foreground population at larger velocity separations from \civ\ absorbers, is shown for reference in grey. The dashed lines highlight the regions where a clear excess signal of LAEs around \civ\ is detected over the uncorrelated background.}
\label{fig:CIVGalConnection}
\end{figure}

To establish the extent with which the detected \civ\ trace ionized gas in proximity of LAEs, we cross-match the catalogue of \civ\ absorbers with the one of the LAEs in redshift space. For this purpose, we center on each absorption system and we search for the galaxies within $\pm 1000\rm\,km\,s^{-1}$ along the line of sight, reserving the definition of a more stringent criterion (i.e., a smaller velocity separation) for the following analysis. In Figure \ref{fig:CIVGalConnection} we show the gas-galaxy connections as a function of the line-of-sight separation (horizontal axis) and as a function of the transverse distance (vertical axis). We distinguish between the LAEs detected with high confidence (red dots) and those observed at the edge at the field of view (blue dots). We also highlight the galaxies associated to \civ\ absorbers that are  connected to Lyman limit systems (LLSs) from the \maggiv\ paper (gold dots). This sample includes 127 LAEs associated to 61 absorption-line systems detected at $S/N\geq5$, 79 detected at $S/N\geq7$. As $\approx70\%$ of these LAEs are in the vicinity of a \civ\ absorber, this fraction is also independently recovered by our search. 

In order to characterize the significance of the gas-galaxy connection, we build a control sample (in grey within Figure~\ref{fig:CIVGalConnection}) which aims at reproducing the typical  distribution of the galaxies around random regions in the Universe not selected in absorption, where we do not necessarily expect \civ\ systems. To do so, we bootstrap over the sightlines, randomly selecting 28 fields from the full MAGG sample for a total of $10^{3}$ realizations. At each iteration and in each selected field, we measure the distance of LAEs to the redshift of a real \civ\ absorber, which is however randomly extracted from all the systems detected in the remaining sightlines. Using this technique, we are effectively building a control sample using real data (both for LAEs and \civ\ absorbers), optimally controlling for systematics like the presence of sky line residuals. 

Examining the distribution of velocities along the line of sight in the upper histogram of Figure \ref{fig:CIVGalConnection}, we observe a clear excess of LAEs near \civ\ absorbers compared to the random sample distribution at offset velocities of $\Delta v = 0-500\rm\,km\,s^{-1}$. Assuming that the galaxies are randomly distributed relative to the line of sight, one would expect to find a relative velocity separation between LAEs and \civ\ that is symmetric around the \civ\ redshift. However, being based on the Ly$\alpha$ emission line, our estimate of the galaxies redshift suffers from the resonant scattering that affects the frequency and spatial distribution of the Ly$\alpha$ photons emitted by the galaxies. To measure this offset, we fit the velocity separation distribution with a Gaussian model plus a background as a function of the velocity separation, obtaining a mean value of $\Delta v_{\rm Ly\alpha}\approx246\rm\,km\,s^{-1}$. 
Similar trends are indeed observed in the literature. \citet{Rakic2011} measured an offset $\Delta v_{\rm Ly\alpha}\approx(295\pm30)\rm\,km\,s^{-1}$ between $z\sim2.3$ LBGs and stacked \hi\ absorption lines. More recently, \citet{Muzahid2020, Muzahid2021} derived a velocity offset of the order of $\Delta v_{\rm Ly\alpha}\approx(171\pm8)\rm\,km\,s^{-1}$ between LAEs and stacked \hi\ absorption at redshift $z\approx3$. They also obtained a similar velocity offset, although with larger uncertainties, between LAEs and stacked \civ\ profile. Lastly, the velocity separation between LAEs and \hi\ absorbers observed in \maggiv\ peaks at $\Delta v_{Ly\alpha}\approx250\rm\,km\,s^{-1}$. In light of this result, in the following we establish associations by correcting the redshift of the LAEs for the mean offset $z_{\rm LAE}^\prime=z_{\rm Ly\alpha}-\Delta v_{\rm Ly\alpha}\,(1+z_{\rm Ly\alpha})^{-1}\,c^{-1}$. 

Since the majority of the LAEs is observed within $0-500\rm\,km\,s^{-1}$ from the absorber, we can restrict the velocity window in which a galaxy is considered associated with an absorber to build a sample of connected systems. In the following, we thus perform our analysis on galaxies that are $\pm500\rm\,km\,s^{-1}$ from a \civ\ absorber, an interval that becomes $-254<\Delta v/(\mathrm{km\,s^{-1}})<746$ once corrected for the Ly$\alpha$ velocity offset. We detect at least one LAE within this velocity separation for 79 \civ\ absorbers, corresponding to a detection rate of $36\pm5$ per cent (79/220).

The right histogram of Figure \ref{fig:CIVGalConnection} shows instead the transverse separation distribution between the \civ\ absorbers and the LAEs. The overall trend is consistent with an increasing number of detections with increasing area of the annulus  ($\propto R^{2}$), and a steep decrease due to edge effects at $R>200\rm\,kpc$. 
Although with a significant scatter, only a very small excess of LAEs is detected at transverse separations $R<100\rm\,kpc$ over the control distribution, suggesting that some, but not many, \civ\ systems preferentially arise from the inner CGM of the associated galaxy. Overall, however, the impact parameter distribution looks similar to the one of the control population.  

In summary, the connection in redshift space between the \civ\ absorbers and the LAEs reveals that a factor of $\approx2.6$ more galaxies are observed within $\pm500\rm\,km\,s^{-1}$ from an absorption line system relative to the number of galaxies detected in random regions of the Universe. This result implies that the \civ\ absorbers do not trace a random region of the Universe, but regions with a preference for hosting LAEs.

\subsubsection{Number density of LAEs associated to \civ\ absorbers}

The detection of a local clustering signal along the line-of-sight is indicative of the existence of a physical connection between the \civ\ absorbers and the LAE population. In this section, we measure statistically the LAE number density by deriving the luminosity function (LF),
$\phi(L)dL$. A comparison of LFs from different environments (e.g., in the field or near \civ\ absorbers) further allows us to study the extent to which the LAE number density depends on the proximity to the absorption-line systems. The procedure we follow in the computation of the LF is described in detail by \citet{Fossati2021}.
We assume any redshift dependence to be negligible in the probed range $3.0<z<4.5$ as, based on the results from \citet{Herenz2019}, no significant redshift evolution is observed in the wide range $2.9<z<6.0$. In the analysis, we compare a non-parametric estimator of the LF, $1/V_{\rm max}$ from \citet{Schmidt1968} and \citet{Felten1976}, with the results obtained from fitting the full sample with no binning using a parametric Schechter function.

The $\/V_{\rm max}$ non-parametric estimator relies on the survey selection function $f_{\rm c}(L,z)$, defined as the probability to find an LAE with a given luminosity at a given redshift. The selection function is derived performing a simulation based on the injection of 500 mock LAEs and measuring the fraction of sources we recover running {\sc CubEx} and applying our search algorithm. The flux, redshift and spatial positions of the mock sources are tabulated. This procedure is repeated for 1000 iterations over all the MAGG fields, with the only exception of J142438$+$225600 that is lensed \citep{Patnaik1992}. 

\begin{figure} 
\centering
\includegraphics[width=\columnwidth]{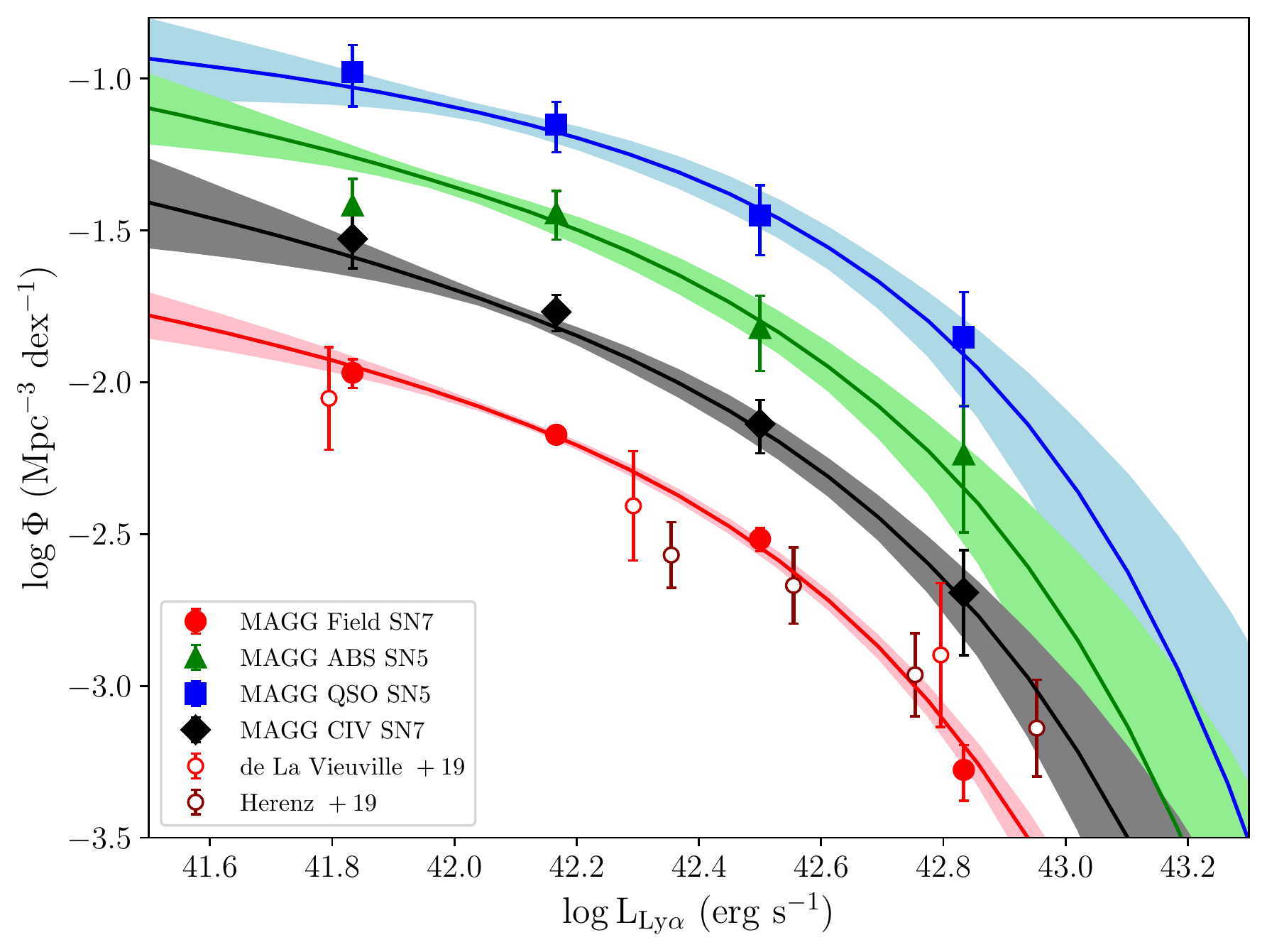}
\caption{Luminosity function derived for the LAEs associated to the \civ\ absorbers (black diamonds) in a luminosity range $41.5<\log(L_{\rm Ly\alpha}/\rm erg\,s^{-1})<43.0$, using non-parametric (data points) and parametric (lines) estimates. The results from this work are compared with the results from a field control sample (red circles), the environment of high-column density \hi\ systems (green triangles) from \maggiv, and the environment around quasars from \citet{Fossati2021} (blue squares). For comparison, we also show the field luminosity function from \citet{Herenz2019} and \citet{deLaVieuville2019}.}
\label{fig:CIV_LF}
\end{figure}

In order to compute the effective co-moving area of the survey, we re-scale the MUSE field area by a factor $\delta_{\rm obs}(z)$ that accounts for the number of fields where we searched for LAEs at a given redshift. Therefore, we center on the redshift of each \civ\ absorber and consider only the contributions from the redshift range enclosed within $\pm500\rm\,km\,s^{-1}$ from the absorber. The factor $\delta_{\rm obs}(z)$ is equal to zero in the ranges excluded from the searched path. The luminosity bins are $0.2\rm\,dex$ wide. We limit the uncertainties on the weight given by $V_{\rm max}$ by excluding from the analysis all the LAEs identified below $f_{c}(L,z)<0.1$. 
The results of the luminosity function using this non-parametric estimate for the LAEs connected to the \civ\ absorbers are shown as black dots in Figure \ref{fig:CIV_LF} with respective uncertainties.  

We also derive a parametric estimate of the LAE luminosity function by fitting a Schechter function \citep{Schechter1976} to the non-binned sample:
\begin{equation}
    \phi(L) = \ln(10)\,\phi^{\star}10^{(\log L-\log L^{\star})(1+\alpha)}\,\exp{[-10^{(\log L-\log L^{\star})}]}\:.
\end{equation}
For the fit, we proceed as described in \citet{Fossati2021}. 
The result from the parametric modelling of the luminosity function is shown as  a solid black line in Figure \ref{fig:CIV_LF} for the LAEs associated to the \civ\ absorbers, with the shaded areas marking the $1\sigma$ confidence intervals. 
In order to study how being in the vicinity of the \civ\ absorbers affects the LAEs number density, we apply the procedure described above to a sample of LAEs  from a random sample (referred to as ``MAGG field'') that is not selected by its proximity to a specific tracer. This field sample is built by including 617 LAEs detected in the range $3.0<z<4.5$, corresponding to the redshift interval in which we observe \civ\ absorption in MAGG. The non-parametric estimate and the parametric Schechter fit are shown in red in Figure \ref{fig:CIV_LF}.

\begin{table*}
\centering
\begin{tabular}{cccccc}
\hline
Sample & S/N limit & $\log [L^{\star}/\rm(erg\,s^{-1})]$ & $\alpha$ & $\log[\phi^{\star}/\rm (Mpc^{-3}\,dex^{-1})]$ & $\log (I/\rm Mpc^{-3})$ \\
\hline
QSO & 5 & 42.573 $\pm$ 0.221 & -1.164 $\pm$ 0.319 & -1.429 $\pm$ 0.276 & -1.239 \\
ABS & 5 & 42.558 $\pm$ 0.242 & -1.350 $\pm$ 0.309 & -1.788 $\pm$ 0.343 & -1.540 \\
\civ& 7 & 42.575 $\pm$ 0.212 & -1.418 $\pm$ 0.323 & -2.175 $\pm$ 0.163 & -1.887 \\
Field  & 7 & 42.471 $\pm$ 0.082 & -1.345 $\pm$ 0.158 & -2.243 $\pm$ 0.123 & -2.262 \\
\hline
\end{tabular}
\caption{Parameters of the Schechter function describing the LAE luminosity function around the different tracers studied in MAGG. The last column is the integral of the Schechter function computed at luminosities $\log[L_{\mathrm{Ly\alpha}}/(\rm erg\,s^{-1})]\geq41.8$.}
\label{tab:fitLF}
\end{table*}

To test the consistency of the field luminosity function with previous determinations, the results are compared to the literature. The estimate from \citet{Herenz2019} is based on a sample of 237 LAEs spread in a redshift range $2.9<z<6.0$ and detected in $\sim1$ hour of MUSE observations. Here we report their results for a sub-sample restricted to the range $3.0<z<4.0$, chosen to be consistent with the \civ\ absorbers detected in MAGG. To extend the comparison at lower luminosities, we consider the results from \citet{deLaVieuville2019}. This sample contains 156 LAEs at $2.9<z<6.7$ with luminosities $39\lesssim\log (L_{\rm Ly\alpha}/\rm erg\,s^{-1})\lesssim43$. As done before, we consider the luminosity function derived for a sub-sample of LAEs detected in the range $2.9<z<4.0$. As we adopt $H_0=67.7~\rm km~s^{-1}~Mpc^{-1}$ while the literature studies chose $H_0=70~\rm km~s^{-1}~Mpc^{-1}$, the luminosity functions derived from the these two samples are scaled to be consistent with our cosmological parameters (a $\lesssim 10\%$ correction). As shown in Figure \ref{fig:CIV_LF}, the luminosity function derived for the MAGG field sample is consistent within the uncertainties with the results from the literature. Furthermore, the large size of the MAGG field sample provides a statistically significant modelling of the luminosity function at $3.0<z<4.5$ for a population of lower-mass star forming galaxies traced by LAEs with halo mass $M_H\approx 10^{11}~\rm M_\odot$ compared to, e.g., continuum detected populations like LBGs with $M_H\approx 10^{12}~\rm M_\odot$ (see Section~\ref{sec:cfrlbg} for additional details).

Having computed the field luminosity function, we derive an estimate of the LAE number density in the vicinity of the \civ\ absorbers extending the comparison with other populations in MAGG. In \maggiv, we studied the luminosity function of 97 LAEs detected within $\pm500\rm\,km\,s^{-1}$ from 45 strong hydrogen absorbers selected as tracers of the dense CGM (the ``MAGG ABS'' sample). We note the MAGG ABS and the MAGG \civ\ samples are partially overlapping (not all \civ\ aborbers are associated to LLSs, see Section~\ref{sec:civandhi}), thus the comparison is not independent. In addition, the ``MAGG QSO'' sample from \citet{Fossati2019} includes 86 LAEs found within $\pm500\rm\,km\,s^{-1}$ from the central quasar of each field. For consistency, these two samples include LAEs detected in a redshift range $3.0\lesssim z\lesssim4.0$ that is comparable to the MAGG \civ\ sample. The use of different $S/N$ thresholds for the various samples ($S/N\geq5$ for MAGG ABS and QSO samples, $S/N\geq7$ for MAGG \civ\ and field samples) is justified by noticing that a higher $S/N$ limit does not change the overall shape either of the non-parametric or the parametric estimate (as expected from the completeness correction based on the survey selection function), but only broadens the $1\sigma$ confidence regions. The best Schechter parameters estimated from the parametric fit are listed in Table \ref{tab:fitLF} for each MAGG sample. 

 The parameters $\log L^{\star}$ and $\alpha$ of the different samples are all consistent within $1\sigma$. This suggests that the overall shape of the LAE luminosity function does not depend on which tracer is used to select the environment within which these galaxies reside. Despite the normalization constant, $\phi^{\star}$, having the units of a LAEs number density, it is not computed at fixed luminosity and thus cannot be directly compared among the different tracers. Therefore, we derive a measure of the relative LAEs overdensity by computing the integral $I$ of the Schechter function over the luminosities $\log[L_{\mathrm{Ly\alpha}}/(\rm erg\,s^{-1})]\geq41.8$. The results are shown in Table \ref{tab:fitLF}, where we find that the cumulative number of LAEs at close separation from the \civ\ absorbers is a factor $\approx2.4$ higher than the field. The MAGG ABS sample is associated to an intermediate LAE overdensity that is a factor $\approx2.2$ higher than the \civ\ sample. Finally, the normalization for the MAGG QSO sample is a factor $\approx4.5$ higher relative to the MAGG \civ\ sample and $\approx2.0$ compared to the MAGG ABS sample, pinpointing a richer overdensity of LAEs.

In conclusion, the analysis of the distribution of LAEs in velocity space and of the luminosity function in different environments underscores a systematic preference for LAEs to cluster around \civ\ absorbers with equivalent width $\gtrsim 0.05~$\AA, implying in turn that \civ\ systems do not always trace average regions of the Universe but, often, regions near LAEs. Indeed, as we will show in detail below, $\approx 36\%$ of the total number of the absorbers is in fact associated to at least an LAE, with $\approx48\%$ of these \civ\ systems being associated to multiple LAEs.


\subsection{Correlation functions}

Having established the presence of an overdensity of LAEs near \civ\ absorbers, we now quantify the connection between the galaxies and the ionized gas statistically, by deriving the LAEs auto-correlation and the LAEs-absorber cross-correlation functions. 
The main goal of this analysis is to establish whether the \civ\ absorbers are physically connected to the galaxies, and actually populate halos, or whether they are preferentially observed in IGM filaments. To do so, following the interpretation of \citet{Adelberger2003, Adelberger2005}, we exploit the Cauchy-Schwarz relation that compares the galaxy-galaxy and the absorber-absorber auto-correlation functions, $\xi_{\rm GG}(r)$ and $\xi_{\rm AA}(r)$ respectively, with the galaxy-absorber cross-correlation function, $\xi_{\rm GA}(r)$:
\begin{equation}
\xi_{\rm GA}^{2}\leq\xi_{\rm GG}\,\xi_{\rm AA}\:.
\label{eq:cauchy_schwarz}    
\end{equation}
This relation can be used to compare the galaxies and absorbers underlying matter density distribution since the equality holds if the \civ\ absorbers and the galaxies originate from density fluctuations that are linearly dependent. A similar approach has been widely adopted in the literature to measure the clustering of the LBGs at redshift $z\sim3$ \citep{Adelberger2003, Adelberger2005, Bielby2011} and establish the correlation of galaxies with various ions, such as \ion{H}{I} and \ion{O}{VI} at redshift $z<1$ \citep{Chen2009, Tejos2014, Finn2016, Prochaska2019} or \civ\ at $z\sim3$ \citep{Adelberger2003, Adelberger2005}. 
Furthermore, by relying on the $k$-estimator \citep{Adelberger2005} that counts the number of pairs without requiring the assembly of a random sample as an alternative to the typical two-point correlation function, \citet{Diener2017} and \citet{Alonso2021} detected, for the LAEs auto-correlation, a clustering signal extended up to $\approx3.0\rm\,pMpc$ in a large sample of emitters at $3.0\lesssim z\lesssim 6.0$. 

\subsubsection{Formalism and random samples}

 The correlation functions are derived as a function of the pairs projected distance $R$ and their separation along the line-of-sight $R_{\rm los}$. The latter is computed as velocity separation and converted into a distance assuming pure Hubble flow. We use the  \citet{Davis1983} estimator to compute both the auto-correlation and the cross-correlation functions, which compares the number of data-data pairs and random-random pairs at a given $R$ and $R_{\rm los}$.
 

\begin{figure} 
\centering
\includegraphics[width=.45\columnwidth]{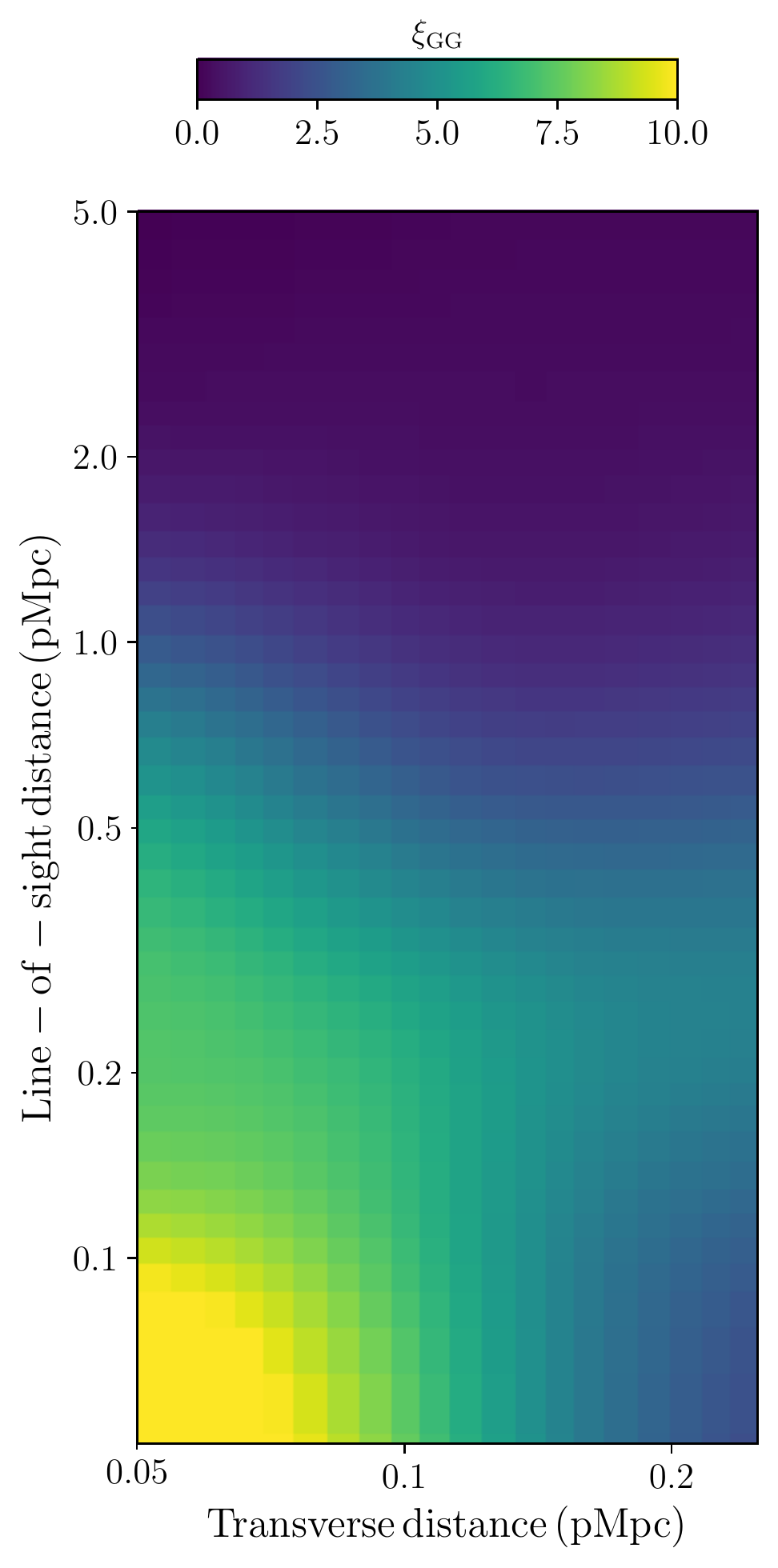}
\includegraphics[width=.45\columnwidth]{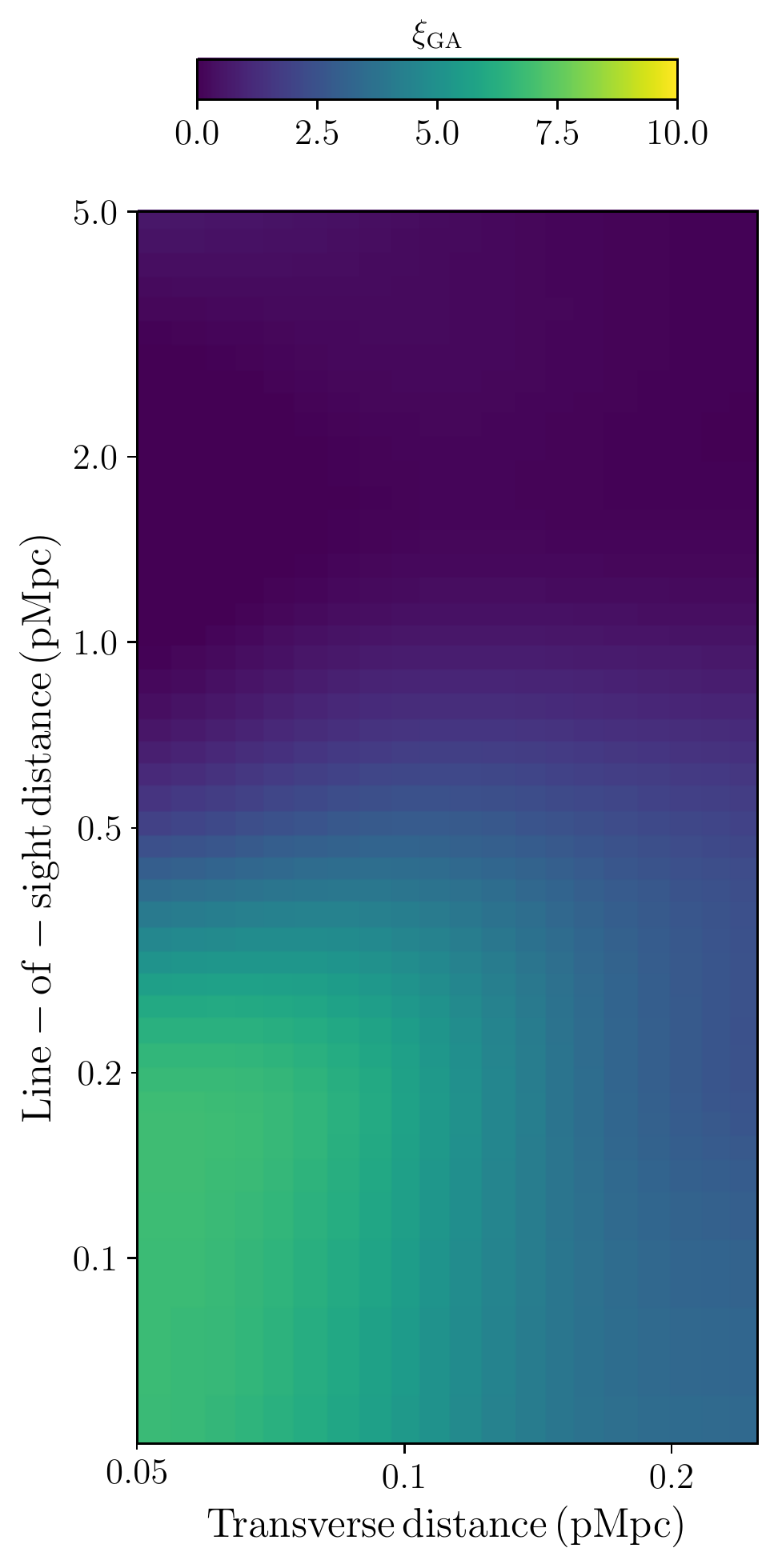}
\caption{Two dimensional auto-correlation function for LAEs (left panel) and LAE-\civ\ cross-correlation function (right panel).}
\label{fig:Corr2D}
\end{figure}


For the computation of the random-random pairs, we proceed as follows. In order to preserve the original geometry of the survey, we perform a random extraction of the galaxy coordinates weighted by the sum of the exposure maps of the 28 MAGG fields. 
The redshift of each mock galaxy is randomly extracted from a flat distribution limited to the range $3.0\leq z\leq4.5$. We require the size of the random sample to be a multiple of the number of real galaxies: $N_{\rm R,g}=\beta_{\rm g}\times N_{\rm D,g}$. The factor $\beta_{\rm g}=3$ is chosen as it is sufficient to populate the coordinates space and optimize the Poissonian fluctuations of the random sample.

 We assemble the random sample of absorbers in a similar fashion. Lying on the line-of-sight, the three-dimensional position of each absorber is defined by the quasar coordinates and the redshift measured from the Voigt fit of the line. To preserve the geometry of the survey for the random sample of absorbers, we avoid the telluric bands and generate the redshift of the mock absorbers by randomly extracting values in the \civ\ redshift path. Given a number $N_{\rm D,a}$ of observed \civ\ systems, we build a sample of $N_{\rm R,a}=\beta_{\rm a}\times N_{\rm D,a}$ random absorbers, where $\beta_{\rm a}=5$ to compensate for the smaller sample size. 
 
\subsubsection{Galaxy auto-correlation}

The galaxy auto-correlation function is derived by counting the number of real galaxies and random pairs in each bin 
in two dimension (transverse distance and line-of-sight separation). The intervals of transverse distances are logarithmically spaced between $0.05-0.50\,\rm pMpc$, where the lower bound accounts for the typical resolved size of the emitters and the upper bound is set by the field dimension. Along the line of sight, we consider pairs with a separation $5\leq\Delta v/\mathrm{km\,s^{-1}}\leq 2500$, corresponding to $0.02\lesssim D_{\rm los}/\mathrm{Mpc}\lesssim8.2$ at $z=3$. To optimally extract the signal, the binned correlation function is oversampled and smoothed with a Gaussian filter of standard deviation $\sigma=5$, corresponding to the oversampling factor. The result is shown in the left panel of Figure \ref{fig:Corr2D}, up to a transverse distance $R=0.25\rm\,pMpc$ to make it comparable with the galaxy-absorber cross-correlation function. A clear clustering signal is detected at a transverse distance $R\lesssim 0.1\,\rm pMpc$ and line-of-sight separation $R_{\rm los}\lesssim 0.1\rm\,pMpc$ as expected for galaxy populations that are not randomly distributed in the Universe. A weaker clustering signal is elongated up to $R_{\rm los}\approx0.5\rm\,pMpc$, likely due to the proper motions of the galaxies.

\subsubsection{Comparison with the galaxy-absorber cross-correlation}

We use the same procedure employed for the galaxy auto-correlation function to derive the LAE-\civ\ correlation function in two dimensions, as a function of the transverse distance and the line-of-sight separation. Since the \civ\ absorbers are detected along the sightline to the central quasar in each field,  we choose the interval $0.05-0.25\rm\,pMpc$ for the transverse separations, where the lower limit accounts for the quasar PSF and the upper limit corresponds to half the size of a field at $z\approx3.0$. The line-of-sight separation between the LAEs and the \civ\ systems are derived by correcting the redshift of the galaxies for the observed offset from the absorbers, $\Delta v\approx246\rm\,km\,s^{-1}$. The cross-correlation function is shown in the right panel of Figure \ref{fig:Corr2D}. At transverse distances $R\lesssim0.1\rm\,pMpc$ and line-of-sight separations $R_{\rm los}\lesssim0.5\rm\,pMpc$, that is at small separations from a \civ\ absorber, the probability to observe an LAE is enhanced  with respect to a random point in the field, reinforcing the evidence that the LAEs are clustered to \civ\ absorbers on this scale. 
Both the galaxy auto-correlation and the galaxy-absorber cross-correlation functions are elongated in redshift space, possibly due to the galaxies proper motions and the gas proper motions relative to the galaxies, as it is typically observed in the literature \citep{Adelberger2003,Adelberger2005,Turner2014,Alonso2021}. In detail, \citet{Turner2014} performed a similar analysis measuring the optical depth as a tracer of metals in different ionization stages (e.g., \hi, \civ\, \ion{O}{VI}) around galaxies and found evidence of absorption enhancement extending up to $\approx 180\rm\,kpc$ in the transverse direction and on scales a factor $\approx5$ larger along the LOS.

A complete analysis involving the Cauchy-Schwarz relation would be possible by deriving the \civ\ absorbers auto-correlation function. In our survey, the absorption systems are observed along the line-of-sight of the central quasar of the MAGG fields which are separated by large transverse distances that exclude any possible physical correlation between the absorbers. Therefore, the \civ\ absorber auto-correlation function can only be computed in one-dimension as a function of the line-of-sight separation. 
The amplitude of the \civ\ absorbers auto-correlation function is almost flat with small fluctuations around $\xi_{\rm AA}\approx0.0-0.2$ at any line-of-sight separation. This suggests that the clustering of \civ\ absorbers may be too weak to be observed with our sample since it collects a high number of absorption systems, but spread across a large redshift range and along 28 different sightlines. For this reason, we can only  derive a $3\sigma$ upper limit for the absorbers auto-correlation function that results in $\xi_{\rm AA}\lesssim2.0$ at any line-of-sight separation in the range $\Delta v=200-2500\rm\,km\,s^{-1}$ (the lower limit accounts for the rest frame separation of the \civ\ doublets and the median width of the lines), corresponding to $R_{\rm los}\approx0.6-7.0\rm\,pMpc$. 

The comparison between the LAEs auto-correlation with the galaxy-absorber cross-correlation function can be used to test whether galaxies and \civ\ systems trace the same underlying matter distribution. Previously, \citet{Adelberger2003, Adelberger2005} detected metals around galaxies up to a transverse distance of $\approx300\rm\,pkpc$ and found evidence of a clustering signal of \civ\ absorption systems near LBGs from the analysis of the galaxy-absorber cross-correlation function at $z\sim3$. They used the analogies in the galaxy auto-correlation and the galaxy-absorber cross-correlation function as a hint to conclude that the galaxies and the absorption systems show a tight correlation and likely arise in the same regions of the Universe. However, a comparison between the LAEs auto-correlation and the galaxy-absorber cross-correlation suggests that the two functions do not show identical shapes nor the same amplitude in MAGG. Based on this result, we do not have direct evidence in support of the fact that LAEs and the \civ\ absorbers selected by MAGG are exclusively tracing the same regions of the Universe. It is of course possible that the \civ\ auto-correlation function at small line-of-sight separations combines with the galaxy auto-correlation function to yield the Cauchy-Schwarz equality, but we are fundamentally limited on measuring the \civ\ auto-correlation below few hundreds of $\rm km~s^{-1}$. 
Therefore, from this analysis, we can conclude that a correlation between the two populations is evident from what is shown so far, but there is no evidence that would exclude that at least a fraction of the \civ\ systems arise  beyond the halos of LAEs, from other regions such as the IGM filaments.

\begin{figure} 
\centering
\includegraphics[width=\columnwidth]{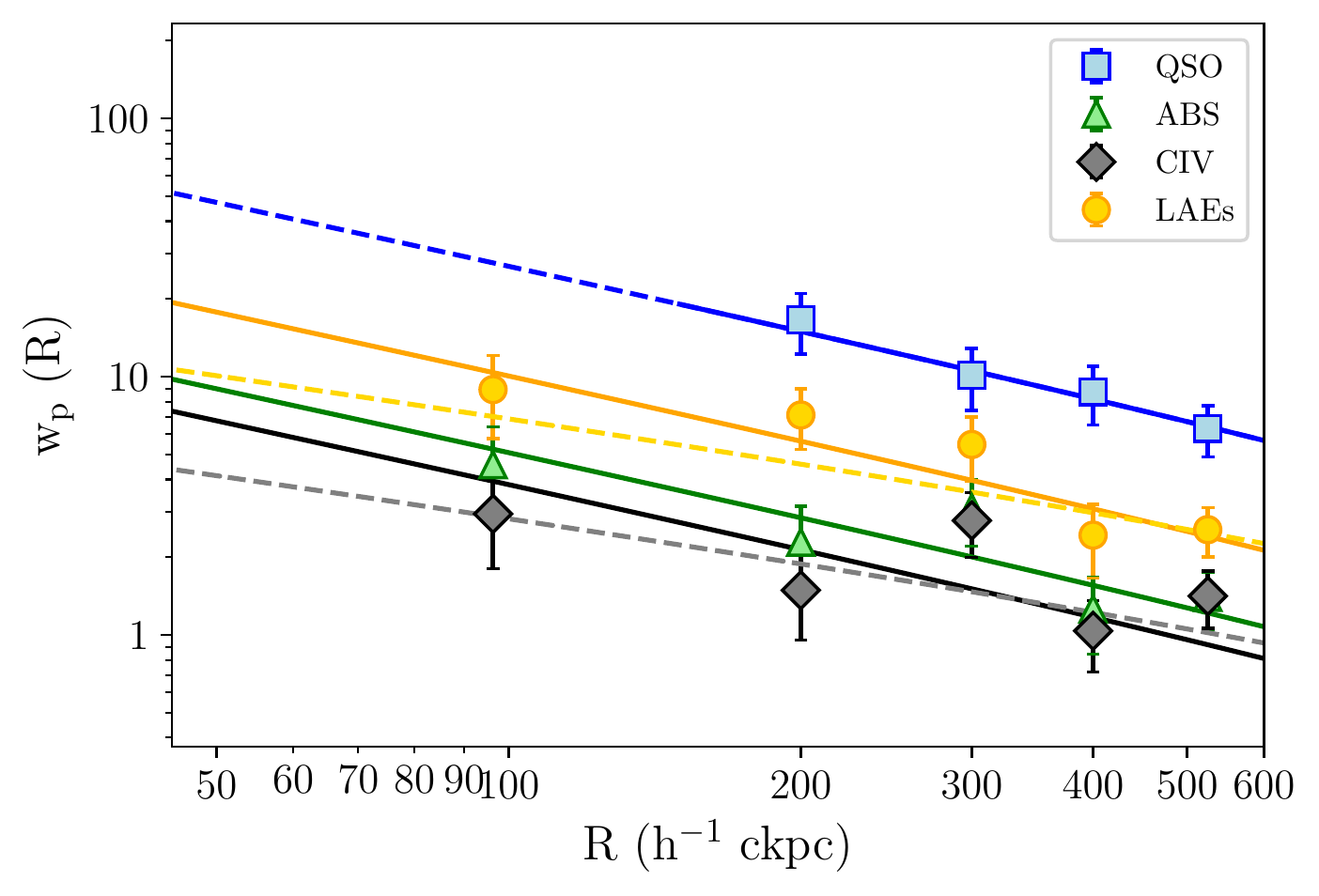}
\caption{Reduced angular cross-correlation function for the LAEs associated to the \civ\ absorbers (black diamonds) with $\rm\gamma=1.8$ (solid black line) and $\rm\gamma=1.5$ (dotted black line) at $S/N>7$. We compare this result with the ABS sample from \maggiv\ (green triangles) and the QSO sample from \citet{Fossati2021} (blue squares), both of which are computed for $S/N>5$ and assuming $\gamma=1.8$. The LAEs auto-correlation function (gold circles) is obtained by integrating the 2-dimensional function along the line-of-sight direction.}
\label{fig:CIV_XCorr}
\end{figure}

\subsubsection{Radial extent of the LAE overdensity near \civ}

As a complement to the analysis of the luminosity function, we now study the LAE-\civ\ clustering signal in terms of the reduced angular cross-correlation function, also comparing with different tracers of the LAE environment (\civ\ and \hi\ absorbers, and quasars). As found before, the two-dimensional cross-correlation function is elongated in the line-of-sight direction (as seen in the right panel of Figure \ref{fig:Corr2D}) due to the uncertainties in the redshift measurement via Ly$\alpha$ combined with the underlying peculiar velocities. Therefore, the line-of-sight velocities are not an ideal measure of the radial distance for the galaxy-absorber cross-correlation. We thus focus next on the angular cross-correlation function, $\omega_{\rm p}(R)$, as a function of the transverse distance $R$ by integrating the original three-dimensional cross-correlation $\xi(r)=(r/r_{0})^{\gamma}$ over a small redshift window. To do so, we follow the method described in \citet{Trainor2012}, also adding the results for the correlation functions derived in MAGG using the central quasars \citep{Fossati2021} and the LLSs (\maggiv) as tracers. The full description of this analysis is detailed in \citet{Fossati2021} and only briefly summarized below.


Our estimate of the angular cross-correlation function in the data is, consistent with what is adopted in other papers of the MAGG series, defined as the number of observed LAEs, $N_{\rm obs}$, in excess with respect to the predicted number, $N_{\rm pred}$:
\begin{equation}
    \omega_{\rm P}(R)=\frac{N_{\rm obs}}{N_{\rm pred}}-1
\label{eq:wp}    
\end{equation}
The estimator in Equation \ref{eq:wp} requires a measure of the comoving mean density of LAEs in the fields. To derive this value, we first compute the number of LAEs within $\pm500\rm\,km\,s^{-1}$ from a random redshift obtained from shuffling the real redshifts of the \civ\ absorbers observed in other fields. We then divide this number by the comoving area of the survey, assuming each field to be modeled as a square with $0.98\rm\,arcmin$ on a side. The measure and the relative uncertainty of $n_{\rm pred}=(35.16\pm6.94)h^{-2}\rm\,cMpc^{-2}$ is derived by bootstrapping the \civ\ redshift with random extractions over $10^{3}$ repetitions and by taking the 16th and 84th percentiles.
Finally, the expected number of galaxies in each interval of distance is derived by multiplying the mean density of galaxies in the fields, $n_{\rm pred}$ by the comoving area of the circular annulus corresponding to the $i$-th radial bin: $N_{\rm pred}=n_{\rm pred}\,\pi(R_{\rm i+1}^{2}-R_{\rm i}^{2})$. 

\begin{table}
\centering
\begin{tabular}{ccc}
\hline
Sample & $r_{0}\rm\,(h^{-1}\,cMpc)$ & $\gamma$ \\
\hline
QSO & $3.58_{-0.39}^{+0.34}$ & 1.8 \\
LAE & $2.07_{-0.24}^{+0.23}$ & 1.8 \\
    & $2.47_{-0.40}^{+0.38}$ & 1.5 \\
ABS & $1.42_{-0.27}^{+0.23}$ & 1.8 \\
\civ& $1.23_{-0.27}^{+0.25}$ & 1.8 \\
    & $1.39_{-0.36}^{+0.32}$ & 1.5 \\
\hline
\end{tabular}
\caption{Parameters of the best power-law function that reproduces the reduced angular cross-correlation function for the tracers studied in MAGG.}
\label{tab:fitXcorr}
\end{table}

The results are shown as black dots in Figure \ref{fig:CIV_XCorr} with the respective uncertainties and summarized in Table \ref{tab:fitXcorr}. Due to the central quasar PSF contamination, we mask the region at distances $R<2~\rm arcsec$ corresponding to $R\lesssim43\rm\,h^{-1}\,ckpc$ at redshift $z\approx3.0$. The solid and dashed black lines are a fit to the 
projected correlation function with free parameter $r_{0}$. Due to the degeneracy of the two parameters $r_{0}$ and $\gamma$, we fix the slope of the power law, and study the implication of this assumption by choosing two different values $\gamma=1.8$ \citep{Hennawi2006, Bielby2011, Diener2017, Garcia-Vergara2017, Garcia-Vergara2019} and $\gamma=1.5$ \citep{Trainor2012}. Consistently with what \citet{Trainor2012} pointed out, we observe that the correlation length decreases with increasing slope of the power law from $r_{0}=1.39^{+0.32}_{-0.36}\rm\,h^{-1}\,cMpc$ for $\gamma=1.5$ to $r_{0}=1.23^{+0.25}_{-0.27}\rm\,h^{-1}\,cMpc$ for $\gamma=1.8$. Despite the model suggesting that a positive correlation extends towards the edge of the MUSE field of view, at distances $R\gtrsim400\rm\,h^{-1}\,ckpc$ the larger uncertainties make the number of LAEs at larger separation from the \civ\ absorbers consistent with expected field number.

The observed LAEs overdensity at small distances from the \civ\ absorbers is compared in Figure \ref{fig:CIV_XCorr} with the excess of LAEs observed in the surroundings of the central quasars of each field \citep{Fossati2021} and the LLSs from \maggiv. Differently from the connection criteria adopted by \citet{Fossati2021}, in order to perform a consistent comparison, we restrict the QSO sample to the LAEs at $S/N>5$ observed at line-of-sight separations within $\pm500\rm\,km\,s^{-1}$ from the quasars (which is half the window in the original analysis). To do this, we correct the LAE mean density from \citet{Fossati2021} as $n_{\rm LAE,pres}^{\rm QSO}=0.5\,(17.15\pm2.28)=(8.58\pm1.14)\rm\,h^{-2}\,Mpc^{-2}$. For the \civ\ and \hi\ MAGG samples, the region at distances $R<50\rm\,h^{-1}\,ckpc$ is masked due to contamination of the central quasars PSF. The only exception is represented by the MAGG QSO sample, for which the masked region is extended to $R<150\rm\,h^{-1}\,ckpc$. In this case, the behaviour of the correlation function on smaller scales is an extrapolation from the best-fit power law (dashed blue line in Figure \ref{fig:CIV_XCorr}). In the entire MUSE FoV, the correlation length with respect to quasars is $r_{0}=3.58^{+0.34}_{-0.39}\rm\,h^{-1}\,cMpc$, which is higher than it is around the LLSs, $r_{0}=1.42^{+0.23}_{-0.27}\rm\,h^{-1}\,cMpc$, and the \civ\ absorbers, $r_{0}=1.23^{+0.25}_{-0.27}\rm\,h^{-1}\,cMpc$.  

A similar analysis has been performed to study the radial profile of \civ\ absorption around LBGs. \citet{Adelberger2003,Adelberger2005} selected \civ\ absorbers with column densities $N_{\civ}\gtrsim10^{12.5}\rm\,cm^{-2}$ (corresponding to a rest-frame equivalent width $W_{r}\approx0.01$\AA, of the same order of magnitude as the $50\%$ completeness limit of this work sample) and derived a \civ-LBGs cross-correlation length that is a factor $\approx2.2$ larger than what we measured for the LAEs in our sample, assuming a power-law slope $\gamma=1.8$. Adopting a different approach, \citet{Turner2014} produced 2D optical depth map showing that  metal absorption is enhanced at small transverse separations from $z\approx2-4$ galaxies. A significant excess of \civ\ absorption is also observed near LAEs at higher redshift $z\gtrsim3$. \citet{Muzahid2021} recovered \civ\ optical depth profiles by stacking lines in the spectra of 8 bright quasars and found a significant enhancement of \civ\ absorption within $\approx500\rm\,km\,s^{-1}$ and $\approx250\rm\,kpc$ from 96 LAEs at $z\approx3.3$.

To complete the analysis, we 
derive the projected LAEs auto-correlation function by integrating the two-dimensional function shown in Figure \ref{fig:Corr2D} along the line-of-sight direction up to a separation of $500\rm\,km\,s^{-1}$. Uncertainties are derived by applying a bootstrap procedure over the assembly of the random galaxy sample. Following the same strategy we applied for the other MAGG tracers, we measure the correlation length from a power-law modelling of the projected function for fixed slopes. The resulting LAEs correlation length $r_{0}=2.07_{-0.24}^{+0.23}\rm\,h^{-1}\,cMpc$ for $\gamma=1.8$ (and $r_{0}=2.47_{-0.40}^{+0.38}\rm\,h^{-1}\,cMpc$ for $\gamma=1.5$) is a factor $\approx 1.5$ and $\approx1.7$ larger compared to that obtained for the \hi\ and \civ\ sampled, respectively and $\approx1.7$ times lower compared to that obtained for LAEs clustering around quasars. Our estimate of the LAEs correlation length is found to be consistent, within the uncertainties, with the values measured in the literature for LAEs at redshift $z\sim3$ (corrected for the cosmology adopted in this paper): $r_{0}=2.5^{+0.6}_{-0.7}\rm\,h^{-1}\,cMpc$ \citep{Gawiser2007}, $r_{0}=1.78^{+0.41}_{-0.48}\rm\,h^{-1}\,cMpc$ \citep{Ouchi2010} and $r_{0}=2.99^{+0.35}_{-0.35}\rm\,h^{-1}\,cMpc$ \citep{Bielby2016}. We also observe that our result is consistent with the correlation length measured by \citet{Alonso2021} for a sample of $\approx700$ LAEs at redshift $3<z<6$. At first, they optimized the $k$-estimator to measure the LAE clustering for a survey with large redshift range, but limited angular coverage, and measured $r_{0}=3.76^{+3.24}_{-0.94}\rm\,h^{-1}\,cMpc$ and $\gamma=1.3^{+0.36}_{-0.45}$. However, employing a more traditional method, they modelled the projected two-point correlation function with a power law and measured $r_{0}=2.34^{+0.26}_{-0.37}\rm\,h^{-1}\,cMpc$ with a slope $\gamma=1.85^{+0.25}_{-0.25}$ that appears to fully agree with our findings.

Altogether, the analysis of the correlation functions, of the luminosity function and of the excess of LAEs in velocity space, provide firm evidence that LAEs cluster differently around the various tracers found across the entire MUSE FoV. The strongest signal is found near quasars and the amplitude of the clustering decreases progressively for LLSs and then for the \civ\ absorbers.
While LLSs and \civ\ absorbers show a comparable level of clustering as expected for an overlapping population (43/220 of the \civ\ selected in this work are in fact associated with the LLSs from \maggiv), there is a sufficient evidence for an excess of LAEs around LLSs compared to \civ, suggesting that high-column density \hi\ absorbers are more prominently associated to the (outer) CGM of halos compared to the \civ\ absorbers with $\gtrsim 0.08~$\AA, which are more likely to also trace gas at larger distances from LAEs.


\subsection{Covering fraction of ionized gas}
\begin{figure*} 
\centering
\includegraphics[scale=0.42]{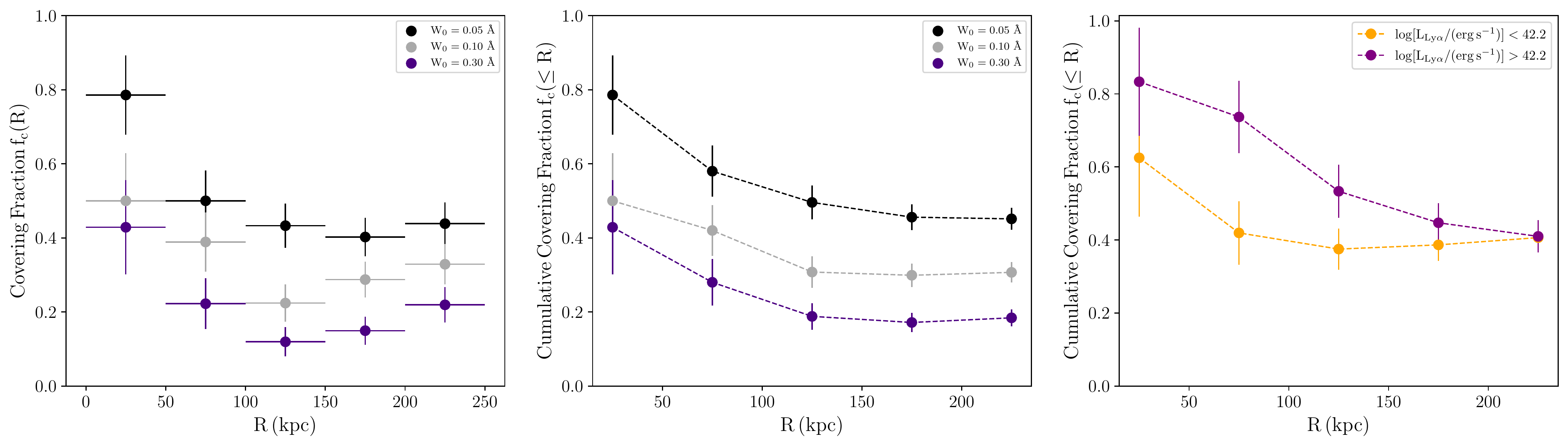}
\caption{Differential (first panel) and cumulative (second panel) covering fraction, $f_{\rm c}(R)$, as a function of the projected distance from the line-of-sight assuming three different equivalent width limits (see legend). The vertical error bars reproduce the $1\sigma$ Wilson-score confidence interval, while horizontal error bars in the left panel mark the width of each radial bin. The third panel shows the cumulative covering fraction dividing the LAEs in two bins of Ly$\alpha$ luminosity, assuming an equivalent width threshold $W_{0}=0.05\,$\AA.}
\label{fig:CIV_CF}
\end{figure*}

In the previous sections, we explored the existence of a physical association between the \civ\ absorbers and the LAEs using an \textit{absorber-centered} approach and searching for galaxies in their surroundings. Carrying on the analysis of the connection between \civ\ absorbers and LAEs with a statistical method, we now turn to a \textit{galaxy-centered} point of view. To avoid proximity effects, we masked all the LAEs observed within a LOS separation $|\Delta v|\leq3000\rm\,km\,s^{-1}$ from the redshift of the central quasar of each field. We thus derive the \civ\ covering fraction as a function of the transverse distance in order to measure the observed occurrence of \civ\ absorbers around the LAEs. The covering fraction is defined as the number of LAEs connected to a \civ\ absorber (according to the above definition, i.e. within a $500~\rm km~s^{-1}$ window centred on the peak seen in Figure~\ref{fig:CIVGalConnection}) with rest-frame equivalent width above a given threshold $W_{\rm r}^{1548}\geq W_{0}$, relative to the the total number of LAEs lying in the \civ\ redshift path:
\begin{equation}
    f_{\rm c}(R) = \frac{N_{\rm obs}(R; W_{\rm r}^{1548}\geq W_{0})}{N_{\rm tot}(R)}\:.
\end{equation}

In Figure \ref{fig:CIV_CF} we show the differential covering fraction derived in radial annuli (first panel), and thus including LAEs with transverse distances in the $i$-th interval $R_{i}<R<R_{i+1}$, and the cumulative covering fraction, at radial separations $R<R_{i+1}$ (second panel). The vertical error bars account for the $1\sigma=68\%$ Wilson-score confidence intervals, while in the left panel the horizontal bars reproduce the width of each radial interval. The covering fraction is computed for three rest-frame equivalent width thresholds$(0.05,0.10,0.30)\,$\AA. As expected from the frequency distribution function, the \civ\ covering fraction decreases with increasing threshold from $f_{\rm c}\approx80\%$ with $W_{0}=0.05\,$\AA, to $f_{\rm c}\approx50\%$ with $W_{0}=0.10\,$\AA\ and $f_{\rm c}\approx35\%$ with $W_{0}=0.30\,$\AA\ at $R\lesssim50\rm\,kpc$. For each \civ\ equivalent width threshold, the covering fraction decreases with increasing transverse separation between LAEs and the connected absorbers from $f_{\rm c}\approx60\%$ at $R<100\rm\,kpc$ with $W_{0}=0.05\,$\AA\ to $f_{\rm c}\approx50\%$ at $R>100\rm\,kpc$. For the strongest \civ\ absorbers in the sample, assuming a threshold $W_{0}=0.30\,$\AA, the covering fraction is smaller at any transverse separation and decreases from $f_{\rm c}\approx30\%$ at $R<100\rm\,kpc$ to $f_{\rm c}\approx20\%$ at $R>100\rm\,kpc$. We also observe that the differential covering fraction (first panel in Figure \ref{fig:CIV_CF}) increases in outer radial interval ($R>200\rm\,kpc$). Although it is not statistically significant, \citet{Dutta2020} observed a similar trend studying \ion{Mg}{II} absorbers at lower redshift and explained it as possibly due to the superposition of individual galaxy halos in case of multiple galaxies associated with the same \civ\ system. In the end, the observed trend suggests that the probability to observe a \civ\ within $\pm500\rm\,km\,s^{-1}$ of an LAE is enhanced at small transverse separations and decreases at higher distances roughly up to $R\approx100\rm\,kpc$ before flattening. Hence, \civ\ is present around LAEs on scales that extend well beyond the virial radius, which typically measures $R_{\rm vir}\approx25-53\rm\,kpc$ for halo mass of the order of $M_{\rm h}\approx10^{10}-10^{11}\rm\,M_{\odot}$ \citep{Alonso2021} respectively. 

Similar trends as a function of the transverse separation and the absorption strength are observed in the literature \citep{Bordoloi2014, Turner2014, Burchett2016, Rudie2019,Dutta2021}. In particular, \citet{Bordoloi2014} derived the covering fraction at $z<0.1$ for \civ\ absorbers detected up to $\approx100\rm\,kpc$. They measure, at transverse separations of $R<50\rm\,kpc$, $f_{\rm c}\approx73\%$ for \civ\ absorber with $W_{\rm r}>0.1\,$\AA\, and $f_{\rm c}\approx36\%$ for $W_{\rm r}>0.3\,$\AA. Both these results are consistent within $1\sigma$ with our findings at $R<50\rm\,kpc$ for the same equivalent width thresholds. At higher redshift, $\rm z\sim2.0$, \citet{Rudie2019} derived the covering fraction by computing the fraction of LBGs with a \civ\ system within line-of-sight separation up to $1000\rm\,km\,s^{-1}$ and within a transverse separation $100\rm\,kpc$. The covering fraction decreases from $f_{\rm c}\approx50\%$ to $f_{\rm c}\approx33\%$ for increasing column density threshold from $\log (N_{\rm 0}/\rm\,cm^{-2})=13.0$\footnote{$\rm W_{0}=0.05\,$\AA\ corresponds to approximately $\log (N_{\rm 0}/\rm\,cm^{-2})=13.5$.}. This result is consistent at $1\sigma$ with the fraction of LAEs for which we detect \civ\ absorption above the threshold $W_{0}=0.05\,$\AA\ within a similar transverse separation $R<100\rm\,kpc$. 
Using a large sample of galaxies up to redshift $z\approx 1.5-1.6$, \citet{Dutta2021} measured a \civ\ covering factor beyond $200~\rm kpc$ of around 40\% for $\rm W_{0}=0.03\,$\AA\, which is just $25\%$ below our determination for $\rm W_{0}=0.05\,$\AA\ at comparable distances. Thus, comparing with our MAGG survey, the analysis of the \civ\ covering fraction over $\approx 10~$Gyr of cosmic time could potentially imply small  evolution with redshift, which appears to be driven by the lowest equivalent width systems.

 In order to study whether the incidence of \civ\ absorbers depends on the properties of the galaxies, we derive the cumulative covering fraction for a threshold of $W_{0}=0.05\,$\AA\ in two bins of Ly$\alpha$ luminosity  above and below  $\log (L_{\rm Ly\alpha,0}/\rm erg\,s^{-1})=42.2$ (see the third panel in Figure \ref{fig:CIV_CF}). We find that the fraction of \civ\  within $\pm500\rm\,km\,s^{-1}$ of LAEs is enhanced up to $\rm f_{\rm c}\approx80\%$ for luminous galaxies relative to $f_{\rm c}\approx63\%$ for the fainter LAEs within transverse separation $R<50\rm\,kpc$. The covering fraction decreases to $f_{\rm c}\approx75\%$ and $f_{\rm c}\approx41\%$ at higher separations $R<100\rm\,kpc$ for the two galaxies population, respectively, before flattening at $f_{\rm c}\lesssim50\%$ and $f_{\rm c}\lesssim40\%$ for distances $R>100\rm\,kpc$. 
 A larger covering factor around LAEs with higher SFR (assuming Ly$\alpha$ as a proxy of star formation activity) is in line with what is observed for \mgii\ at $z\approx 1$ by \citet{Dutta2020} in MAGG and by \citet{Dutta2021} at $z<2$ in QSAGE survey. As for lower redshift, however, it is difficult to discern whether this difference arises because of the different star formation or whether -- assuming LAEs lie on the star formation main sequence -- it is a mere reflection of the different size of the halos probed as a function of Ly$\alpha$ luminosity. Additional information, and in particular independent mass estimates from IR observations, are required to investigate this trend further. 

\subsection{Detailed analysis of the \civ-LAE associations}

\begin{figure*} 
\centering
\includegraphics[width=\columnwidth]{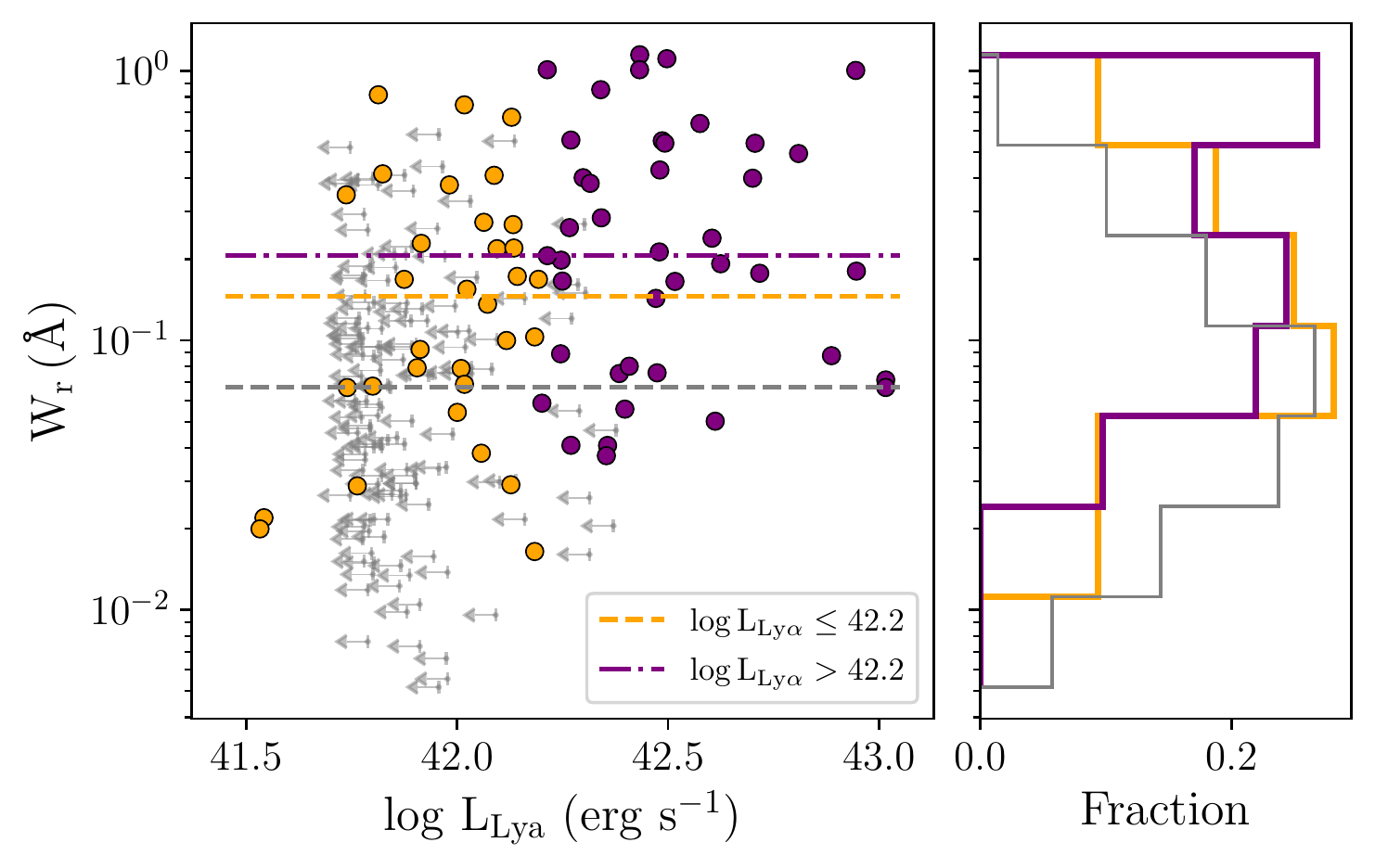}
\includegraphics[width=\columnwidth]{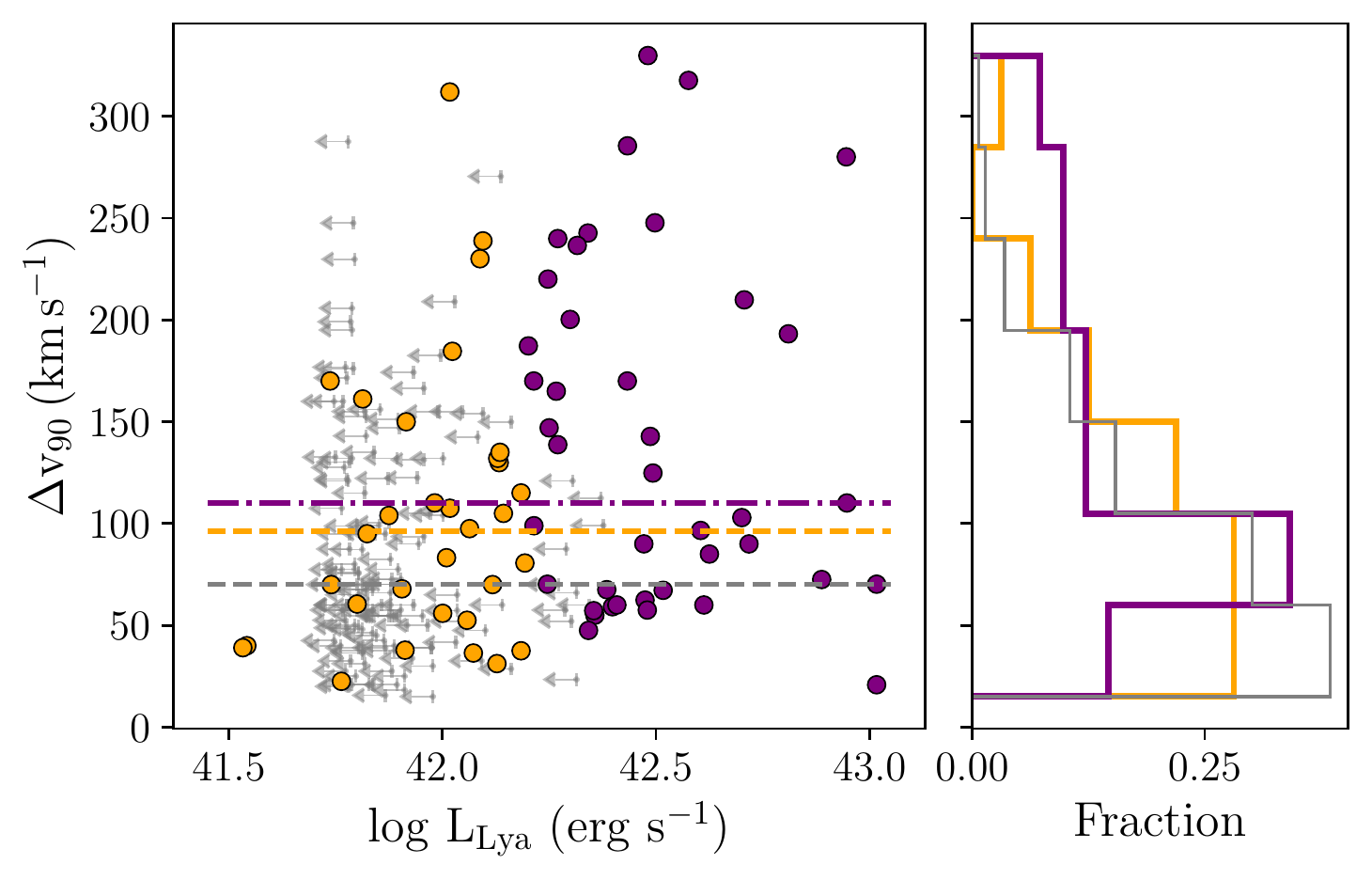}
\includegraphics[width=\columnwidth]{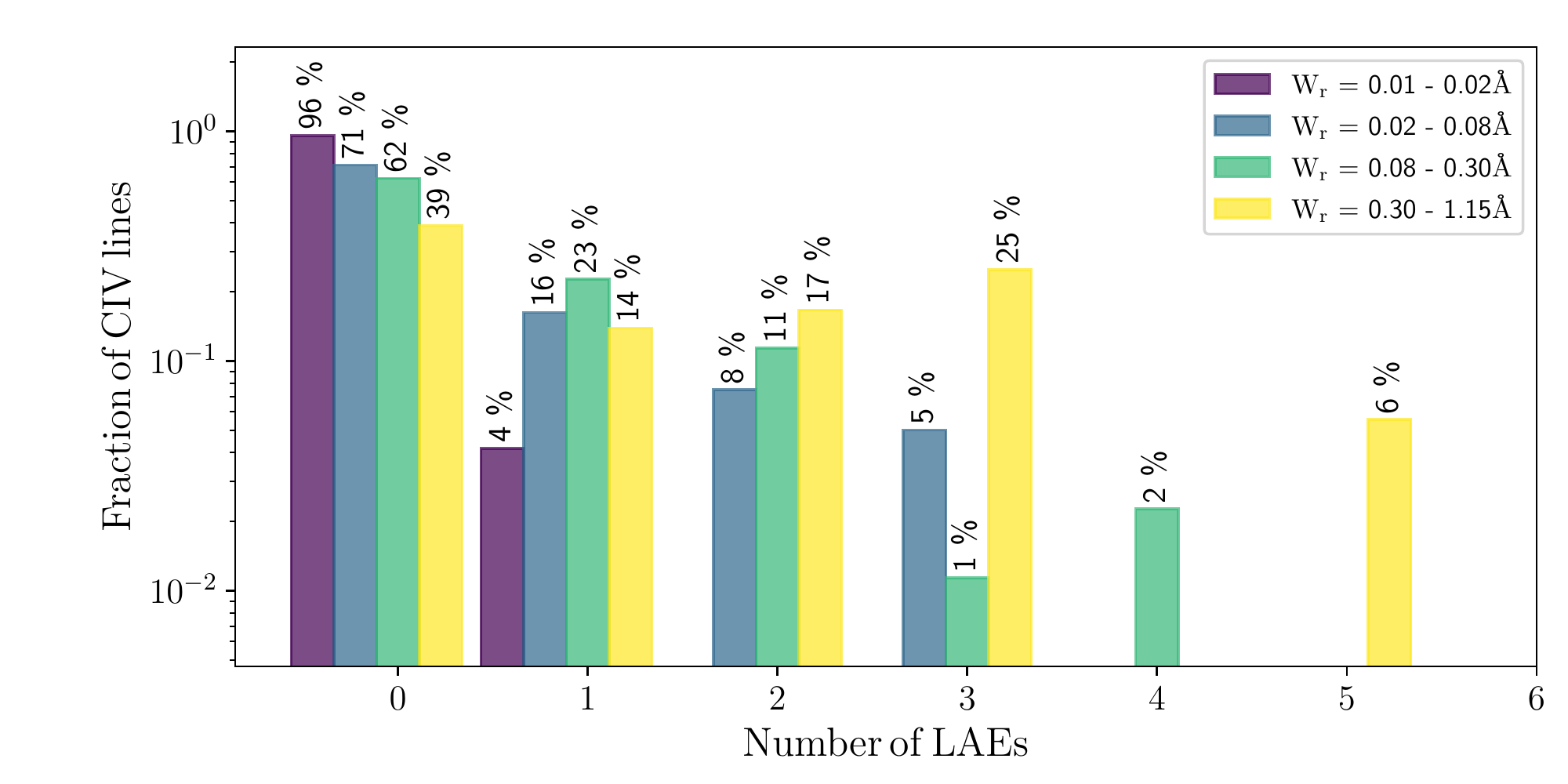}
\includegraphics[width=\columnwidth]{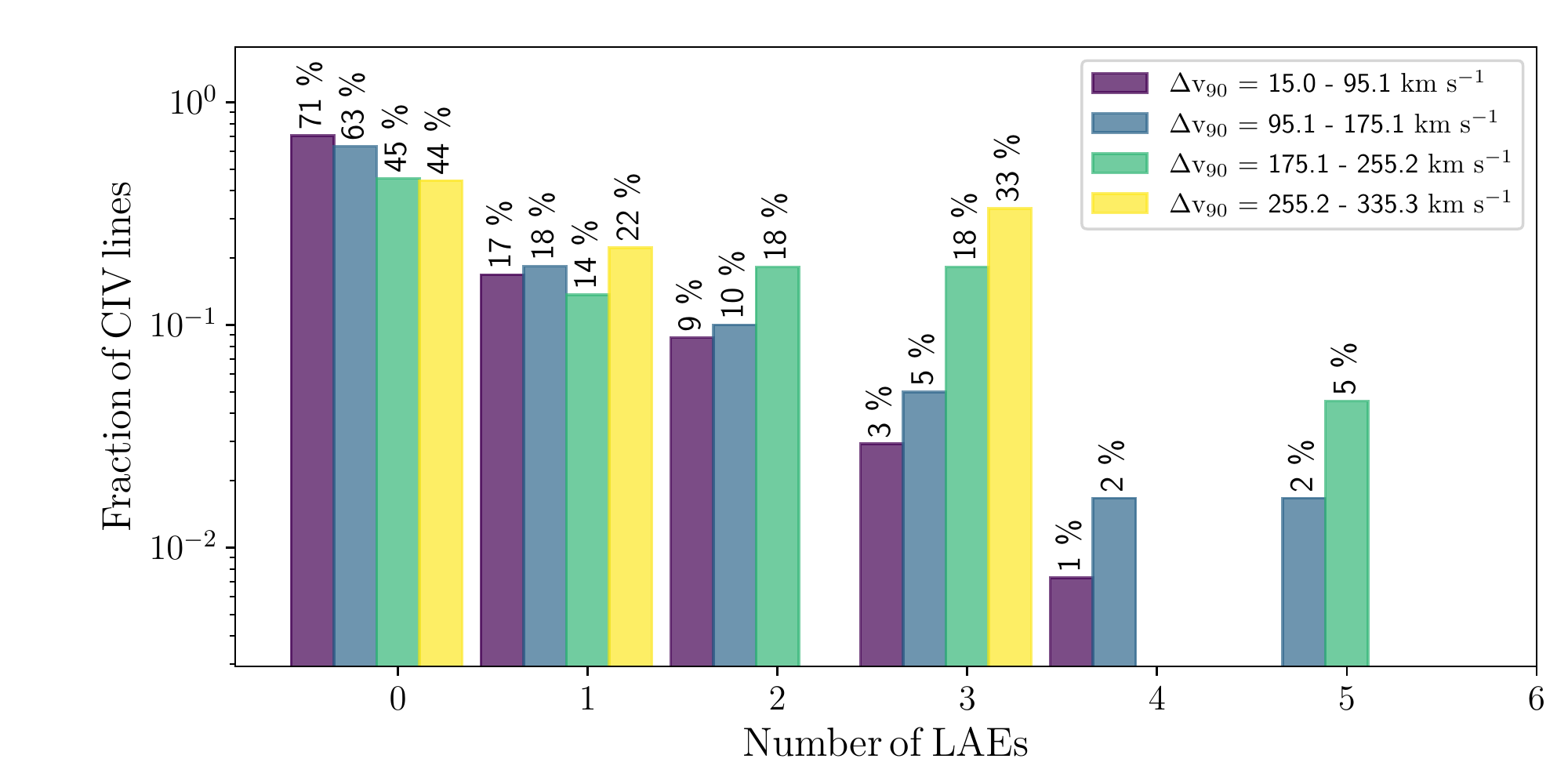}
\caption{Upper panels: \civ\ absorbers' equivalent width (left) and velocity width (right) as a function of the Ly$\alpha$ luminosity of the galaxies found within $\pm500\rm\,km\,s^{-1}$, divided in two sub-samples below (orange) and above (purple) a threshold of $L_{\rm Ly\alpha}=10^{42.2}\rm\,erg\,s^{-1}$. Dashed lines mark the median of each distribution, shown as histograms along the right hand-side axis. The grey arrows mark instead the Ly$\alpha$ luminosity upper-limit for the \civ\ systems without any associated LAE. Lower panels: distribution of the equivalent width (left) and velocity width (right) as a function of the number of galaxies connected to each absorber.}
\label{fig:ScattPlot}
\end{figure*}

Since the link between the \civ\ absorption-line systems and LAEs encodes useful information about the enrichment of the CGM, we now explore the possible correlations between the properties of the ionized gas and those of the galaxies. 
The upper panels in Figure \ref{fig:ScattPlot} show the absorbers' rest-frame equivalent width and velocity width as a function of the Ly$\alpha$ luminosity of the associated galaxies. Here, we divide the LAEs in two sub-samples based on a threshold on the Ly$\alpha$ luminosity ($L_{\rm Ly\alpha}=10^{42.2}\rm\,erg\,s^{-1}$) and we compare the resulting distributions to highlight any intrinsic difference in the two sub-populations. For those \civ\ absorbers that are not connected to any LAEs within $\pm500\rm\,km\,s^{-1}$, we take an upper-limit corresponding to the Ly$\alpha$ luminosity at which our search is  $10\%$ complete. The brightest LAE of each group is considered in the analysis. 

In the upper-left panel in Figure \ref{fig:ScattPlot} we observe that the distribution of the equivalent width of the \civ\ absorbers not connected to a galaxy reveals that a higher number of systems is weak (with $W_{\rm r}\lesssim0.05$\AA) compared to those with LAEs detected, and drops off at larger equivalent widths. The equivalent width distribution of the absorbers connected to bright LAEs is significantly skewed toward stronger systems with $W_{\rm r}\gtrsim0.1$\AA, suggesting that the strongest absorbers show a preference to be connected to luminous galaxies. Weak \civ\ systems with $W_{\rm r}\lesssim0.02$\AA\ are completely absent in this more luminous sub-sample. To quantify the statistical difference among these distributions, we perform a Kolmogorov–Smirnov test and derive the probability that the $W_{\rm r}$ distribution of \civ\ absorbers with no associated LAEs and that of \civ\ absorbers connected to bright ($L_{\rm Ly\alpha}>10^{42.2}\rm\,erg\,s^{-1}$) or faint ($L_{\rm Ly\alpha}\leq10^{42.2}\rm\,erg\,s^{-1}$) LAEs are drawn from the same parent distribution. The results, shown in Table \ref{tab:KS_prop}, corroborate on statistical grounds the observation of stronger \civ\ absorbers being connected to brighter galaxies. Likewise, \civ\ absorbers that are not connected to any galaxy within $\pm500\rm\,km\,s^{-1}$ are weaker, with large significance, than those associated to faint LAEs. However, we also note the distribution for the absorbers with no associated LAEs is not limited to the weakest systems, but extends up to $W_{\rm r}\approx1.0$\AA\ covering the full range of equivalent-width we measured. LAEs connected to these strong absorbers may not be detected because they are outside the MUSE FOV or they may be heavily obscured by dust.

\begin{table}
\centering
\begin{tabular}{ccc}
\hline
Samples & $W_{\rm r}$ (\AA) & $\Delta v_{90}\,(\rm km\,s^{-1})$ \\
\hline
Bright LAEs - Faint LAEs  & 0.22  & 0.12 \\
Bright LAEs - LAE non-detections & 5.12$\times10^{-5}$ & 4.20$\times{10^{-3}}$ \\
Faint LAEs - LAE non-detections  & 0.01 & 0.22 \\
\hline
\end{tabular}
\caption{$p$-values resulting from the KS test that measures the probability that the distribution of \civ\ properties for different Ly$\alpha$ luminosity of the associated galaxy is drawn from the same parent distribution.}
\label{tab:KS_prop}
\end{table}

A similar analysis, focusing instead on the \civ\ velocity width, is presented in the second panel of Figure \ref{fig:ScattPlot}. Here, comparing the two sub-samples, we find that the LAE non-detections are mostly limited to the narrowest \civ\ systems with $\Delta v_{90}\lesssim100\rm\,km\,s^{-1}$ with a sharp drop-off at $\Delta v_{90}\gtrsim200\rm\,km\,s^{-1}$. The distribution of the velocity width of the \civ\ absorbers connected to bright LAEs shows a tail extending at $\Delta v_{90}\gtrsim 150\rm\,km\,s^{-1}$ relative to those associated to less luminous galaxies.
In Table \ref{tab:KS_prop} we show the $p$-values resulting from the KS test, which again support on  statistical grounds the differences observed between the two samples.

Finally, in the lower panels of Figure~\ref{fig:ScattPlot} we investigate the \civ\ absorbers equivalent width (left figure) and velocity width (right figure) as a function of the number of associated LAEs within $\pm500\rm\,km\,s^{-1}$. To do so, we divide the absorbers into 4 intervals of $W_{\rm r}$ and $\Delta v_{90}$. We derive the fraction of \civ\ absorbers connected to a certain number of LAEs and with equivalent-width (velocity width) in a certain range, over the total number of systems with $W_{\rm r}$ ($\Delta v_{90}$) in that interval. The bottom-left panel of Figure \ref{fig:ScattPlot} supports the existence of a marked correlation between the strength of the \civ\ absorption and the number of connected galaxies. We do not detect any LAE at close separation from the $\approx96\%$ of the weakest absorbers with $W_{\rm r}\lesssim0.02$ \AA, while the remaining $\approx4\%$ is connected to isolated galaxies. Conversely, $\approx48\%$ of the strongest absorbers ($W_{\rm r}\gtrsim0.3$ \AA) are found to be associated with $\geq2$ LAEs, with the few rich groups hosting 4 and 5 galaxies being found exclusively near high equivalent width \civ\ absorbers.

The same analysis is then repeated for the absorbers' velocity width in the bottom right panel of Figure \ref{fig:ScattPlot}. In this case, we do not find a correlation that is as significant as above. Indeed, we do not observe any LAE in the vicinity of $\approx71\%$ of the narrow systems with $\Delta v_{90}\lesssim95\rm\,km\,s^{-1}$, but  $\approx13\%$ of them are connected to $\geq2$ galaxies. However, $\approx44\%$ of the broadest systems with $\Delta v_{90}\gtrsim255\rm\,km\,s^{-1}$ are not associated with any LAE, $\approx22\%$ is found in proximity to 1 galaxy. Only $\approx33\%$ of this sample is connected to 3 LAEs. A significant fraction of both narrow and wide absorbers, $\approx19\%$ of the systems with $\Delta v_{90}\sim95-175\rm\,km\,s^{-1}$ and  $\approx41\%$ with $\Delta v_{90}\sim175-255\rm\,km\,s^{-1}$, is associated to $\geq2$ galaxies.

In conclusion, splitting the sample of \civ-LAE associations below and above the galaxies luminosity $L_{\rm Ly\alpha}=10^{42.2}\rm\,erg\,s^{-1}$ reveals that the \civ\ absorbers found within $\pm500\rm\,km\,s^{-1}$ from a bright galaxy are on average stronger and more kinematically complex compared to the systems connected to faint galaxies or without any LAE detected in the field-of-view. A high fraction ($\approx44\%$) of the strongest absorbers with $W_{\rm r}\gtrsim0.30$ \AA\ is found to be connected to $\geq2$ galaxies and the number of associated LAEs decreases with decreasing equivalent width. The same analysis applied to the absorbers velocity width revealed no such clear correlation, with only the $\approx33\%$ of broad systems with $\Delta v_{90}\gtrsim255\rm\,km\,s^{-1}$ connected to multiple galaxies.

\subsubsection{Radial profile of the \civ\ absorption strength}

\begin{figure*} 
\centering
\includegraphics[width=\columnwidth]{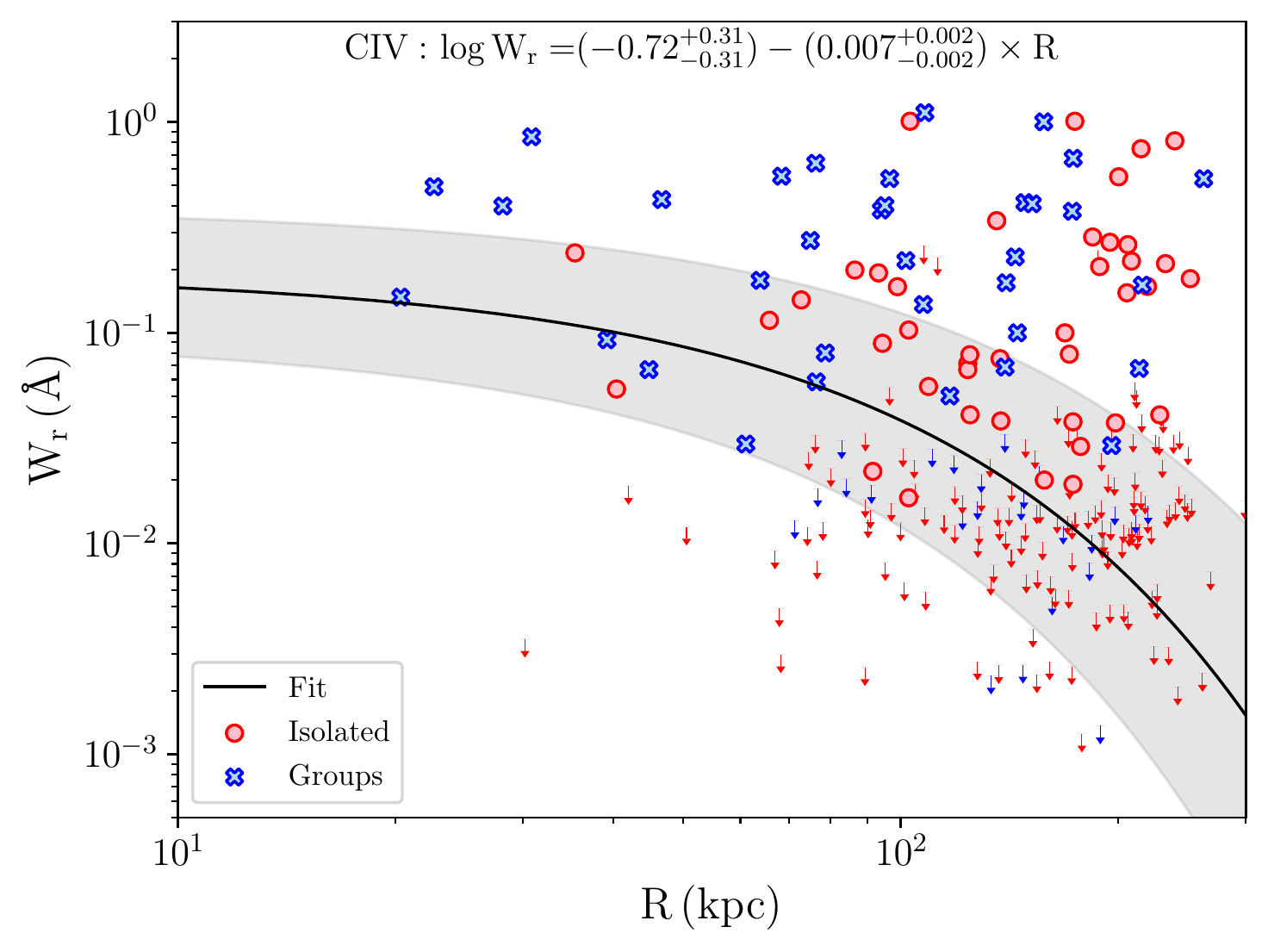}
\includegraphics[width=\columnwidth]{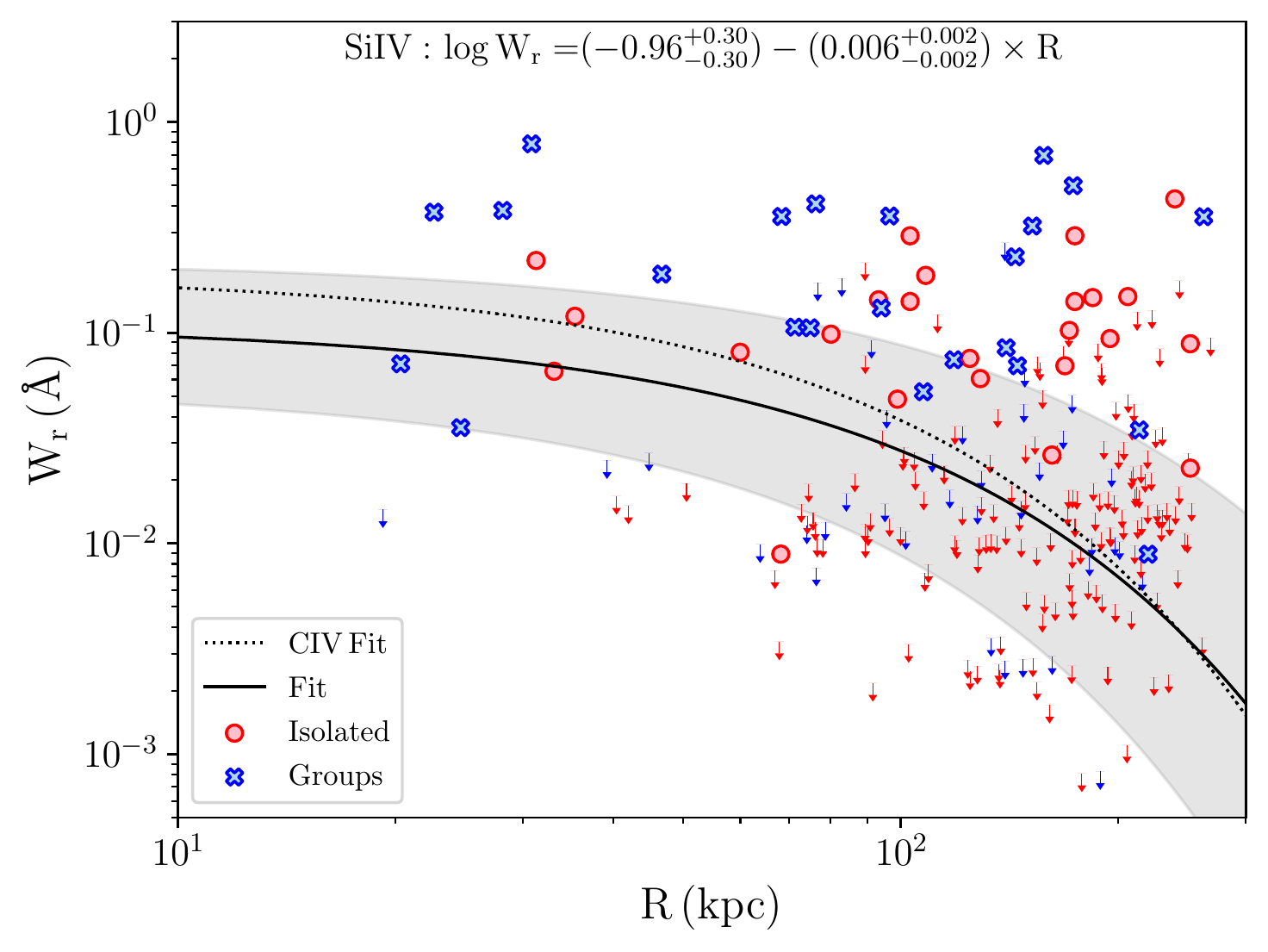}
\caption{Rest-frame equivalent width as a function of the transverse separation from the associated galaxies for the \civ\ (left panel) and the \siv\ (right panel) absorbers. The isolated LAEs are shown in red, while the closest galaxy of a group is shown in blue. For the galaxies without any associated absorber we derive a $3\sigma$ upper limit (arrows with the same color-coding). The observed relationship is fitted with a log-linear function taking upper limits into account. The best fit result is drawn as a solid black line and the $1\sigma$ uncertainties are marked by shaded regions. In the right panel, the best fit obtained for the \siv\ absorbers (solid line) is compared to the best fit derived for the \civ\ (dotted line).}
\label{fig:EWvsR}
\end{figure*}

To complete the picture of the distribution of the ionized gas around LAEs, we turn to investigate how the rest-frame equivalent width depends on the projected distance from the galaxy associated to each \civ\ absorber. The results for \civ\ are shown in the left panel of Figure~\ref{fig:EWvsR}. 

Many studies in the literature support the evidence of an anti-correlation between the rest-frame equivalent width of the \civ\ absorbers and the transverse separation from the associated galaxy. Most of these are focused on galaxies at low redshift $z\lesssim1$ \citep{Chen2001, Bordoloi2014, Burchett2016} and $z\sim1-1.5$ \citep{Dutta2021}. Similar results are observed for LBGs at higher redshift $z\sim2-3$ with a statistical approach based on spectral stacking and optical depth analysis \citep{Steidel2010, Turner2014}. Motivated by these findings, we search for an anti-correlation following the procedure described by \citet{Dutta2020} and briefly summarized here. The rest-frame equivalent width is expected to decrease with increasing distance from the associated galaxies following a log-linear relation that models an expected steep decrease of the strength of the absorption regulated by a scale factor that is linked to the galaxy virial radius:
\begin{equation}
    \log\,(W_{\rm r}\,/\text{\AA})=a + b \cdot (R\,/\rm kpc)\:.
\label{eq:EW_R}
\end{equation}

For those LAEs that lie in \civ\ redshift path, but that are not connected to any \civ\ absorber within $\pm500\rm\,km\,s^{-1}$, we derive a $3\sigma$ upper limit (arrows in Figure \ref{fig:EWvsR}). We fit the full sample with the function in Eq. \ref{eq:EW_R} applying a Bayesian method based on a likelihood that takes both measurements and upper limits into account (see for more details and applications \citealp{Chen2010,Rubin2018,Dutta2020,Dutta2021}). The result is shown in Figure \ref{fig:EWvsR} as solid black line, where the shaded grey region marks the $1\sigma$ confidence interval. We notice that the modelling is strongly driven by the upper limits at large distances where a significant fraction of the data is scattered upward the relation and the equivalent width profile appears to be flatter. 

\citet{Dutta2021} performed a similar analysis on 123 \civ\ absorption-line systems at lower redshift, $z\sim0.1-2.4$. The best log-linear model resulting from their fit is shifted significantly upward compared to the one derived in this work and parametrized by $a=-0.42^{+0.25}_{-0.28}$ and $b=-0.002^{+0.001}_{-0.001}$. The QSAGE \citep{Bielby2019, Stott2020} dataset explored by \citet{Dutta2021} results in a $3\sigma$ sensitivity limit of $W_{\rm r}^{\civ}\approx0.1$\AA\ for the detection of \civ\ absorbers, dominated by the measurements based on UV spectra with a lower resolution and S/N compared to the optical spectra available in MAGG. In this work, the upper limits at transverse distances $R\lesssim100\rm\,kpc$, that are absent in the sample from \citet{Dutta2021}, follows from the higher resolution of the MAGG spectra and pushes down the fit towards lower equivalent width values. Moreover, the size of the field-of-view at low redshift is about a factor of $\approx2$ larger than it is at high redshift, making it possible to detect galaxies up to transverse separations $R\approx750\rm\,kpc$, far beyond the typical virial radius of a dark matter halo with $M_{\rm H}\approx10^{11}\rm\,M_{\odot}$. Extending the analysis to larger distances lead to a fit that is flatter compared to the one at high redshift. Thus, the differences in the two datasets make it difficult to directly compare our findings with low redshift results and explain the discrepancies as due to a physical evolution with redshift. We note, however, that detections occupy a comparable region of parameter space, all clustering between $0.1-1~$\AA.  

\subsection{Dependence on galaxy environment}

\begin{figure*} 
\centering
\includegraphics[width=.65\columnwidth]{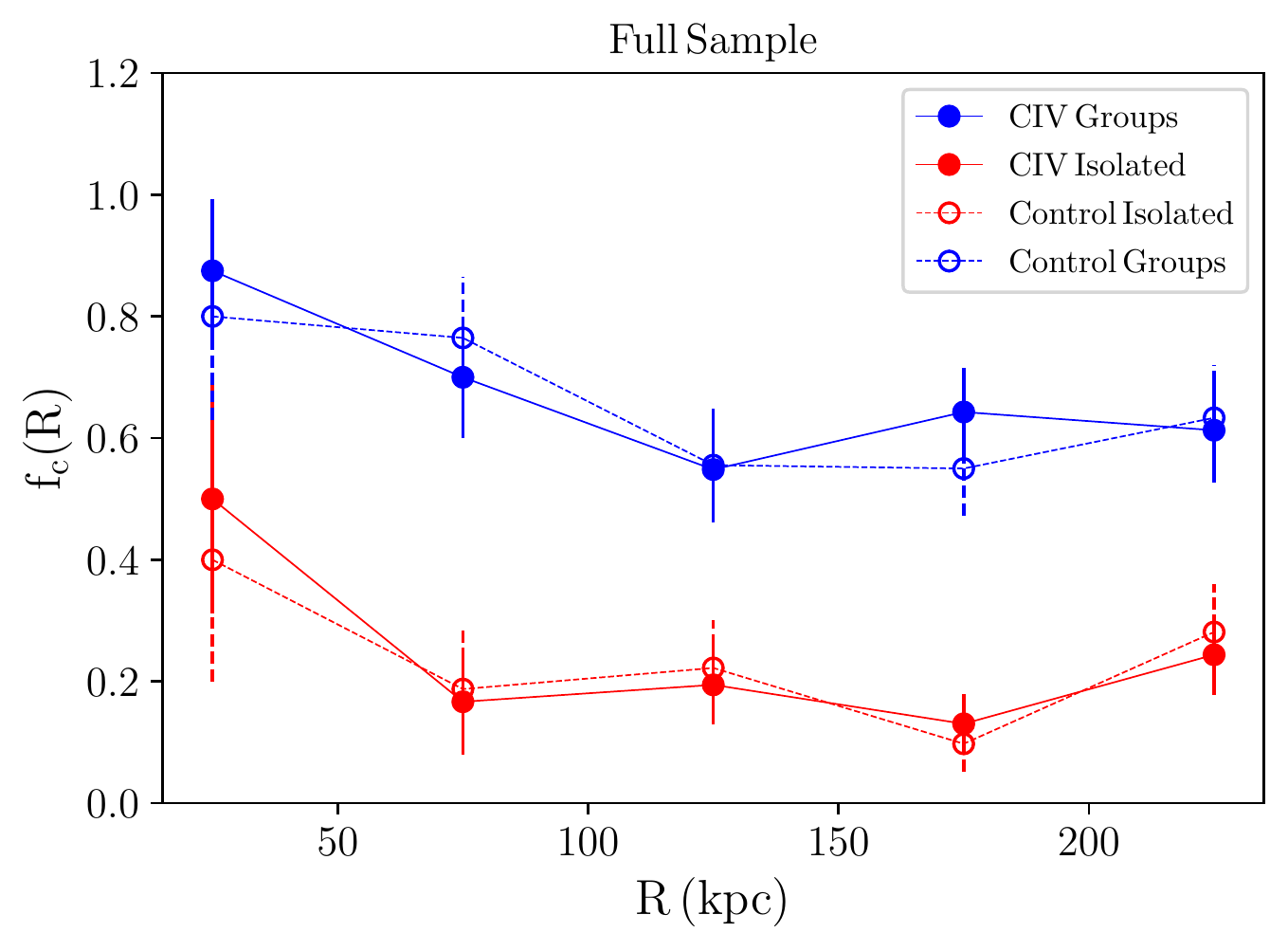}
\includegraphics[width=.65\columnwidth]{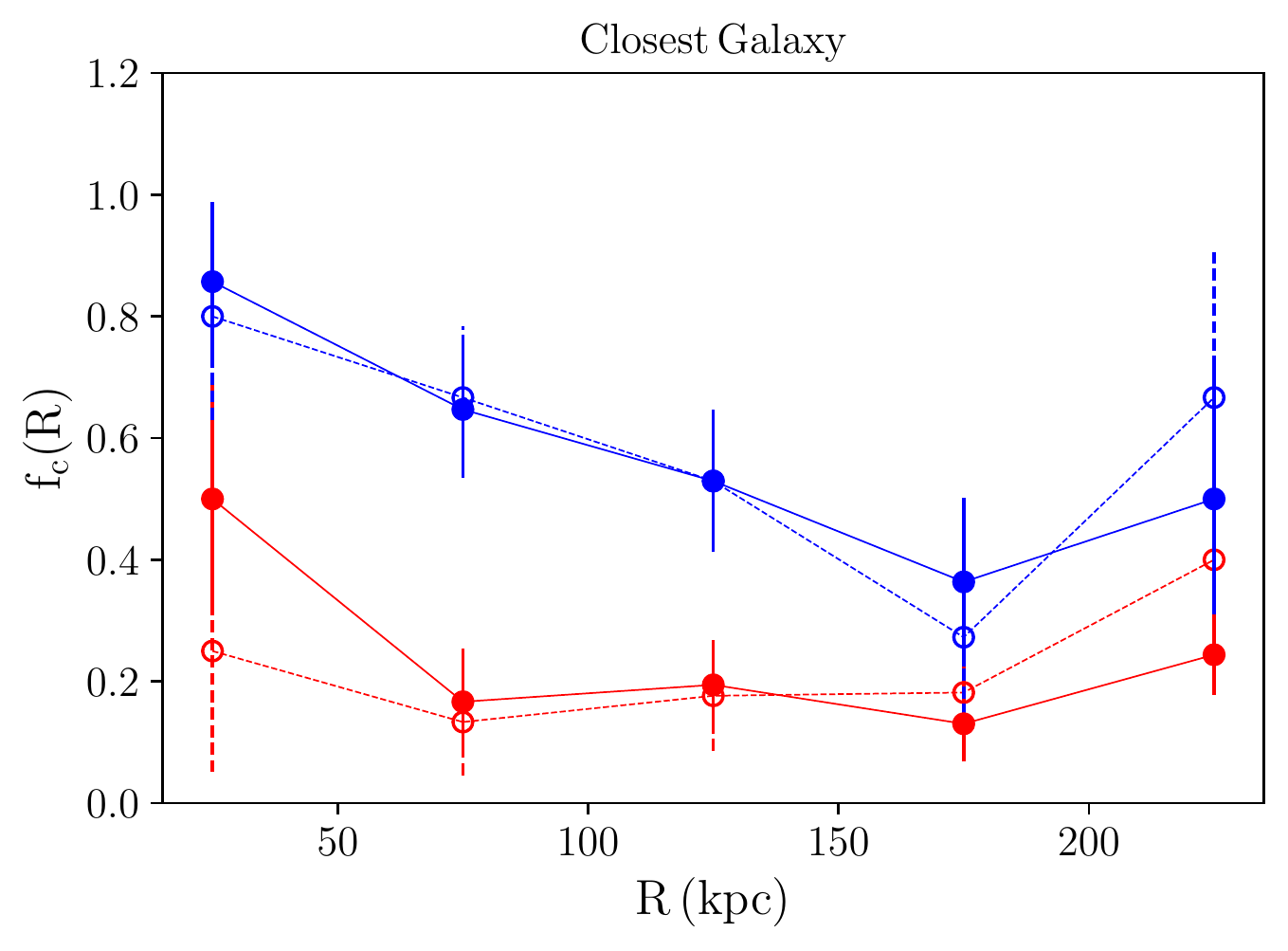}
\includegraphics[width=.65\columnwidth]{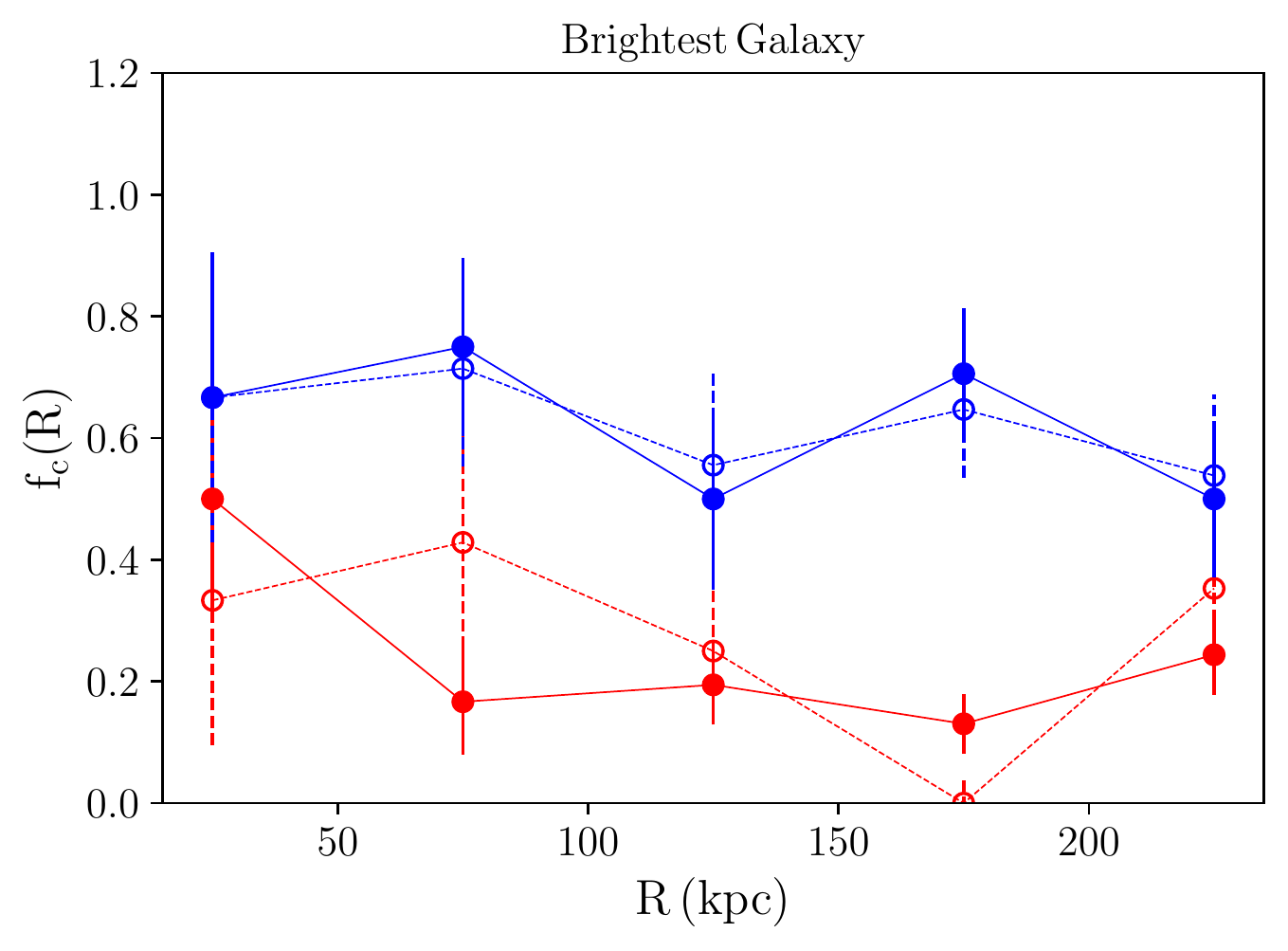}
\includegraphics[width=.65\columnwidth]{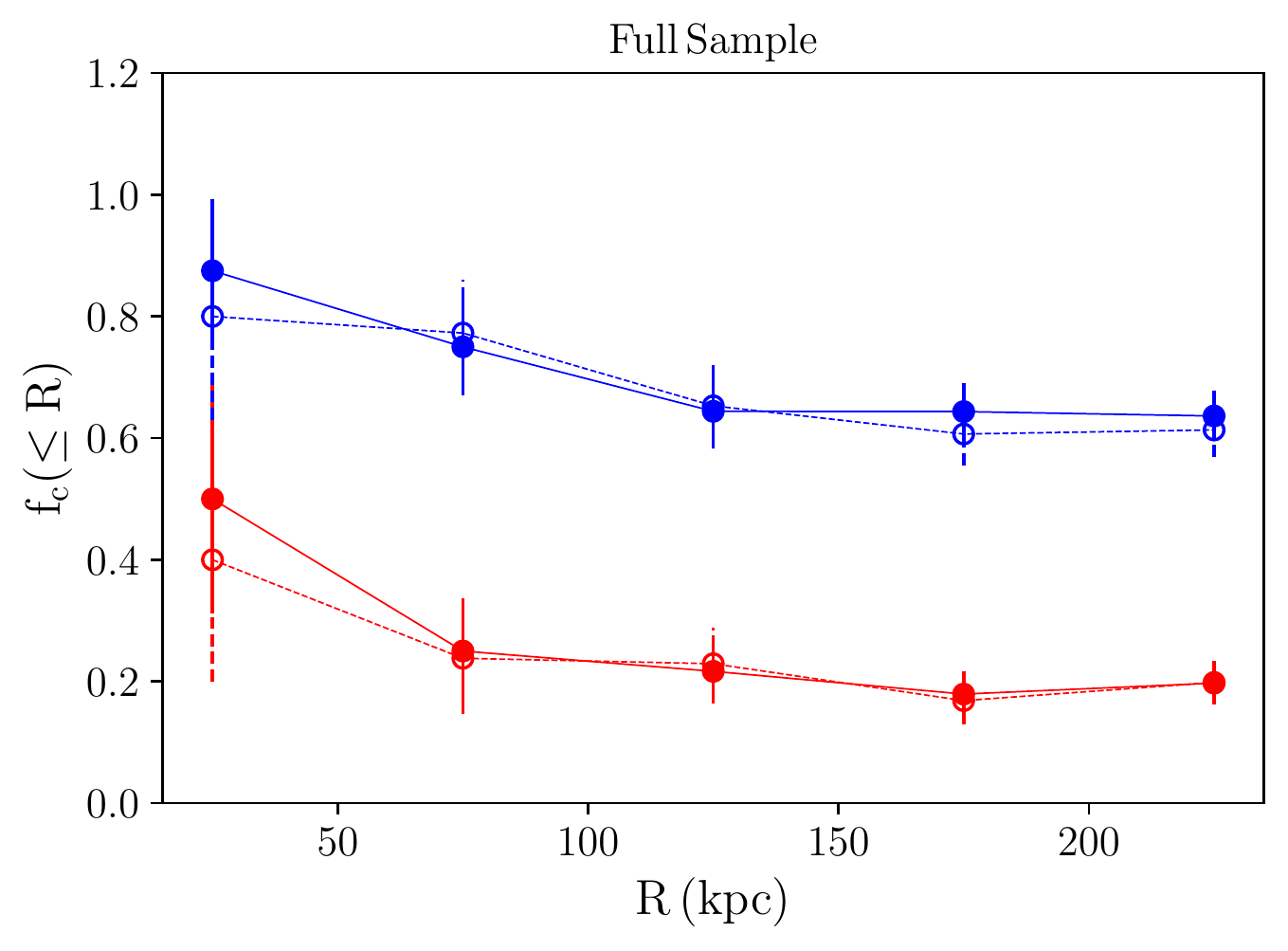}
\includegraphics[width=.65\columnwidth]{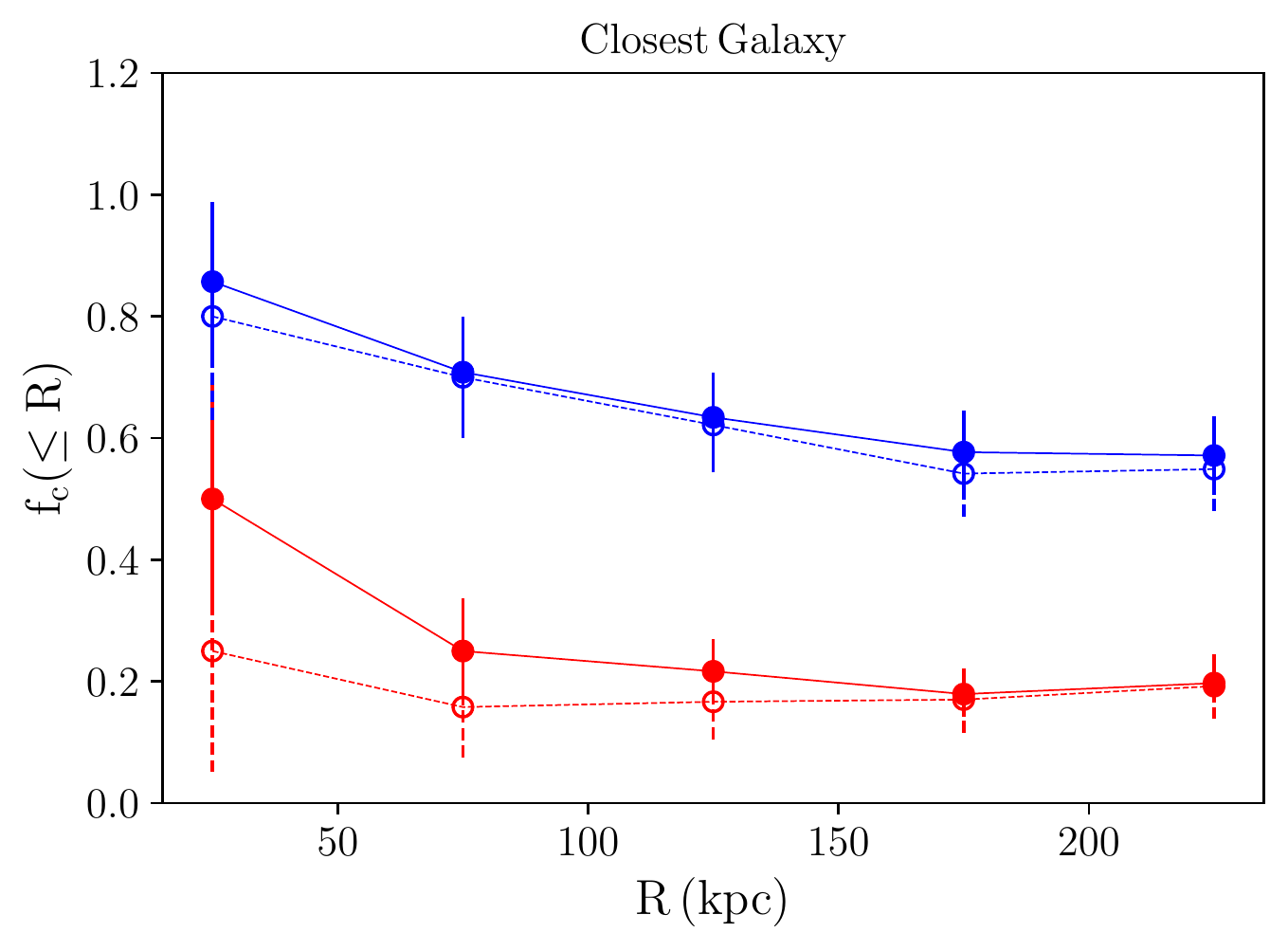}
\includegraphics[width=.65\columnwidth]{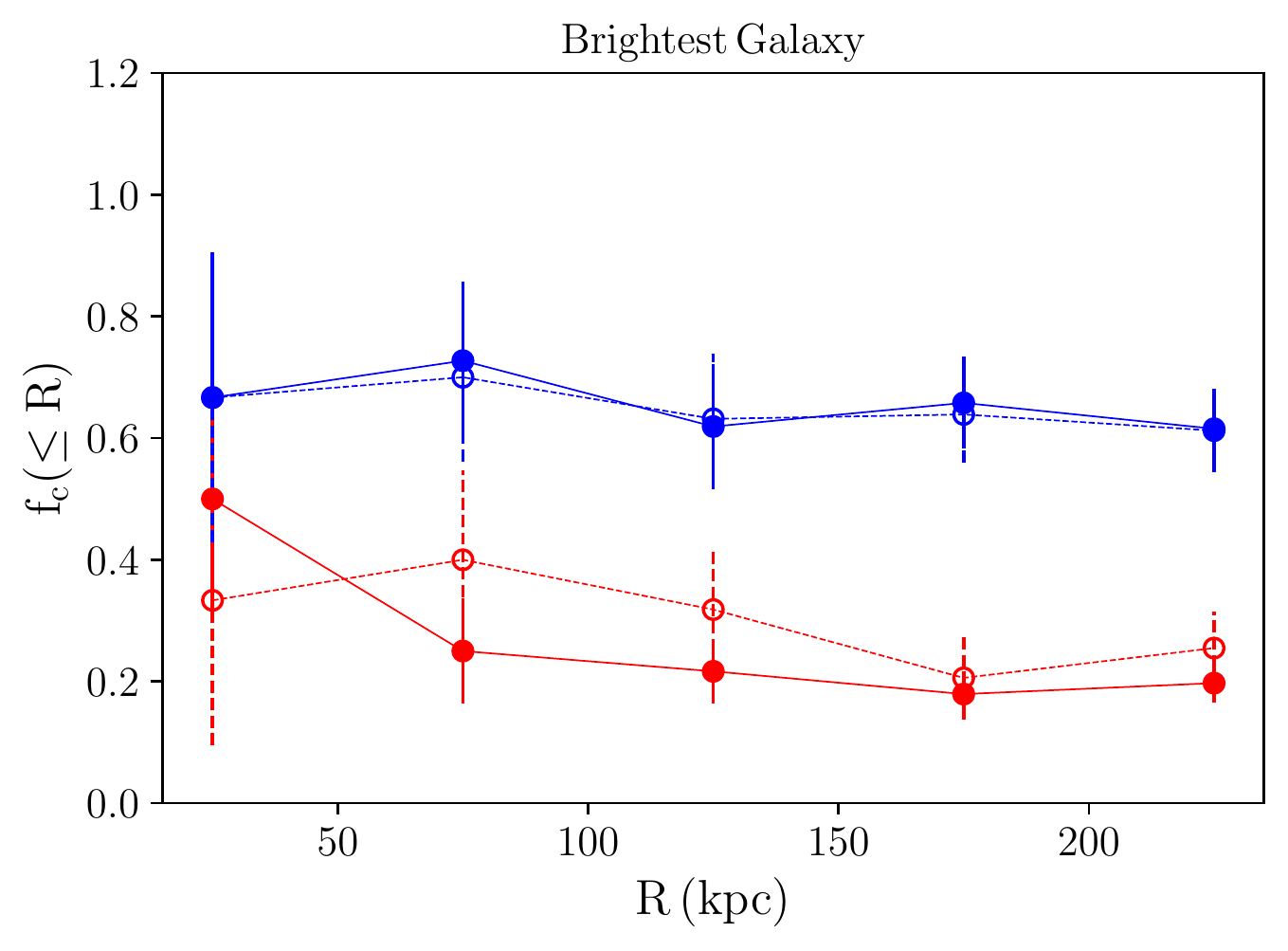}
\caption{Differential (top row) and cumulative (bottom row) covering fraction of \civ\ around isolated galaxies and galaxies in groups that are connected to \civ\ absorbers within $\pm500\rm\,km\,s^{-1}$, as a function of the transverse separation. All the galaxies in groups are considered in the \textit{Full Sample}. The analysis is reproduced by taking only the closest (second column) and the brightest (third column) component of each group. These measurements (solid lines) are compared to the covering fraction estimated for control samples (dashed lines) assembled by requiring the properties of isolated galaxies and groups to match.}
\label{fig:GRP_CF}
\end{figure*}

The ability of MUSE observations to obtain a complete survey inside the FoV reveals that a large fraction of the observed galaxies are part of a group (see Sect.~\ref{sec:groups} for the definition and identification of groups), allowing us to study the multiphase circumgalactic gas as a function of galaxy environment  (see e.g. \citealp{Bielby2017, Fossati2019, Muzahid2021, Dutta2021}).

In Figure \ref{fig:EWvsR}, we distinguish between isolated galaxies (red) and groups (blue), considering only the closest galaxy of each group, and we place an equivalent width upper-limit at $3\sigma$ for the LAEs that are detected in the \civ\ redshift path, but are not associated to any absorber within $\pm500\rm\,km\,s^{-1}$ (arrows). An excess of galaxies in groups, relative to the number of isolated LAEs, is observed at small transverse separations for $R\lesssim60\rm\,kpc$. The observed \civ\ absorbers connected to groups show, on average, a rest-frame equivalent width that is $\approx2.5$ times larger than that associated to isolated galaxies at any separation, consistent with the trend shown in third panel in Figure \ref{fig:ScattPlot}. This difference is suppressed approaching the edges of the field-of-view at $R\gtrsim200\rm\,kpc$. Galaxies without \civ\ absorption detected within $\pm500\rm\,km\,s^{-1}$ are mostly found at large distances ($R\gtrsim100\rm\,kpc$) from the sightlines. We recall that our classification of isolated galaxies and groups is limited by the size of the FoV and thus it is not possible to know if an isolated galaxy, especially those connected to strong absorbers, does not have any companion outside the observed footprint.

Next, we derive the differential and cumulative covering fraction for the isolated LAEs (to the completeness limit of our survey) and the galaxies that are part of a group (i.e. with at least a companion  within $\pm500\rm\,km\,s^{-1}$ and within the MUSE FoV). The results are shown Figure \ref{fig:GRP_CF} for the full sample of galaxies in groups (first column) and for sub-samples assembled including only the closest or the brightest component of each group (second and third columns respectively).
The cumulative covering fraction of isolated and group galaxies decreases with increasing transverse distance and flattens above $R>100\rm\,kpc$. We observe, however, that the fraction of galaxies in groups with detected \civ\ absorption within $\pm500\rm\,km\,s^{-1}$ is on average a factor $\approx 3$ higher relative to the fraction of isolated galaxies for any transverse separation up to $R\approx250\rm\,kpc$.

Our result is, at small separations, consistent with the findings from \citet{Dutta2021} at $z<2$ on a sample of $\approx750$ galaxies detected in MAGG and QSAGE. Indeed, their analysis revealed that the covering fraction of \civ\ absorbers around group galaxies is a factor $\approx2$ higher relative to that of the isolated galaxies up to $R<100\rm\,kpc$. In their analysis, the difference is less pronounced than in our sample for larger impact parameters, where we observe a persistent difference up to the edge of the FoV. Exploiting the large spectral coverage of MAGG and QSAGE, they also compared the covering fraction of the intermediate ionized gas phase, traced by the \civ\ absorbers, with that of a low ionized phase traced by the \ion{Mg}{II} absorption, both for groups and isolated galaxies, finding that the covering fraction is enhanced by a factor $\approx2-3$ for the \ion{Mg}{II} gas in group galaxies. A substantial difference is, however, observed once control samples are built for the isolated galaxies by requiring them to match the impact parameter, the stellar mass and the redshift of the galaxies in groups. For these control samples, the enhancement of the covering fraction in group galaxies relative to that of the isolated galaxies is preserved only for the \ion{Mg}{II} absorption, but it is suppressed for the \civ\ gas. 

In light of this result, we define a control sample to study in detail whether the excess in covering factor of group galaxies is an intrinsic property or depends on the properties of the sample. We build control-matched samples from groups by selecting the members of the groups with properties resembling those of the isolated galaxies. Specifically, we select group galaxies within $\pm20\rm\,kpc$ from the impact parameter and $\pm0.15\rm\,dex$ from the log of the Ly$\alpha$ luminosity of the isolated LAEs, without allowing for repetitions. The same strategy is then applied to assemble a control sample of isolated galaxies. The result is shown in Figure \ref{fig:GRP_CF} (blue and red dashed line for groups and isolated galaxies, respectively). In order to test the quality of the control samples, we perform a KS test comparing the properties of the control samples to the ones of groups and isolated galaxies and measure a $p$-value $>0.95$ for all the cases shown in Figure \ref{fig:GRP_CF}.  In building these control samples, we notice (see Table \ref{tab:groups}) that isolated galaxies and groups do not differ significantly in their properties, but stronger differences are observed only if considering the closest or the brightest component of each group. In these cases, we assembly control samples of groups and isolated galaxies with the same strategy described above to reproduce the properties of the isolated sample and of the closest or the brightest component of each group respectively. The results are shown in the middle and right panels in Figure \ref{fig:GRP_CF}.
The comparison to the covering fraction of group galaxies (blue-dashed line) and that of the isolated galaxies (red-dashed line) in matched samples show that the enhancement in groups is preserved for the control samples.
  
A similar study on the effect of environment was also conducted by \citet{Muzahid2021} who performed a statistical analysis on the strength of \civ\ absorption along the line of sight based on spectral stacking. In line with our findings, they report a stronger \civ\ absorption connected to galaxies in groups, relative to those observed in the vicinity of isolated galaxies, up to velocities of a few hundreds $~\rm km/s$.
These authors interpret the observed excess as a consequence of galaxies being embedded in a large-scale structure which give rise to the \civ\ absorption, and not simply due to the properties of the two samples which appear comparable (see their table 4).

\begin{table}
\centering
\begin{tabular}{ccccc}
\hline
Property & Isolated & Groups & Groups ($R$) & Groups ($L$) \\
\hline\hline
\multicolumn{5}{|c|}{LAEs in \civ\ redshift path} \\\hline
$R\rm\,(kpc)$ & 163.54 & 169.42 & 128.45 & 162.88 \\
$\log [L_{\rm Ly\alpha}/\rm\,(erg\,s^{-1})]$ & 42.12 & 42.08 & 42.07 & 42.20 \\\hline\hline
\multicolumn{5}{|c|}{LAEs with \civ\ within $\pm500\rm\,km\,s^{-1}$}  \\\hline
$R\rm\,(kpc)$ & 168.62 & 168.41 & 105.53 & 168.10 \\
$\log [L_{\rm Ly\alpha}/\rm\,(erg\,s^{-1})]$ & 42.20 & 42.03 & 41.97 & 42.13 \\\hline
\end{tabular}
\caption{Median properties of LAEs in the \civ\ redshift path (rows $1-2$) and of LAEs with \civ\ absorbers detected within $\pm500\rm\,km\,s^{-1}$ (rows $3-4$). Columns $4-5$ are referred to the samples containing only the closest (R) and the brightest galaxy (L) of each group respectively.}
\label{tab:groups}
\end{table}

\subsection{Comparison with \siv\ as a tracer}\label{sec:siv}

\begin{figure*} 
\centering
\includegraphics[width=\columnwidth]{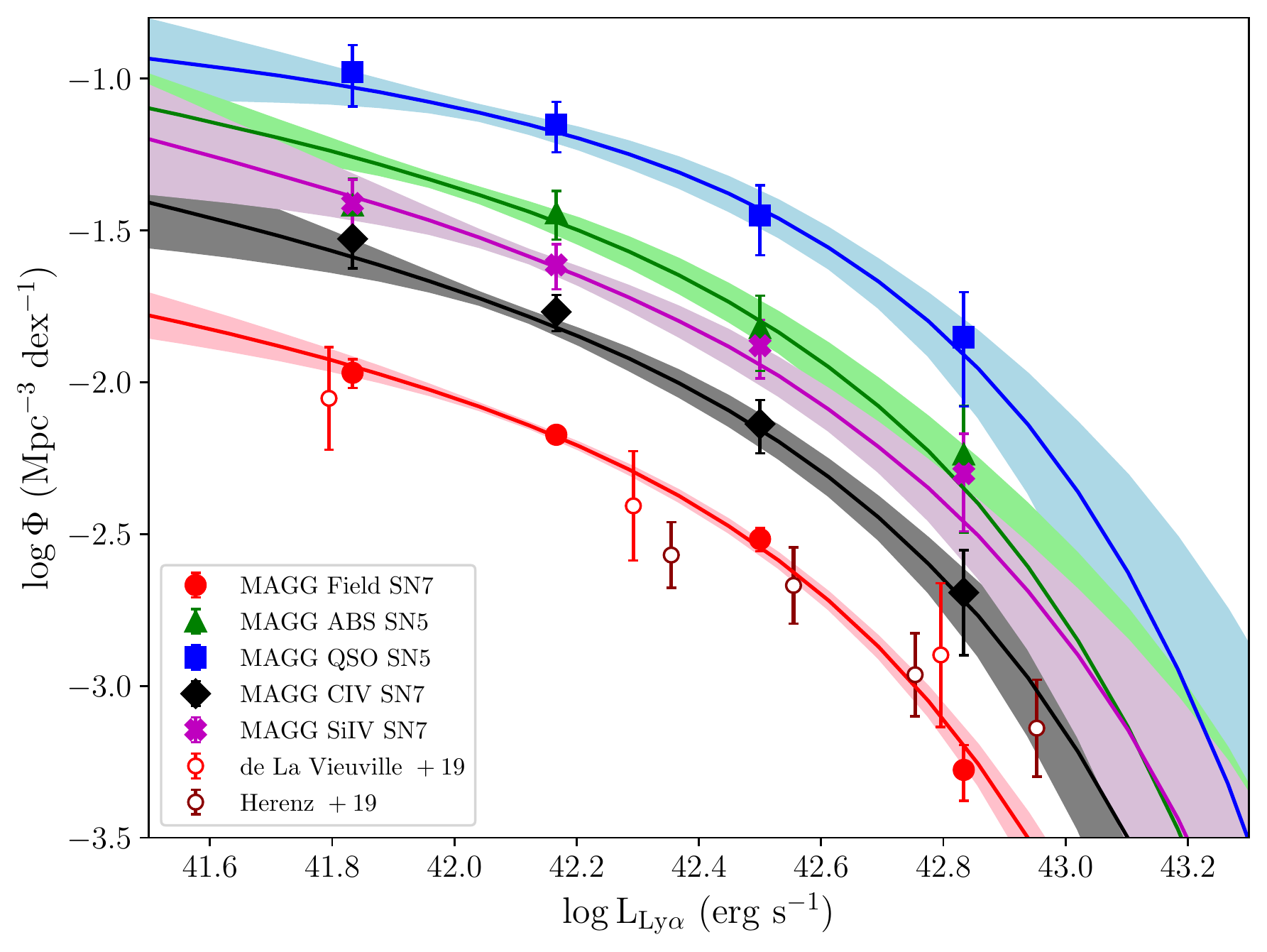}
\includegraphics[width=\columnwidth]{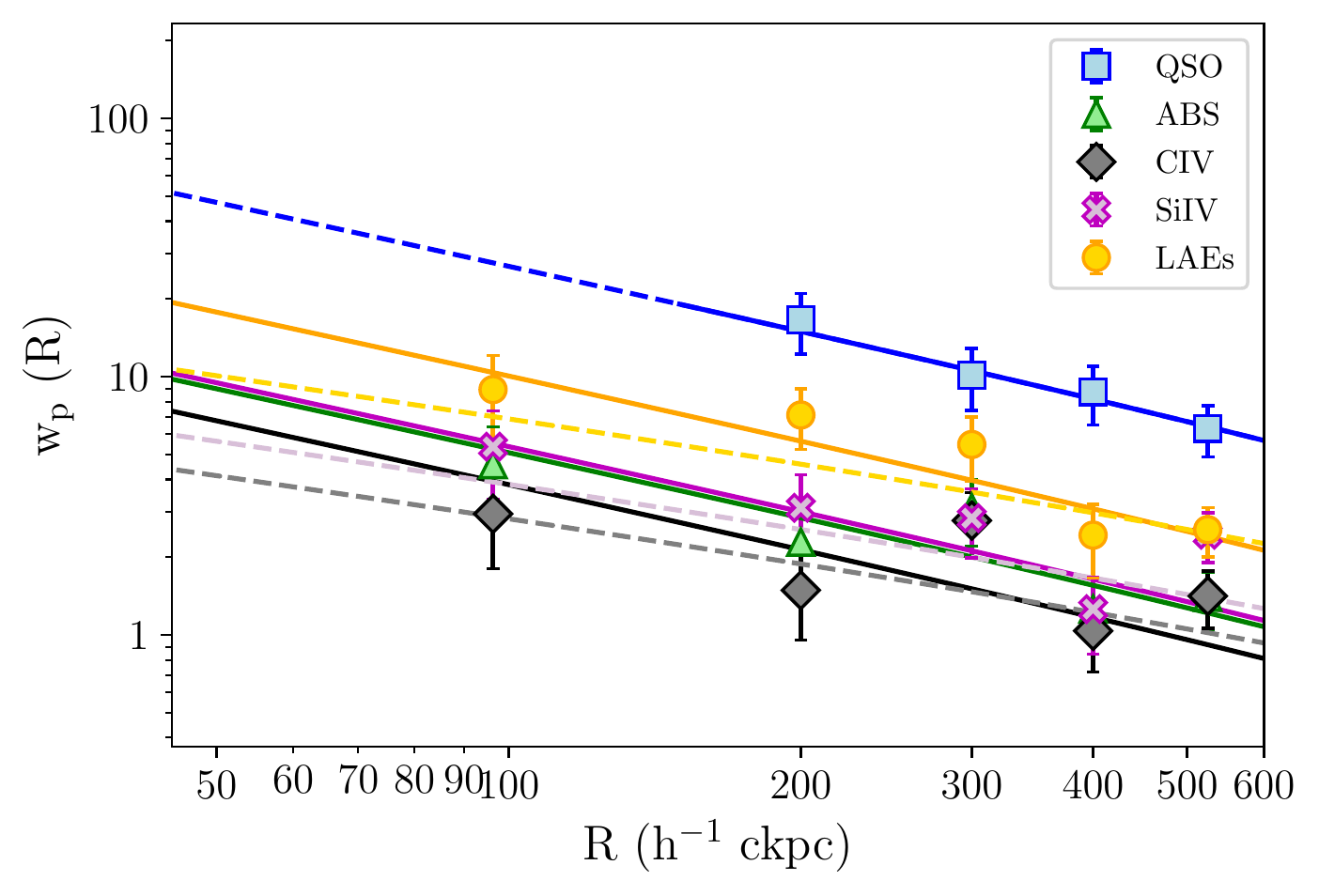}
\caption{Measure of the amplitude and the radial profile of the LAEs overdensity observed within $\pm500\rm\,km\,s^{-1}$ from \siv\ absorbers using the luminosity function (left panel, purple markers) and the projected cross-correlation (right panel, purple markers). Markers and color-coding are the same as Figures \ref{fig:CIV_LF} and \ref{fig:CIV_CF}, respectively, for the other tracers studied in MAGG.}
\label{fig:SiIV_LAE}
\end{figure*}

The analysis presented in the previous sections has focused on \civ\ as tracer of moderately ionized and possibly warm gas. An alternative tracer of gas in the same ionization stage that is accessible for observation with MAGG spectroscopy is the \siv\ doublet, which has a slightly lower but overall comparable ionization potential ($\approx 45.1$~eV for \ion{S}{IV} and $\approx 64.5$~eV for \ion{C}{IV}).
Making use of the detection of both \civ\ and \siv\ absorption-line systems in MAGG, we now test explicitly whether differences in the correlation between \siv\ and LAEs exist in comparison to \civ\ (arising, for instance, from differences in the nucleosynthesis or the shape of the ionizing spectrum), or whether the results presented above largely apply for \siv\ as well. The latter scenario would support the hypothesis that   \civ\ and \siv\ absorbers are both tracers of the same gas phase in the CGM or IGM.
 To this end, we derive the galaxies luminosity and cross-correlation functions of LAEs and \siv\ systems in comparison with what is found with \civ. We perform a blind search of LAEs within $\pm500\rm\,km\,s^{-1}$ from a \siv\ absorber with the same procedure adopted for the \civ\ analysis. We find that 49 \siv\ absorbers are connected to 86 LAEs, corresponding to a detection rate of $46\pm8$ per cent ($49/108$) that is a factor $\approx1.4$ higher compared to the \civ\ detection rate but still consistent within $1\sigma$ uncertainty. Once we identified the LAEs associated to \siv\ absorbers, we compute the LAE luminosity function and the projected cross-correlation to measure the amplitude and the radial profile of the galaxy overdensity. The results are shown in Figure~\ref{fig:SiIV_LAE}. We find that the luminosity function of LAEs around \siv\ absorbers is modelled by a Schechter function whose best-fit parameters are the characteristic luminosity $\log [L^{\star}/(\rm erg\,s^{-1})]=42.645\pm0.273$, slope $\alpha=-1.466\pm0.370$ and normalization $\log[\phi^{\star}/(\rm Mpc^{-3}\,dex^{-1})]=-2.058\pm0.431$. These estimates are all consistent with the \civ\ results within $1\sigma$ uncertainties. Furthermore, the projected cross-correlation function is modelled by a power law with correlation length $r_{0}=1.48^{+0.29}_{-0.32}\rm\,h^{-1}\,cMpc$ assuming a slope $\gamma=1.8$ and increases to $r_{0}=1.71^{+0.40}_{-0.42}\rm\,h^{-1}\,cMpc$ for $\gamma=1.5$. All these parameters are consistent within $1\sigma$ with the results derived for the \civ-LAE cross-correlation. 

Complementary to this empirical analysis, we further support these findings by adopting a more theoretical approach, detailed in Appendix \ref{app:cloudy}, to show that \civ\ and \siv\ absorbers likely arise from the same gas.

 In conclusion, this analysis provides evidence that the \civ\ and \siv\ systems are comparable tracers of ionized gas, and that the results obtained in studying the galaxy overdensity around these absorption-line systems are not strongly sensitive to the tracer adopted.

\section{Discussion}\label{sec:discussion}

\subsection{Properties of \civ\ absorbers near galaxies as a function of mass and redshift}

\subsubsection{Comparison of the MAGG and MUSEQuBES samples}

Following the pioneering survey by \citet{Adelberger2003,Adelberger2005}, several studies have reported evidence of metal enrichment in the form of \civ\ in the CGM of star-forming galaxies at the mass scale traced by $z\approx 2-3$ LBGs \citep[$M_{H}\approx 10^{12}~\rm M_\odot$;][]{Steidel2010, Turner2014, Rudie2019}. Only recently, this analysis has been extended to the lower-mass galaxy population traced by LAEs at $z\approx 3$, with $M_{H}\approx 10^{11}~\rm M_\odot$. Besides MAGG, the other major survey that studies the distribution of 
\civ\ near LAEs is MUSEQuBES \citep{Muzahid2021}. By observing samples of LAEs at comparable redshift and sensitivity to our study, these authors followed a complementary approach to ours by stacking quasar spectra at the wavelength of \hi\ and \civ\  near 96 LAEs at redshift $2.9<z<3.8$. Through this technique, they observe an excess of \hi\ and \civ\ absorption within a line-of-sight separation $\Delta v\leq\pm500\rm\,km\,s^{-1}$, i.e., on comparable scales to the velocity clustering  seen in MAGG (Figure~\ref{fig:CIVGalConnection}). 
Consistently with the covering fraction shown in Figure~\ref{fig:CIV_CF}, the excess of \civ\ absorption is extended at large projected distances ($R\geq250\rm\,kpc$), corresponding to $R\approx7\,R_{\rm vir}$ for halo masses $M_{\rm H}\approx 10^{11}\,M_{\odot}$ with virial radius $R_{\rm vir}\approx35\rm\,kpc$. However, through stacking, \citet{Muzahid2021} report no evident difference in the line-of-sight optical depth at different impact parameters, while our detailed analysis of the equivalent width as a function of impact parameter suggests a mild decline in the absorption strength as one moves away from galaxies. Compared to the results by \citet{Muzahid2021}, MAGG data reveal a higher fraction of stronger absorbers near brighter LAEs. This observation hints at a correlation between SFR and \civ\ equivalent width, that is only observed in MUSEQuBES in \hi. We note, however, that \citet{Muzahid2021} rely on rest-frame UV continuum emission for measuring SFRs, while we use Ly$\alpha$ emission.

These authors also report that the \civ\ absorbing gas shows a strong dependence as a function of the different galaxy environments connected to the absorbers. In their sample, $\approx33\%$ of the LAEs are in groups (i.e., not isolated within $\pm500\rm\,km\,s^{-1}$) and this fraction is even higher ($\approx44\%$) in the MAGG sample. They found that both \ion{H}{I} and \civ\ absorption are enhanced around galaxies in groups with respect to the isolated LAEs. According to our estimate of the covering fractions (Figure \ref{fig:GRP_CF}), the \civ\ absorbers are preferentially detected around groups rather than isolated galaxies, at least up to the edges of the field of view. Furthermore, they observe that the \hi\ absorption is stronger and broader in groups, but the weakness of the \civ\ absorption prevents from deriving a significant difference between isolated galaxies and groups. The larger size of the MAGG sample allows us to detect, on average, $\approx2.5$ times stronger and $\approx1.5$ times broader \civ\ absorbers around groups compared to those around isolated galaxies, suggesting that the galaxy environment does also affect the strength of the absorption and, to a lesser extent, the kinematics of the gas. 
 
In summary, the study of both the MAGG and MUSEQuBES samples lead to mostly similar conclusions with independent data and different analysis techniques: there is clear clustering of LAEs and \civ\ absorbers on scales up to $\pm 500~\rm km~s^{-1}$ along the line of sight and $\approx 200~\rm kpc$ in the transfer direction, and the environment appears to modulate this signal. Owing to the study of individual associations, MAGG adds further insight into this analysis, unveiling stronger/wider \civ\ absorption near brighter LAEs, and a correspondence between the strongest absorbers and the presence of groups. Finally, by studying the LAE population near absorbers (and not the strength of absorption near LAEs), we confirm that only a small fraction of \civ\ is in fact associated to galaxies ($\approx 36\pm5$ per cent) and this fraction increases with the strength of the absorption.

\subsubsection{Comparison of MAGG galaxies and $z\approx 2-3$ LBGs}\label{sec:cfrlbg}

Many studies in literature have shown there is a physical connection between continuum detected galaxies with the surrounding ionized gas detected in absorption in background quasars or galaxy pairs \citep{Turner2014,Bordoloi2014,Burchett2016,Dodorico2016,Rudie2019,Schroetter2021}. Compared to LBGs, our study extends this investigation to LAEs at $z\approx 3-4$ in lower-mass halos ($M_H\approx 10^{11}~\rm M_\odot$). We now investigate analogies or differences between these two galaxy populations and the properties of the ionized and enriched gas around them.

Considering first galaxies alone, \citet{Bielby2016} offer a direct comparison between LAEs and LBGs  using data from the VLT Lyman break galaxies (LBG) redshift survey. From their measurement of the LAEs correlation length, $r_{0}=(2.99\pm0.35)\rm\,h^{-1}\,cMpc$, they derive a halo mass $M_{\rm H}=10^{11\pm0.3}\rm\,M_{\odot}$, which is lower compared to the typical values measured for LBGs, $r_{0}\approx4.18\rm\,h^{-1}\,cMpc$ and $M_{\rm H}\approx 10^{12}\rm\,M_{\odot}$ \citep{Adelberger2005,Bielby2011}. However, the correlation length of the two population are found to be consistent at the fainter magnitudes, suggesting that LAEs inhabit on average low-mass haloes but that the two populations are overlapping \citep[see, e.g.,][]{Steidel2011}. Thus, statistics of \civ\ near LAEs and LBGs are not  independent. 

Turning to the association between \civ\ and LBGs, \citet{Adelberger2003,Adelberger2005} studied the spatial distribution of the ionized gas by means of the galaxy auto-correlation, $\xi_{\rm gg}$, and the galaxy-absorbers cross-correlation,  $\xi_{\rm ga}$, functions. Modeling the two-point correlation functions as a power law with a slope $\gamma_{\rm ga}\approx1.6$, their measurement of the cross-correlation length $r_{\rm ga}\approx3.34\rm\,h^{-1}\,cMpc$ is a factor $\approx2.4$ higher compared to the value $r_{0}\approx1.39\rm\,h^{-1}\,cMpc$ estimated for LAEs, assuming a fixed slope $\gamma=1.5$, and a factor $\approx2.7$ higher compared to the estimate $r_{0}\approx1.23\rm\,h^{-1}\,cMpc$ with a fixed slope $\gamma=1.8$ (Figure \ref{fig:CIV_XCorr}). They also found that the shape of the galaxy-absorber cross-correlation is consistent with the galaxy auto-correlation down to \civ\ systems with column density $N_{\rm\civ}\gtrsim10^{12.5}\rm\,cm^{-2}$ suggesting that LBGs and \civ\ absorbers inhabit similar regions of the Universe. However, this is not completely the case for the sample of LAEs and \civ\ systems studied in this work. As mentioned above, our results (Figure \ref{fig:Corr2D}) show that the galaxy auto-correlation and the galaxy-absorbers two dimensional cross-correlation functions have different shapes and amplitudes, suggesting that the LAEs and the ionized \civ\ gas are unlikely to trace exclusively the same underlying matter distribution and thus, do not always inhabit the same regions.

The enhancement of \civ\ absorption around LBGs is also observed by \citet{Turner2014}, who adopted the optical depth method to measure the absorption strength and produce 2D maps of the average distribution of metal absorption around these galaxies. These maps reveal a strong enhancement of metal absorption at small transverse separations from the galaxies at $R\lesssim180\rm\,kpc$ , which is elongated in the LOS direction up to $\Delta v\lesssim\pm240\rm\,km\,s^{-1}\,(\approx1\rm\,Mpc)$ due to the gas peculiar motions.  

The analysis performed on larger scales suggests that the \civ\ absorbing gas is typically extended far beyond the virial radius of LBGs, generalizing the finding of \citet{Rudie2019} who measured the covering fraction of strong ($N\geq10^{13.5}\rm\,cm^{-2}$) \civ\ systems to reach a value $f_{c}\approx50\%$ up to $R\approx100\rm\,kpc$, that is $\approx R_{\rm vir}$. Indeed, \citet{Turner2014} detected \civ\ absorption up to the survey limit ($\approx2\rm\,Mpc$) in the transverse direction, corresponding to $\approx20\rm\,R_{\rm vir}$ for an average halo mass $M_{\rm H}\approx10^{12}\rm\,M_{\odot}$. The \civ\ absorbing gas shows a similar tendency to be extended far beyond the typical virial radius ($R_{\rm vir}\approx25-35\rm\,kpc$ for halo mass $M_{\rm H}\approx10^{10}-10^{11}\rm\,M_{\odot}$ respectively) also around LAEs. The covering fraction in Figure \ref{fig:CIV_CF} shows that $\approx50\%$ of the LAEs are connected to \civ\ absorbers within $\pm500\rm\,km\,s^{-1}$ at $R>100\rm\,kpc$, reaching as far as the survey limit of $R\approx250\rm\,kpc$ which corresponds to $\approx10\rm\,R_{vir}$.  

As the separation between the ionized and enriched gas and the associated LBGs approaches the virial radius, the strength of the absorption is commonly observed to decrease monotonically with increasing distance. The spectral stack  performed by \citet{Steidel2010} shows a steep decrease of the rest-frame equivalent width of strong absorbers (with $\rm W_{r}\geq0.1$ \AA) up to the galaxy virial radius, which can be extrapolated leading to a sharp drop-off at larger distances. Considering weaker absorption, \citet{Turner2014} showed that the equivalent width decreases more slowly with the transverse separation relative to the predictions from \citet{Steidel2010} and flattens around $W_{r}\approx0.1$ \AA\ at $R>100\rm\,kpc$, up to the edges of their survey. 

Turning to the LAEs population at $z\sim3$, \citet{Muzahid2021} observed that the absorption strength as a function of the transverse distance, normalized by the virial radius, is at the level found near LBGs by \citet{Turner2014}, but without any significant monotonic decrease. In Figure \ref{fig:EWvsR}, the log-linear modeling of the equivalent width as a function of the distance shows that the absorption strength decreases at $R\gtrsim30\rm\,kpc$, corresponding to the typical virial radius in which LAEs reside. However, this trend is strongly driven by the non detections, with measured values remaining somewhat flat between $0.1-1~$\AA\ also beyond $\approx 100~\rm kpc$.

In this context, \citet{Hasan2021} studied the statistical distribution of \civ\ gas relative to different galaxy populations and its evolution across the redshift range $0\leq z\leq5$ by means of an absorption model. Their model predicts that all the \civ\ absorbers are confined within the virial radius of low mass halos, while weak systems with $W_{\rm r}\geq0.05$ \AA\ may live beyond the virial radius of the most massive galaxies at redshift $z>2.5$. These predictions are consistent with the observed extension of \civ\ gas around massive galaxies, with strong $W_{\rm r}\geq0.3$ \AA\ absorbers detected in the inner regions of the CGM and the weakest systems spread at distances larger than the virial radius. However, the model does not reproduce the large extension of weak and strong \civ\ absorbing gas around low mass haloes that host LAEs. A possible explanation may reside in the assumptions the model is based on, i.e. each individual \civ\ absorption line system is co-spatial with a single halo that hosts a single galaxy. We found that $\approx20\%$ of the \civ\ absorbers in our sample are connected to a multiple galaxies within $\pm500\rm\,km\,s^{-1}$. Given the limited size of the FoV, it is also possible that isolated galaxies have companions at larger separations. Thus, the galaxy environment might play a role in shaping the differences between the observed and predicted \civ\ extension around low mass galaxies.

Altogether, this comparison highlights how both LAEs and LBGs coexist with a significant distribution of \civ\ gas which is clearly associated to the galaxy themselves on small scales, but that appears more weakly correlated as one moves to larger distances and lower equivalent widths.

\subsubsection{Comparison with lower redshift results}
The MAGG survey has been also employed to investigate the properties of the cool gas, probed by \ion{Mg}{II} absorption, around  galaxies at $z\sim0.8-1.5$ \citep{Dutta2020} for which the authors infer solar masses of $M_{\star}\approx10^{9}\rm\,M_{\odot}$. Moreover, they combined results with the QSAGE survey, to trace both the more neutral and ionized gas phases around galaxies through \ion{Mg}{II} and \civ\ absorption at $z<2$ \citep{Dutta2021}. Their findings allow us to compare the properties of the cooler \ion{Mg}{II}  gas with the more ionized \civ\ gas both at low and high redshift.

At $z<2$, the strength of the absorption is found to decrease with increasing distance. The anti-correlation is steeper for the \ion{Mg}{II} absorbers compared to the \civ\ gas. However, the radial profile of both the \ion{Mg}{II} and \civ\ absorption strength shows a significant scatter and flattening once galaxies in groups are included in the analysis (see also \citealp{Bordoloi2011,Fossati2019}), with strong absorption detected even at large transverse separations around groups. This is consistent with our findings shown in Figure \ref{fig:EWvsR}, where the anti-correlation is sensitive to the upper limits, but strong systems are observed at any separations. A direct comparison 
as a function of redshift is made difficult by the different size of the FoV and the different sensitivities of these surveys. 
However, a general trend appears: a significant contribution to the flattening in the $W_{\rm r}-R$ relation of \civ\ absorbers may come at all redshift from the larger extension of the highly-ionized gas which, as argued at lower redshift \citep{Dutta2021}, is possibly embedding the cooler phase.
Similar results are indeed found around LBGs at $z\sim2$ by \citet{Rudie2019}, with singly-ionized ions commonly detected at smaller distances from the galaxies relative to higher ionization species.

Finally, the galaxy environment is observed to play an important role in shaping the properties of the absorbers both at low and high redshift. At $z\lesssim 2$, the \ion{Mg}{II} systems connected to multiple galaxies are $\approx5$ times stronger and show a $\approx3$ times higher covering fraction up to $R\approx200\rm\,kpc$ ($\sim2\rm\,R_{\rm vir}$) compared to those associated to isolated galaxies. \citet{Fossati2019} found stronger \ion{Mg}{II} absorption around 4 groups at redshift $z\sim0.5-1.5$ and invoked gravitational interactions between the components of a group as a possible mechanism responsible for the enhanced \ion{Mg}{II} absorption. 
Moreover, \citet {Dutta2021} found that the \civ\ covering fraction, both connected to groups and isolated galaxies, is $\approx2$ times higher relative to \ion{Mg}{II} systems, and enhanced in groups up to $R\approx750\rm\,kpc$ ($\approx5\rm\,R_{\rm vir}$). However, once a control sample of isolated galaxies is assembled to reproduce the properties of the groups, the excess is still significant for \ion{Mg}{II} absorbers, but is suppressed for the \civ\ systems. In contrast, at $z\gtrsim 3$ (Figure \ref{fig:GRP_CF}), we notice that the \civ\ covering fraction of control samples of galaxies in groups is confidently detected to be higher compared to isolated galaxies. Our findings are also supported by similar results from \citet{Muzahid2021}. Therefore, there seems to be a hint of a possible redshift evolution in how more neutral and ionized gas phases respond to different galaxy environments. However, the \civ\ sample at $z<2$ suffers from more limited statistics compared to $z\approx 3$. Likewise, there is currently no systematic study of the \mgii\ distribution near $z\approx 3$ LAEs.
This evolution scenario should therefore be tested further with better spectroscopic coverage for \civ\ at $z<2$ and for \mgii\ at $z\gtrsim 3$. The latter effort is the objective of an ongoing X-Shooter campaign (PID 0109.A$-$0559; PI M. Galbiati).

\subsection{\civ\ and \ion{H}{I} around $z>3$ LAEs}\label{sec:civandhi}

As a complement to the analysis presented here for ionized gas traced by \civ\ absorbers around LAEs, in the \maggiv\ paper we have studied the link between these galaxies and the neutral gas traced by optically-thick absorption line systems. The \maggiv\ sample is a collection of 61 strong \hi\ absorbers ($N_{\rm HI}\gtrsim10^{16.5}\rm\,cm^{-2}$), $\approx84$ per cent of which are associated to 127 LAEs within $\pm1000\rm\,km\,s^{-1}$ at redshift $z\approx 2.9-4.2$. Reducing the velocity separation to the limit we chose to associate absorbing gas and galaxies to the one adopted in this work, 45 \hi\ absorbers are connected to 97 LAEs within $\pm500\rm\,km\,s^{-1}$ centered on the peak of the velocity separation distribution (found at $\Delta v\approx 250\rm\,km\,s^{-1}$ in \maggiv). These numbers correspond to an LAE detection rate of $\approx74\pm15$ per cent around \hi\ gas, a factor $\approx2$ higher relative to the fraction of \civ\ systems connected to at least one galaxy within the same velocity window. 

Searching for the presence of metal absorption-lines as a probe of the metallicity of \hi\ systems, in \maggiv\ we identify a sample of 58 \civ\ absorbers that have been analysed independently from this work.  Of these \civ\ absorbers, excluding upper limits for non detections and systems falling within telluric regions (5/58), we found that all these absorbers have a counterpart in the sample assembled in this work within a redshift separation $\Delta z=0.0045$ ($\Delta v\approx300\rm\,km\,s^{-1}$ at $z=3.5$) with $\approx81$ per cent (43/53) detected at $W_{\rm r}^{1550}>3\sigma$. We observe that $\approx51$ per cent (22/43) of these systems are connected to at least 1 galaxy, with $\approx33$ per cent (14/43) connected to galaxies that are part of a group.
Based on this different detection rate, we conclude that if LLSs trace both the CGM and the filaments connecting several LAEs as argued in \maggiv, \civ\ absorbers only partially trace these same structures. 
This conclusion is reinforced by the statistical estimates of the LAE luminosity function (Figure \ref{fig:CIV_LF}) and the galaxy-absorber cross-correlation function (Figure \ref{fig:CIV_XCorr}). Our analysis shows that the LAE number density is higher around strong \hi\ absorbers relative to \civ\ systems both as a function of the galaxy luminosity and the transverse separation, indicating a stronger connection between strong \hi\ and LAEs than between \civ\ and LAEs.

\subsubsection{The role of cosmic filaments}

\begin{figure} 
\centering
\includegraphics[width=\columnwidth]{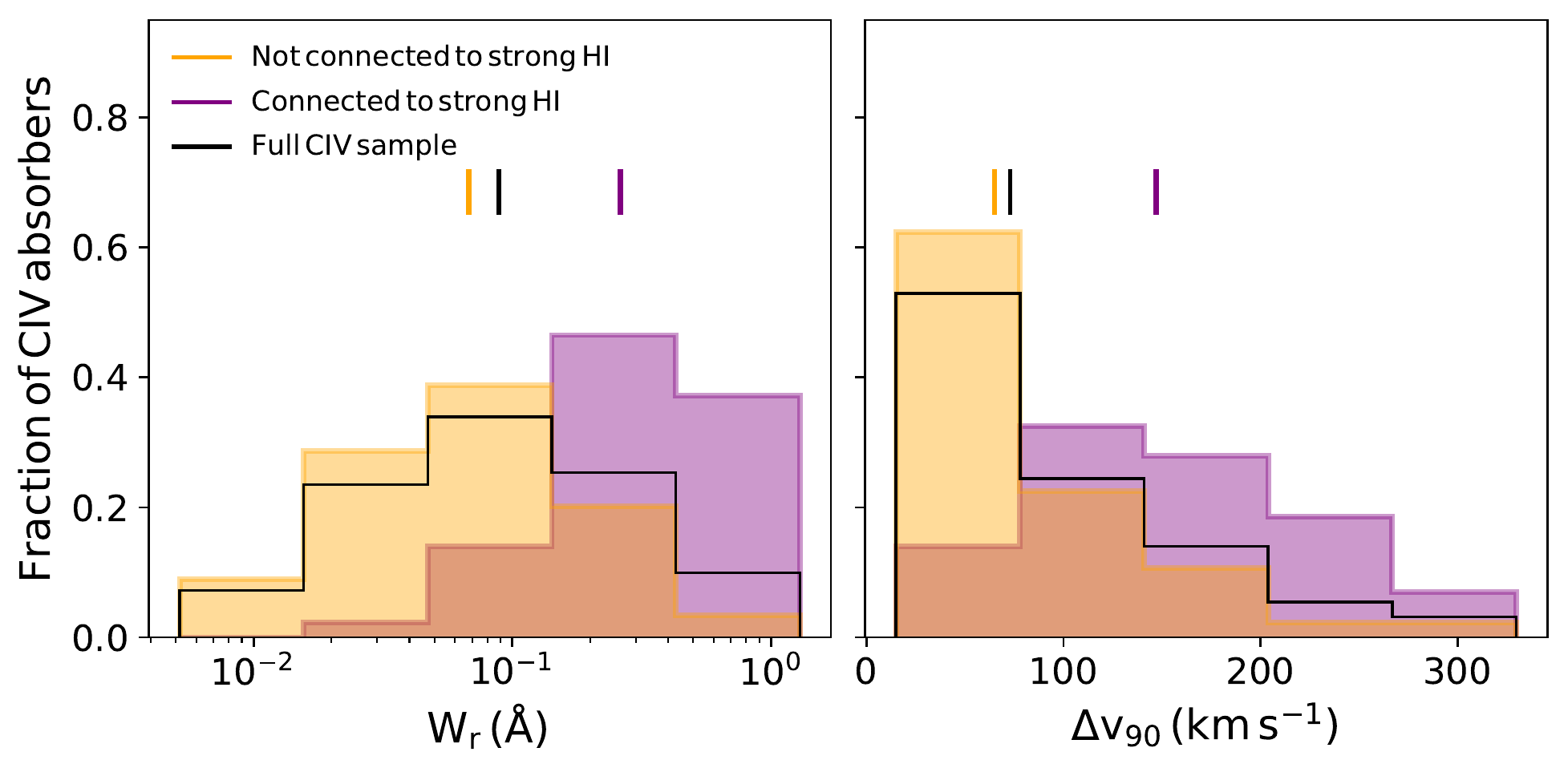}
\includegraphics[width=\columnwidth]{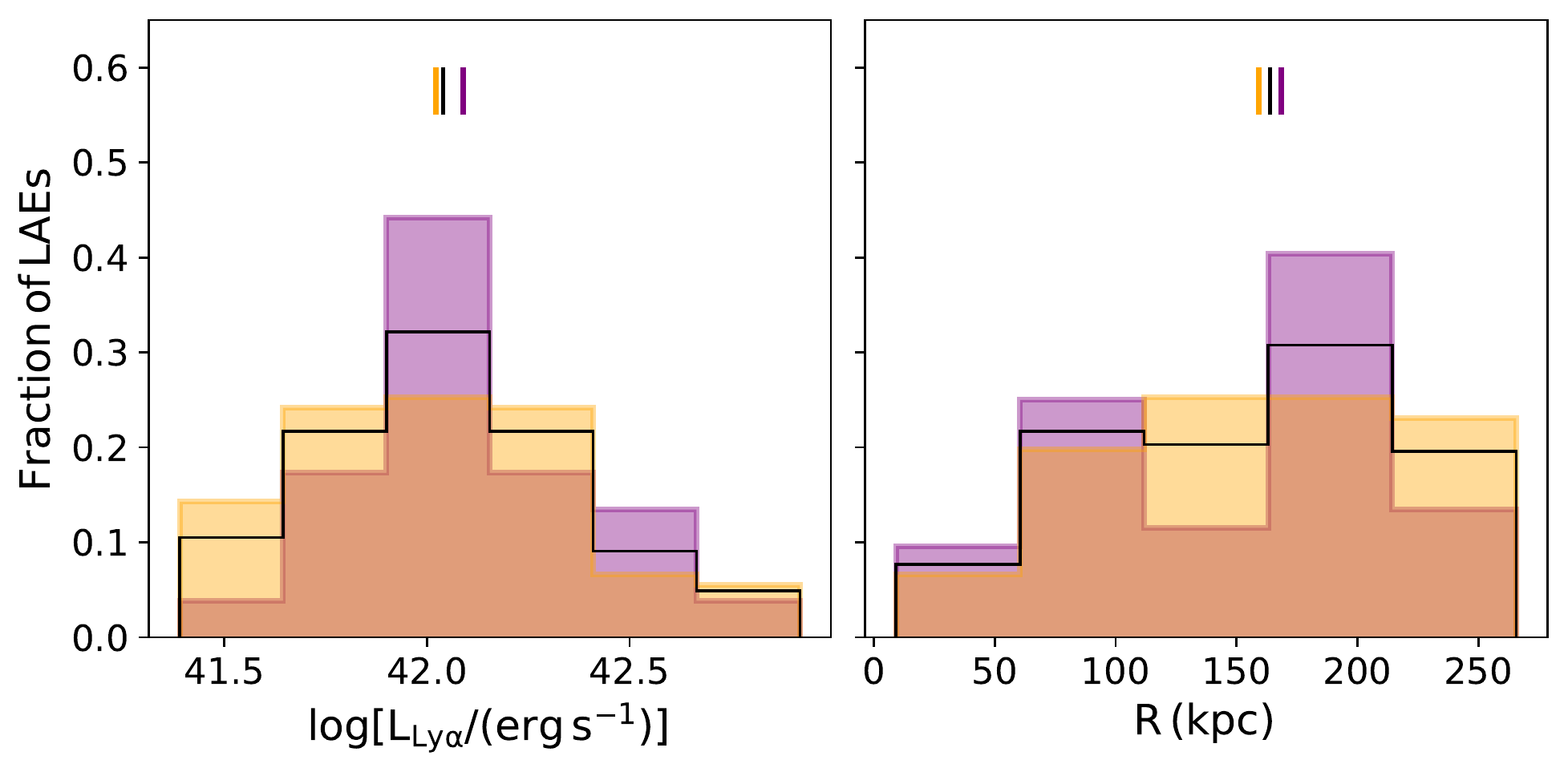}
\caption{Upper panels: comparison between the distribution of the rest-frame equivalent width (left panel) and velocity width (right panel) of the full \civ\ sample (black) with the sub-sample overlapping \maggiv\ catalogue (purple) and the remaining systems (orange). Lower panels: same as above, but comparing the Ly$\alpha$ luminosity (left panel) and the transverse separation (right panel) of the LAEs associated to the different \civ\ samples. Median values are shown as ticks.}
\label{fig:LLS_comparison}
\end{figure}

\begin{table*}
\centering
\begin{tabular}{ccccc}
\hline
Samples & $W_{\rm r}\,$(\AA) & $\Delta v_{90}\,(\rm km\,s^{-1})$ & $\log [L_{Ly\alpha}/(\rm erg\,s^{-1})]$ & $R\,(\rm kpc)$ \\
\hline
With \hi\ - Without \hi\   & 8.69$\times10^{-6}$ & 1.12$\times10^{-5}$ & 0.52 & 0.90 \\
With \hi\ - Full Sample    & 1.52$\times10^{-4}$ & 6.59$\times10^{-3}$ & 0.66 & 0.87 \\
Without \hi\ - Full Sample & 0.46 & 0.57 & >0.99 & >0.99 \\
\hline
\end{tabular}
\caption{Results of KS test ($p-$values) that measures the probability that the distribution of \civ\ and LAEs properties (equivalent width, velocity width, LAE luminosity, impact parameter) for systems that are (With \hi) and are not (Without \hi), connected to strong \hi\ absorbers are drawn from the same parent population.}
\label{tab:KS_HI}
\end{table*}

We investigate whether differences exist between the full sample of \civ\ absorbers relative to the sub-samples of \civ\ systems that are  connected to strong \hi\ systems or not. The results are shown in the upper panels of Figure \ref{fig:LLS_comparison}, while the $p-$values resulting from a KS test are listed in Table \ref{tab:KS_HI}. We observe a lack of weak ($W_{\rm r}\lesssim0.01$ \AA) \civ\ systems from \maggiv\ and a small excess of strong absorbers $W_{\rm r}\gtrsim0.1$ \AA, with $\approx40\%$ of the $W_{\rm r}\geq0.1$ \AA\ \civ\ systems included in this work connected to strong HI absorbers within $\pm500\rm\,km\,s^{-1}$. This suggests that the \civ\ systems connected to \hi\ absorbers are, on average, a factor $\approx3.7$ stronger compared to the \civ\ that is not associated to \hi\ absorbing gas. The KS test suggests that the systems with \hi\ detection, compared to those without \hi\ absorption and the full \civ\ sample, show an equivalent width that is unlikely to be drawn from the same parent distribution, with $p-{\rm value}<10^{-6}$ and $p-{\rm value}<10^{-4}$ respectively. On the other hand, the strength of the \civ\ absorbers without \hi\ detection is more similar to that of the full \civ\ sample ($p-{\rm value}>0.46$). 
The proximity to \hi\ absorbers seems to affect also the kinematics of the gas, since the \civ\ systems have a velocity width that is a factor $\approx2.3$ broader if connected to \hi\ absorbing gas. As found before, the KS test suggests that the $\Delta v_{90}$ of the \civ\ absorbers with \hi\ detection, compared to \civ\ without \hi\ absorption and the full \civ\ sample, are unlikely to be drawn from the same parent population ($p-{\rm value}<10^{-5}$ and $p-{\rm value}<10^{-3}$ respectively). \civ\ absorbers without \hi\ detection do not significantly deviates ($p-{\rm value}>0.57$) from the full \civ\ sample.

A similar analysis is performed to compare the transverse separation and the Ly$\alpha$ luminosity of the LAEs connected to the full \civ\ sample and to the systems that are or are not associated to \hi\ absorbers. From the results shown in the lower panels of Figure \ref{fig:LLS_comparison}, we notice that the \civ\ systems are associated to the population of LAEs that does not show any difference ($p-{\rm value}>0.50$) in the Ly$\alpha$ luminosity they emit and in their transverse separation from the absorbing gas. Thus, this comparison suggests that \civ\ systems matched to strong \hi\ absorbers have higher equivalent widths and are more broad kinematically, but this change in absorption properties does not influence significantly the emission properties of the associated LAEs. As mentioned above, \maggiv\ results support the picture in which LAEs are clustered within large scale structures and strong \hi\ absorbers are tracers of both the galaxies CGM and of the dense optically thick gas contained in the filaments connecting LAEs.

\begin{figure} 
\centering
\includegraphics[width=\columnwidth]{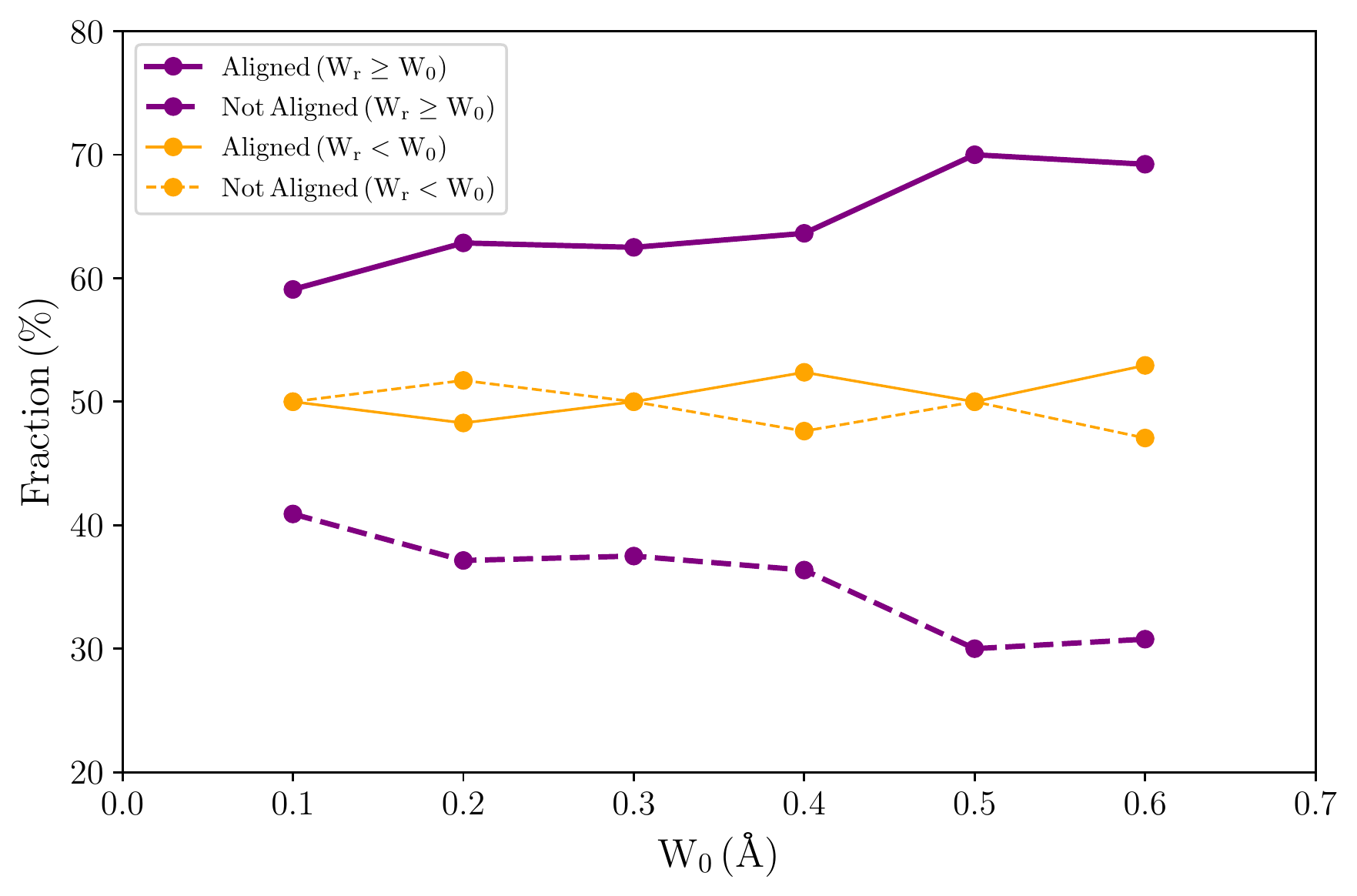}
\caption{Fraction of LAEs aligned ($\ang{0}-\ang{45}$ or $\ang{135}-\ang{180}$, solid line) and not aligned ($\ang{45}-\ang{135}$, dashed lines) with the axis defined by the absorber and the closest LAE, as a function of the absorbers equivalent width. We show results for LAEs connected to $W_{\rm\civ}\geq W_{0}$ and $W_{\rm\civ}< W_{0}$ as purple and orange lines, respectively.}
\label{fig:alignement_frac}
\end{figure}

We have shown so far that strong \civ\ systems are connected to \hi\ absorbers, suggesting that these systems might be tracers of the same large-scale structure as high-density neutral hydrogen. In support of this hypothesis, we would expect the galaxies to be aligned with respect to the positions of the \civ\ absorbers, similarly to what found when analysing strong \hi\ absorbers (see \maggiv). We thus follow the procedure adopted in \maggiv\ and measure the offset between the galaxy positions and the axis connecting the \civ\ absorbers to the closest LAE. Only the 38 \civ\ systems connected to 2 or more LAEs are included in this analysis. Results are shown in Figure \ref{fig:alignement_frac} as a function of \civ\ equivalent width threshold, $W_{0}$. We found an excess of aligned ($<\ang{45}$ and $>\ang{135}$) LAEs connected to strong ($W_{\rm r}\geq W_{0}$) \civ\ absorbers. The fraction of aligned LAEs increases with increasing absorption strength, while LAEs connected to weaker \civ\ ($W_{\rm r}<w_{0}$) does not show any preference to be aligned with the absorbers. 

\begin{figure*} 
\centering
\includegraphics[width=2\columnwidth]{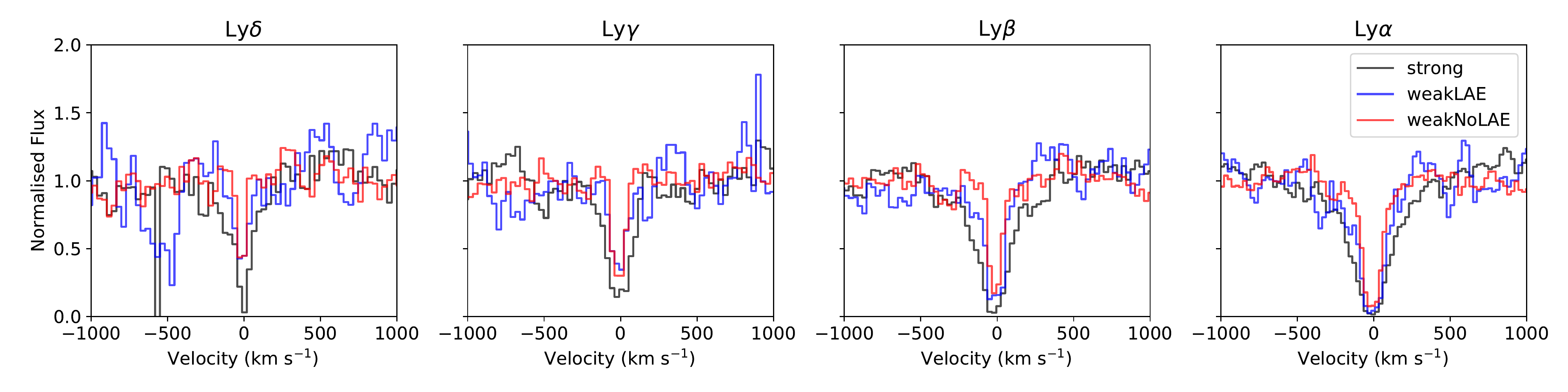}
\caption{Fraction of LAEs aligned (Stacks of Ly$\alpha$, Ly$\beta$, Ly$\gamma$ and Ly$\delta$ absorption at the redshift of each \civ\ system for strong absorbers ($W_{\rm r}\geq0.1$ \AA, black), weak absorbers ($W_{\rm r}<0.1$ \AA, blue) connected to at least one LAE and weak absorbers that are not connected to any galaxies within $\pm500\rm\,km\,s^{-1}$ (red).}
\label{fig:Hstacks}
\end{figure*}

While strong \civ\ absorbers appear to overlap at least in part with optically-thick \ion{H}{I} gas, the above analysis suggest that weaker \civ\ systems are
tracer of a more homogeneous medium, found also at larger distances from galaxies where the \ion{H}{I} column density is expected to decline \citep{Rudie2012}.
We thus investigate if there is any difference in the average \ion{H}{I} profiles associated with different `types' of \civ\ absorbers. To do so, we stacked the Ly$\alpha$, Ly$\beta$, Ly$\gamma$ and Ly$\delta$ absorption at the redshift of each \civ\ system (where data were available) and compare the results we obtained for strong \civ\ absorbers ($W_{\rm r}\geq0.1$ \AA), weak \civ\ absorbers ($W_{\rm r}<0.1$ \AA) connected to at least one LAE and weak \civ\ absorbers that are not connected to any galaxies within $\pm500\rm\,km\,s^{-1}$. 
 The result is shown in Figure \ref{fig:Hstacks} and suggests that strong \civ\ absorbers arise from higher column density regions, likely close to the spine of the cosmic filaments where both galaxies and optically-thick absorbers reside (see \maggiv), while weaker absorbers arise from lower column density regions away from the central parts of the filaments, where the presence of galaxies is more stochastic. Figure~\ref{fig:Hstacks} further hints at a progressive less significant hydrogen absorption in weak \civ\ systems far from galaxies compared to a similar population of metal absorbers near LAEs.

In force of this analysis, we propose a new picture of the distribution of metals around LAEs in which strong \civ\ absorbers, as well as \hi\ absorbers, trace dense regions of the filaments connecting galaxies. On the contrary, weaker \civ\ absorbers seem to lie progressively farther from the galaxies and hence farther from the densest regions of the filaments, pinpointing to the existence of an enriched, but more diffuse and homogeneous medium that degrades towards the IGM.

\subsubsection{The role of outflows}

\begin{figure} 
\centering
\includegraphics[width=\columnwidth]{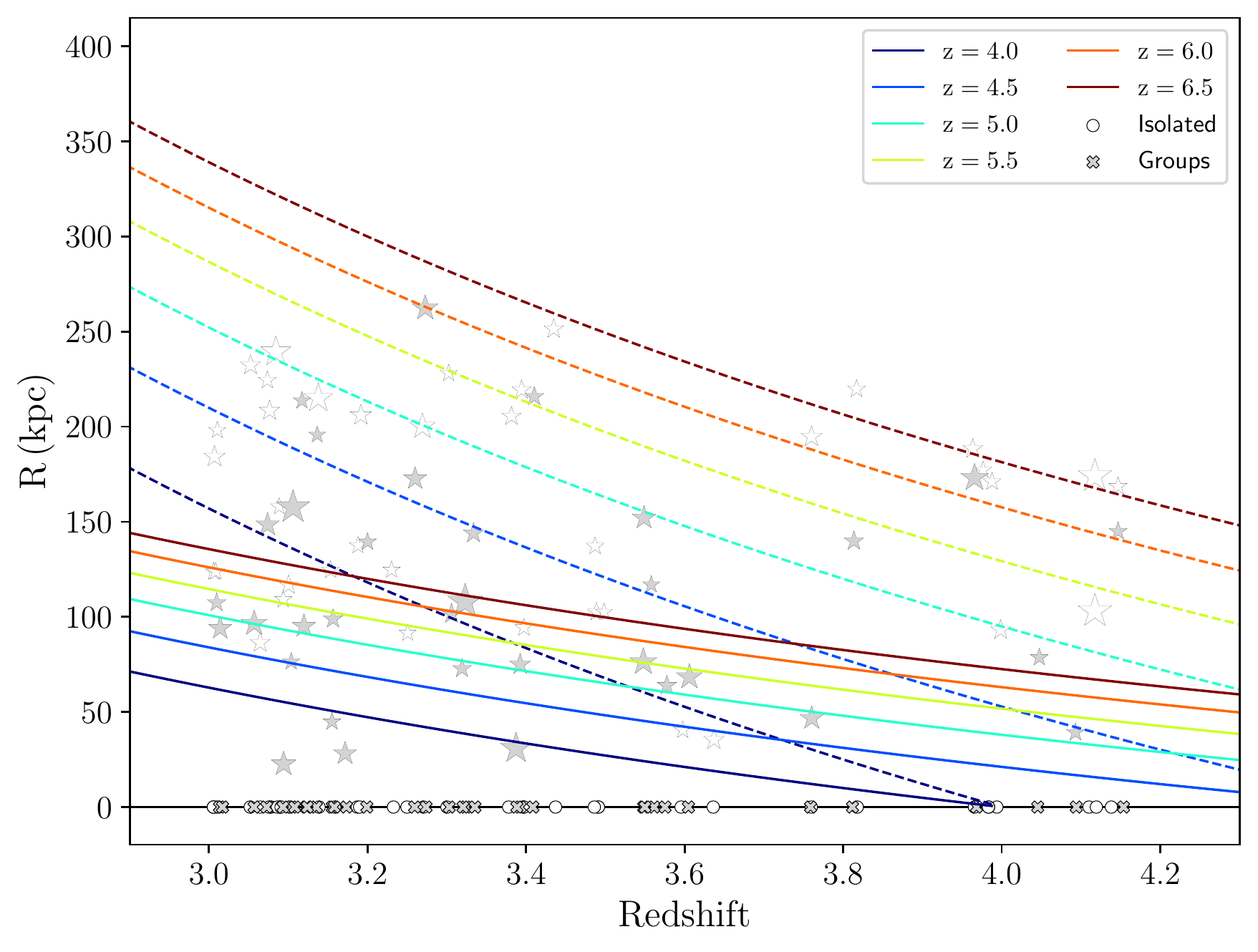}
\caption{Comparison between the distance traveled by outflows ejected during episodes of recent star formation activity (e.g. launched at redshift between $z=4$ and $z=6.5$ with steps $\Delta z=0.5$) in the galaxy and the transverse separation between the LAEs (shown at $R=0~\rm\,kpc$) and the connected \civ\ absorbers (stars), assuming the outflows travel with average velocities $<v>=100, 250\rm\,km\,s^{-1}$ (solid and dashed lines respectively). We distinguish between \civ\ associated to isolated galaxies (white-filled markers) and groups (gray-filled markers). The size of the marker is weighted by the rest-frame equivalent width.}
\label{fig:CIVoutflows}
\end{figure}

A variety of models has been studied in the literature in order to understand the physical processes that lead to the \civ\ enrichment of the CGM and IGM, spanning from Population III stars at high redshift ($\rm z\sim25-30$), to more recent ejections of galactic winds and feedback powered by star formation activity in galaxies at $\rm z\sim3-10$ \citep[e.g.][]{Simcoe2004}. The \civ\ absorbers we observe at redshift $z\sim3-4$ may be the result of cumulative feedback processes that occurred at higher redshift and ongoing feedback ejections at the epoch near the peak of the star formation rate density towards $z\approx 2$. 

We attempt to disentangle these two effects by exploring if it is plausible that the bulk of outflows powered by recent star formation in galaxies has instantaneously enriched all the observed absorption systems near LAEs. To this end, in Figure~\ref{fig:CIVoutflows} we compare the transverse separation between the LAEs and the \civ\ absorbers found within $\pm500\rm\,km\,s^{-1}$ with the distance traveled by outflows, assuming they are launched at redshifts between $z=4$ and $z=6.5$ with steps $\Delta z=0.5$ and travel at average velocities $<v>=100, 250\rm\,km\,s^{-1}$ (solid and dashed lines respectively). We found that to reach \civ\ absorbers at separations $R\gtrsim150\rm\,kpc$, outflows must be launched at redshift $z\gtrsim6.5$ if travelling at $<v>=100\rm\,km\,s^{-1}$, and are thus linked to episodes of star formation that happened in the galaxies $\Delta t\gtrsim1\rm\,Gyr$ before the epoch of observations.  Outflows launched $\Delta t\lesssim750\rm\,Myr$ before the observations and linked to more recent star formation activity in the galaxies must instead travel at average velocities $<v>\gtrsim250\rm\,km\,s^{-1}$ to reach the \civ\ absorbers up to the edge of the FoV ($R\approx250\rm\,kpc$). 

In a similar analysis, \cite{Diaz2021} interpreted  the connection between 11 \civ\ absorbers and LAEs at $z>4.7$ as potentially due to the relatively recent ($\Delta t\lesssim700\rm\,Myr$ at $z\approx5$) galaxy activity, modelled as outflows started at $z=10$ and travelling at average velocities $<v>\lesssim400\rm\,km\,s^{-1}$ for $89\%$ (8/9) of the LAE sample.
This scenario suggests that only outflows traveling at average high velocities ($<v>\gtrsim250\rm\,km\,s^{-1}$) are able to enrich the gas around a fraction of the LAEs if powered by recent ($\Delta t\lesssim500\rm\,Myr$) star formation activity in galaxies.

\section{Summary and conclusions}\label{sec:summary}

In this work, we explored the connection between an ionized gas phase, traced by absorption-line systems in quasar spectra, and a population of Ly$\alpha$ emitting galaxies identified in the MUSE Analysis of Gas Around Galaxies (MAGG) survey, which combines medium-deep VLT/MUSE observations of 28 quasar fields with high $S/N$ and high resolution spectroscopy of the central sources. We traced the ionized gas by assembling a sample of 220 \civ\ and 108 \siv\ absorption-line systems at redshift $3.0\lesssim z \lesssim4.5$. To extend the analysis of gas-galaxies connection to the lower end of the galaxy mass function, we identified  $>1000$ LAEs in the MUSE cubes at $3.0\lesssim z\lesssim6.0$, 292 of which lie in the \civ\ redshift path and are candidates for an association with the absorbing gas. We found 143 LAEs within a velocity separation $|\Delta v|\leq500\rm\,km\,s^{-1}$ from \civ\ absorbers. 
Given the large size of this sample, we were able to study, for the first time with significant statistics, both the global correlation between LAEs and \civ/\siv\ and the properties of the individual associations. 
The main findings of this work are summarized next.

\begin{itemize}

\item[--] We measured, for the first time, the luminosity function of LAEs connected to \civ\ absorbers. Compared to an updated version of the field luminosity function derived from  MAGG, we found a normalization $\approx2.4$ higher for LAEs associated with the \civ\ absorbing gas. This analysis revealed that the LAE number density is sensitive to the proximity to different tracers studied in MAGG and progressively increases from the field, to the surrounding of \civ\ gas, strong \hi\ absorbers and central quasars.

\item[--] We measured the two-point LAE auto-correlation and LAE-\civ\ cross-correlation functions and the respective reduced angular auto- and cross- correlation functions. We found a clustering length $r_{0}=2.07^{+0.23}_{-0.24}\rm\,h^{-1}\,cMpc$ for LAEs and $r_{0}=1.23^{+0.25}_{-0.27}\rm\,h^{-1}\,cMpc$ for LAE-\civ\ absorbers, assuming a fixed slope $\gamma=1.8$. Employing the Cauchy-Schwarz relation, we found that a fraction of the \civ\ absorbers is likely to arise not only in the halo of the LAEs they are connected to, but also from other regions, such as locations in the IGM not connected to galaxies. The LAE-\civ\ cross-correlation function lies below the cross-correlation function measured between LAEs and optically-thick \hi\ absorbers, indicating a larger contribution from more underdense regions in the \civ\ case. 

\item[--] Moving to a galaxy-centered approach (i.e., centering on the galaxies and exploring the surrounding gas distribution) we measure the LAE covering fraction as a tracer of the probability to observe \civ\ absorbers within $\pm500\rm\,km\,s^{-1}$ from a galaxy. The covering fraction is observed to decrease with increasing \civ\ absorption strength and increasing impact parameter up to $\approx 100-150~\rm kpc$ from galaxies, at which point it is observed to flatten with elevated values far beyond the virial radius and up to the edge of the FoV. We also found a higher covering fraction for galaxies with luminosity $\log[L_{\rm Ly\alpha}/(\rm erg\,s^{-1})]>42.2$ compared to fainter LAEs. 

\item[--] Among the galaxies connected to the \civ\ absorbers, $\approx30$ per cent are isolated within $\pm500\rm\,km\,s^{-1}$, while the remaining $\approx70$ per cent are part of a group.  The absorbers connected to 1 or at least 2 LAEs show, on average, a rest-frame equivalent width that is a factor $\approx1.5$ and $\approx4.6$ higher than the median of the full sample, respectively. However, the effect on the gas kinematics is less significant, with only the absorbers connected to more than 1 LAE exhibiting a $\Delta v_{90}$ that is a factor $\approx1.7$ above the median of the full \civ\ sample. \civ\ absorbers with larger equivalent width and $\Delta v_{90}$ tend to be associated with brighter LAEs. The probability to observe a \civ\ absorber near an LAE is a factor $\approx3$ higher around galaxies in groups compared to the isolated ones at any transverse separation. As opposed to what is observed at $z\lesssim2$, the excess of \civ\ absorption around galaxies in groups is present even if controlling for the properties of isolated galaxies and groups.

\item[--] Comparing the statistics of \civ\ and \siv\ absorbers, we found that $\approx29$ per cent of the \civ\ systems show \siv\ absorption with the same number of components aligned with the \civ\ ones.
The fraction of \siv\ systems' matched to \civ\ decreases with decreasing equivalent width, and the ratio between the equivalent width of \siv\ and \civ\ is found to be less than unity, as expected from ionization models for densities $n_{\rm H}\lesssim 10^{-2}~\rm cm^{-3}$. Thus, we conclude that the two ions, with a similar ionization potential and in the same ionization stage, mostly arise from the same gas phase when detected. This is further supported by the fact that an independent analysis of gas-LAE connection leads to similar results if using either \civ\ or the \siv\ ions as a tracer.

\item[--] Despite the large projected distance from LAEs at which \civ\ absorption is detected (up to the edge of the FoV, corresponding to $R\approx250\rm\,kpc\gtrsim 7\rm\,R_{vir}$ for a dark matter halo $M_{\rm H}\approx10^{11}\,M_{\odot}$), we have found that at least a fraction of the absorbers can be instantaneously enriched by outflows traveling at $<v>\gtrsim250\rm\,km\,s^{-1}$ and launched $\Delta t\lesssim750\rm\,Myr$ before the observations. This makes the CGM and IGM, and in particular an intermediate ionized gas-phase, at least partially linked to the recent star formation activity in galaxies. 

\item[--] Compared to similar studies about the connection of metals and LBGs, LAEs and (strong) \civ\ absorbers are less clustered, with a correlation length that is a factor of $\approx2$ lower than LBGs at similar redshifts. For both galaxy populations, the ionized gas is extended far beyond the typical virial radius of the halo with a mild (if any) decline in the equivalent width beyond a few hundred kiloparsecs from the galaxy position. 

\item[--] We found that 43 out of 220 \civ\ systems are also connected to strong \hi\ absorbers within $\pm500\rm\,km\,s^{-1}$ in MAGG. We found that these systems show, on average, higher rest-frame equivalent width and more complex kinematics compared to those that are not associated to \hi\ absorption. We did not observe any difference between the properties of the LAEs connected to \civ\ absorber with or without \hi\ detection, but we found that it is more likely ($\approx 2\times$) for strong \hi\ absorbers to be associated with LAEs than \civ\ systems.  
\end{itemize}

The combination of this study on gas at intermediate ionization near LAEs and the findings presented in our \maggiv\ paper on the link between LAEs and optically-thick \hi\ absorbers, together with literature results, allows to paint a clearer picture for how hydrogen and metals are distributed around $z\approx 3-4$ star-forming galaxies, for the first time extending to lower mass scales $(M_{\rm H}\approx10^{11}\,\rm M_{\odot})$ with a complete and coherent survey. As argued in \maggiv, data support a model in which star-forming galaxies lie within large-scale structures (i.e., cosmic filaments) and strong \hi\ absorbers are a tracer of the neutral phase both within the CGM of the embedded star-forming galaxies and along the filaments connecting halos. This paper confirms a similar picture for the metals traced by \civ, but only when considering relatively strong absorbers, with $W\gtrsim 0.1~$\AA, which are more clearly correlated to galaxies and the LLS themselves. For weaker absorbers, our MUSE observations support the idea of a more widespread and homogeneous medium that extends far beyond the CGM of galaxies, also reaching regions away from the densest parts of the filaments where optically-thick \hi\ absorbers are found. 

We thus propose a `three-component model' to fit all the principal features uncovered by MAGG near LAEs, which agree with the findings and conclusions by \citet{Muzahid2021}. The first component is represented by the CGM of individual systems that gives rise to the strongest \hi\ and \civ\ absorbers observed at close separation from galaxies and which accounts for the elevated covering factors of hydrogen and metals near LAEs. On these scales, it is more likely that galaxies have a direct impact on the properties of the CGM, as seen for instance in some correlations between emission and absorption properties (e.g., \civ\ velocity width and absorption strength with Ly$\alpha$ luminosity), although as argued in \maggiv\ the patchy nature of this gas leads to high scatter in the observed relations. 
The second component describes the overdense gas along filaments, which accounts for a large fraction of LLSs and for the remaining fraction of high equivalent-width absorbers. This gas, together with the outer CGM components, is what drives the excess of galaxies located at distances greater than a few times the virial radius and sets the normalization and shape of the luminosity and cross-correlation function for strong \hi\ and \civ\ systems. The third component is represented by the lower equivalent-width but enriched medium, farther from the denser CGM and the central parts of the filaments connecting galaxies. This medium, which is not easily selected by LLSs, i) accounts for the substantial fraction of \civ\ for which no LAE counterparts are found; ii) lowers the normalization of the luminosity function and the correlation length of the cross-correlation function; iii) leads to a  high covering factor extending to large distances from galaxies; iv) yields a weak dependence of equivalent width with impact parameter. This more diffuse gas is harder to detect with less sensitive tracers like \siv, which explains the stronger clustering/higher normalization of LAEs near \siv\ than \civ\ (in other words, \siv\ is a better tracer of the first two components of this model). 

Finally, environment shapes the distribution of gas, particularly the stronger absorbers that seem to be more affected by the presence of groups. Currently, however, it is not particularly clear how environment affects different gas phases (e.g. ionized and neutral components) differently as a function of redshift, and which physical mechanisms are primarily responsible for the observed trends. Refining this picture is the subject of ongoing observational efforts.
Observations of other galaxy populations that reside in the same structures are expected to yield comparable results. 
For instance, LBGs, which are not completely independent from LAEs and relatively more clustered, are too embedded within these structures and carry their own CGM. Hence, most of the trait discussed for LAEs apply to LBGs (as observed), with the exception of a higher clustering of this population. 

Biases against the detection of particular systems (e.g., dusty or passive galaxies) are surely present and will inevitably lead to changes in this picture. As argued in \maggiv, however, the number density of LAEs is sufficiently high that this population becomes a very good tracer of the large-scale structures surrounding galaxies, within which additional undetected populations will reside. We therefore speculate that it is implausible that populations at much lower number density will drastically alter the model put forward here. Future multiwavelength surveys, particularly at infrared/millimetre wavelengths, are required to settle this point. Sustained effort should also be invested in theoretical work to develop a more quantitative model, e.g. using numerical simulations, that can be compared more quantitatively with our observations to test this `three-component model' rigorously.

\section*{Acknowledgements}
This project has received funding from the European Research Council (ERC) under the European Union's Horizon 2020 research and innovation programme (grant agreement No 757535) and by Fondazione Cariplo (grant No 2018-2329). SC gratefully acknowledges support from the European Research Council (ERC) under the European Union’s Horizon 2020 research and innovation programme grant agreement No 864361. This work is based on observations collected at the European Organisation for Astronomical Research in the Southern Hemisphere under ESO programme IDs 
197.A-0384, 
065.O-0299,
067.A-0022,
068.A-0461,
068.A-0492,
068.A-0600,
068.B-0115,
069.A-0613,
071.A-0067,
071.A-0114,
073.A-0071,
073.A-0653,
073.B-0787,
074.A-0306,
075.A-0464,
077.A-0166,
080.A-0482,
083.A-0042,
091.A-0833,
092.A-0011,
093.A-0575,
094.A-0280,
094.A-0131,
094.A-0585,
095.A-0200,
096.A-0937,
097.A-0089,
099.A-0159,
166.A-0106,
189.A-0424.
This work used the DiRAC Data Centric system at Durham University, operated by the Institute for Computational Cosmology on behalf of the STFC DiRAC HPC Facility (www.dirac.ac.uk). This equipment was funded by BIS National E-infrastructure capital grant ST/K00042X/1, STFC capital grants ST/H008519/1 and ST/K00087X/1, STFC DiRAC Operations grant ST/K003267/1 and Durham University. DiRAC is part of the National E-Infrastructure. This research made use of Astropy \citep{Astropy2013}.

\section*{Data Availability}
The VLT data used in this work are available from the European Southern Observatory archive (\url{https://archive.eso.org/}) either as raw data or phase 3 data products. Spectroscopy obtained at the Keck telescopes is available via the Keck Observatory Archive (KOA, \url{https://www2.keck.hawaii.edu/koa/public/koa.php}).
Cubextractor can be obtained upon request by contacting Sebastiano Cantalupo, and other codes used in this paper have been made available at http://www.michelefumagalli.com/codes.html.

\section*{Online supporting information}
We provide as online material the fits of all the \civ\ and \siv\ absorption lines included in the samples used in this work along with tables listing the properties for each system including redshift, equivalent width and velocity width. We also include the full list of $\rm Ly\alpha$ identified in the MAGG survey at $SNR>7$ and the properties of each galaxy including redshift, $\rm Ly\alpha$ luminosity and impact parameter.



\bibliographystyle{mnras}
\bibliography{reference} 




\appendix

\section{Theoretical comparison between \civ\ and \siv\ absorbers as tracers of the ionized gas} \label{app:cloudy}

\begin{figure} 
\centering
\includegraphics[width=\columnwidth]{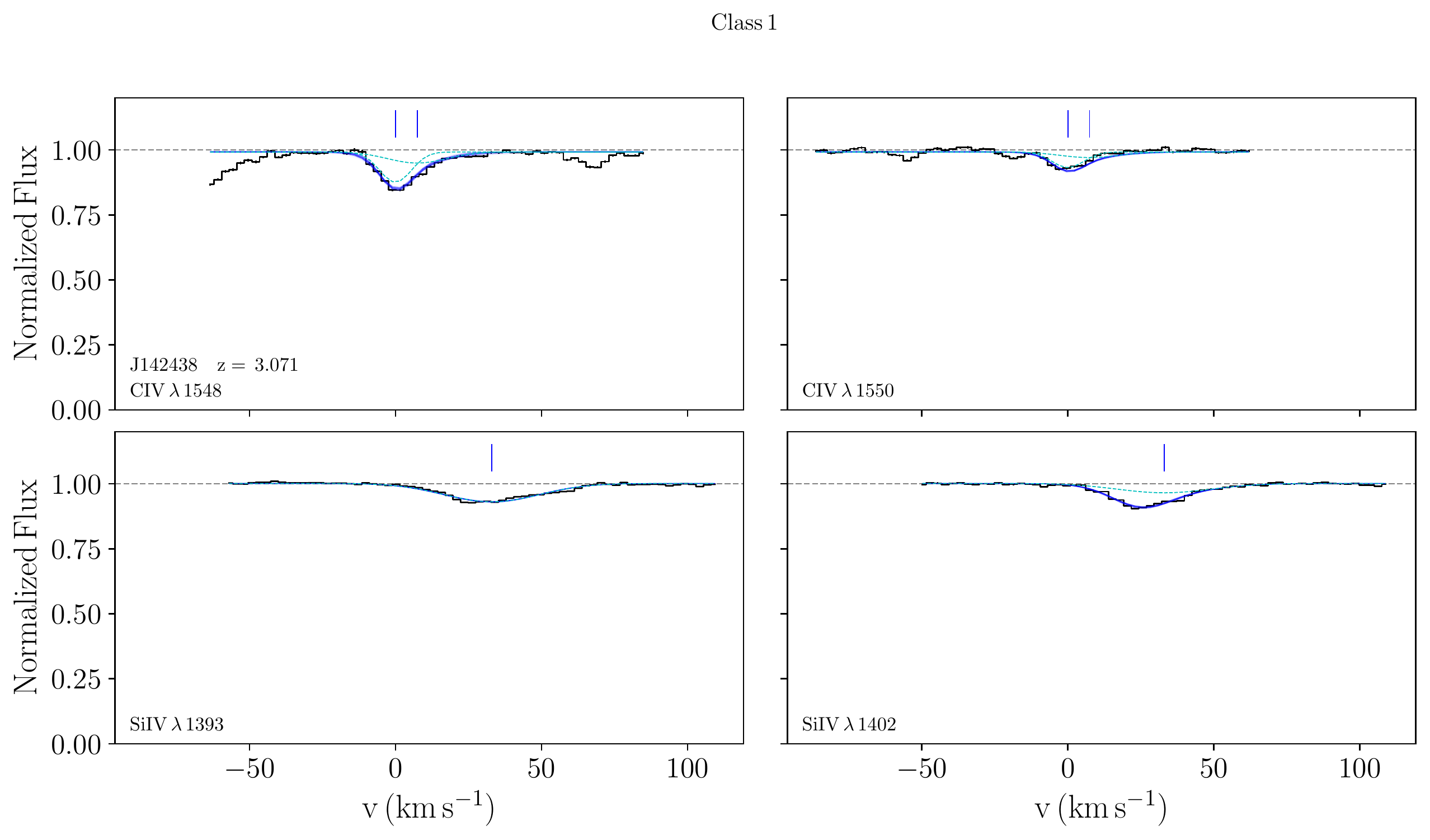}
\includegraphics[width=\columnwidth]{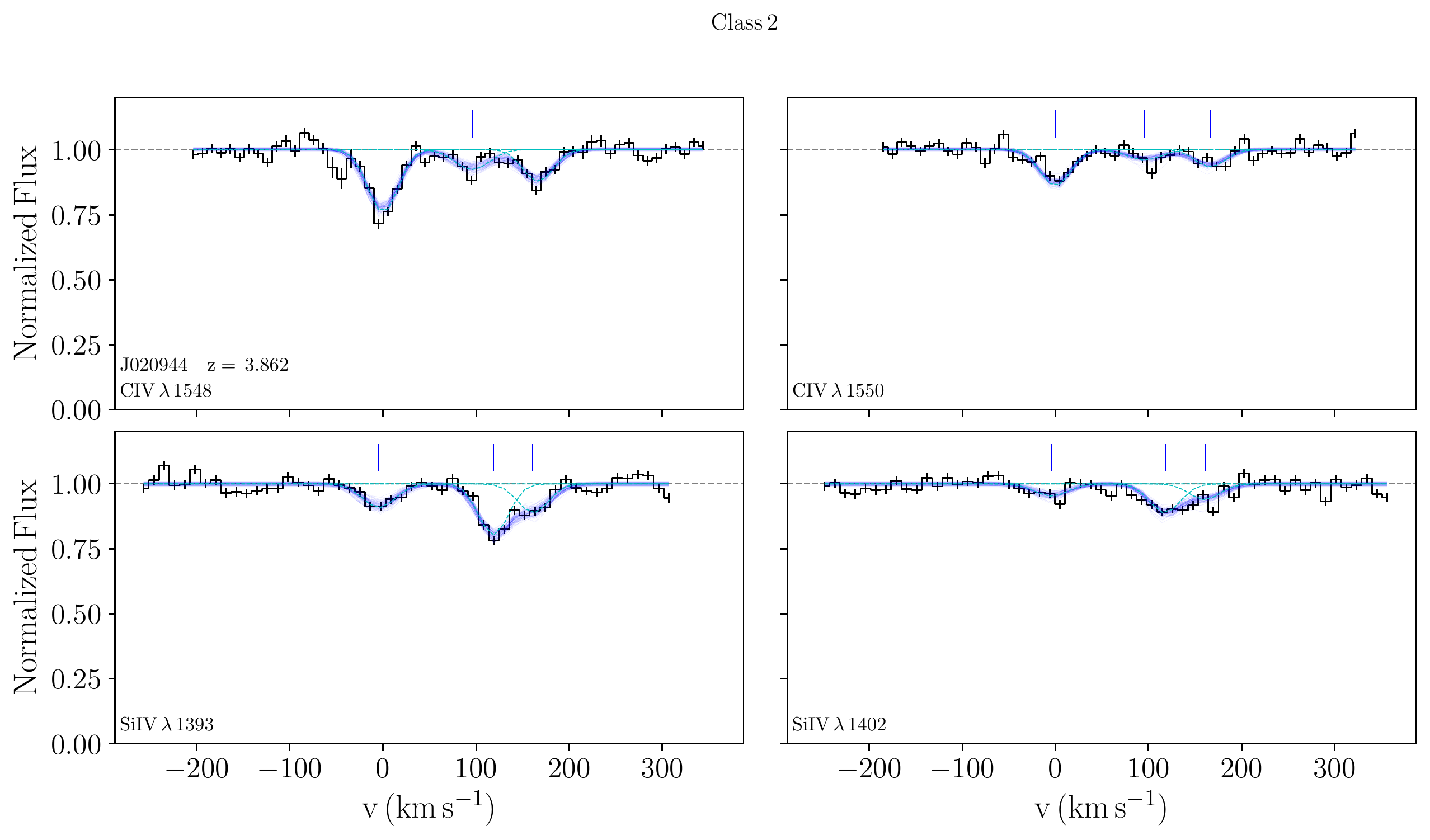}
\includegraphics[width=\columnwidth]{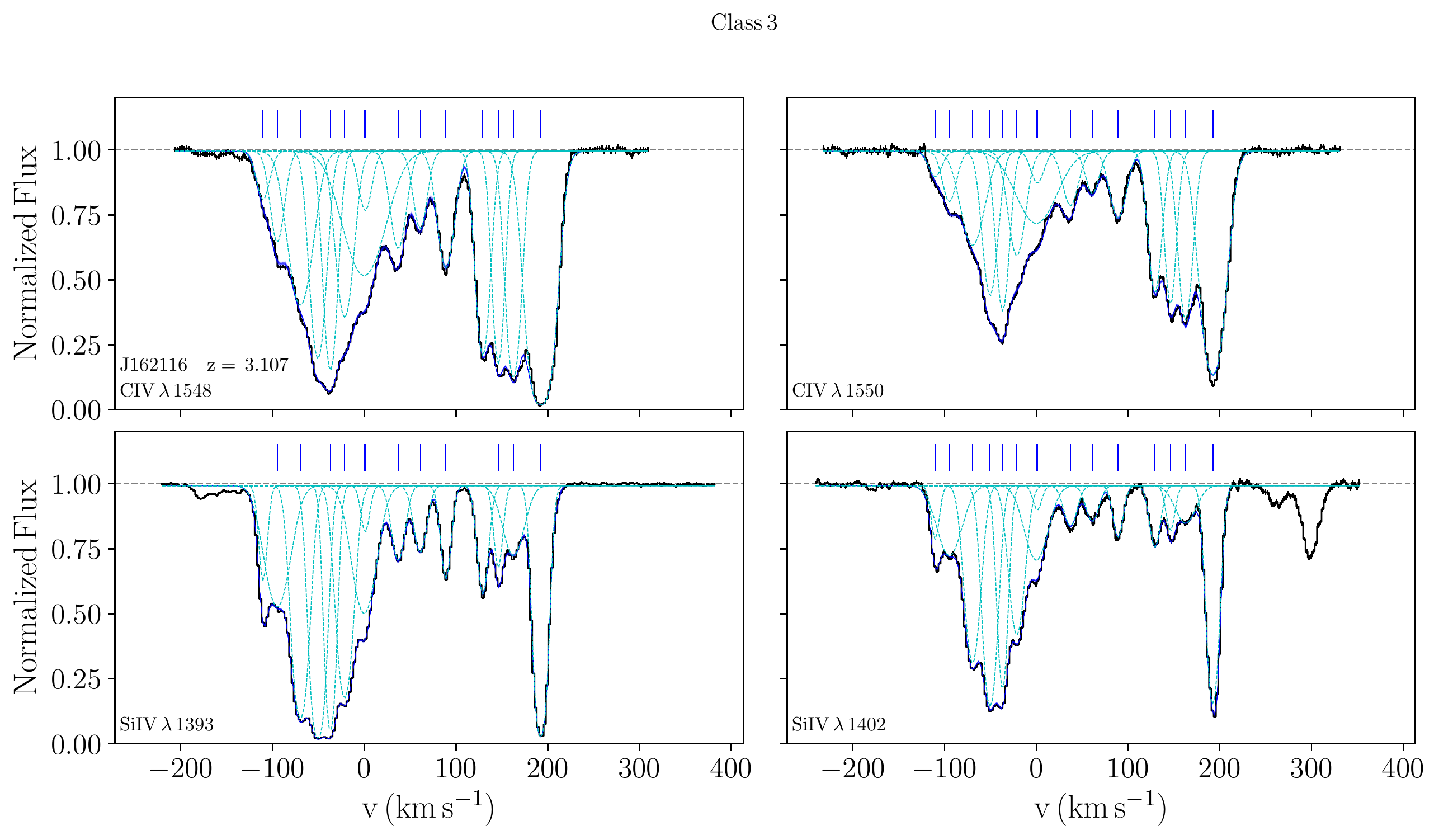}
\caption{Examples of \civ\ and \siv\ systems matched within $\pm150\rm\,km\,s^{-1}$ in velocity space and classified according to the number and the alignment of Voigt components (class 1-3 from top to bottom, see text for details).}
\label{fig:SiIV_Class}
\end{figure}

\begin{figure*} 
\centering
\includegraphics[width=1.8\columnwidth]{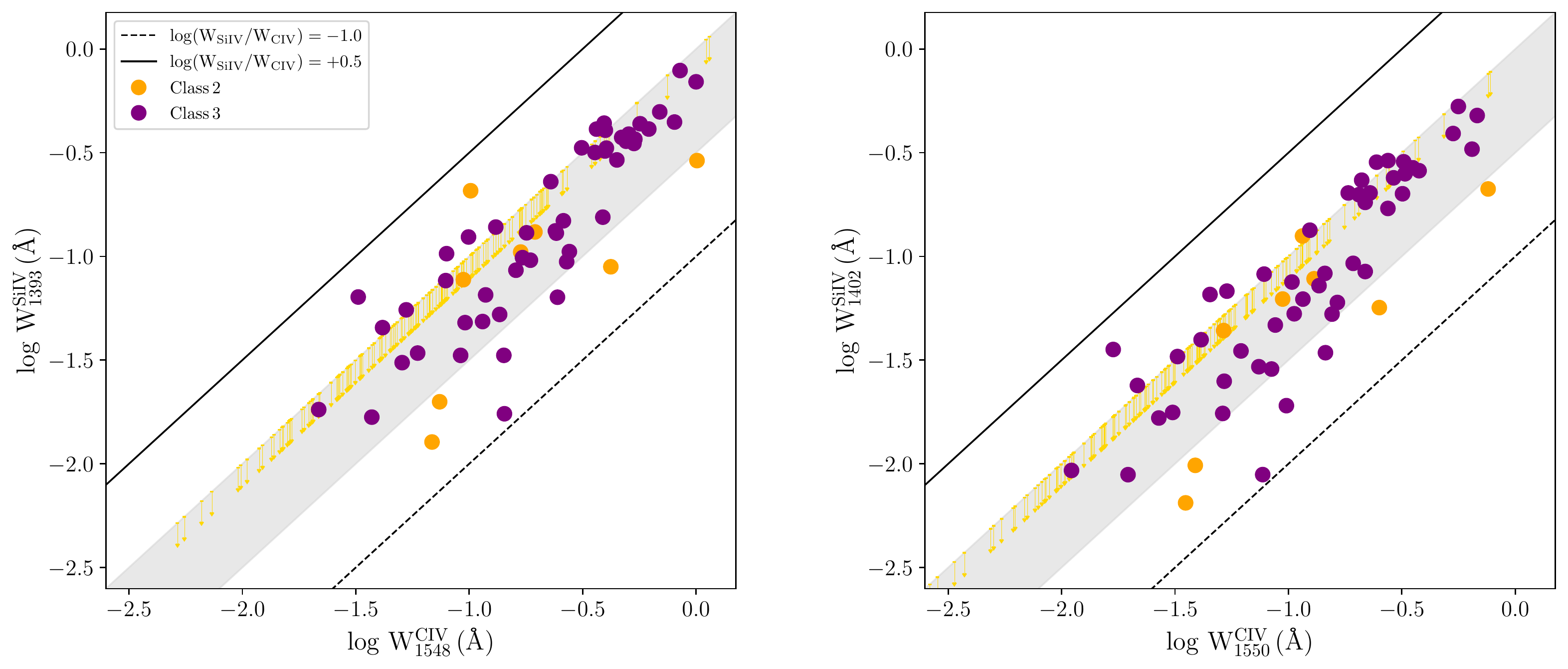}
\caption{Comparison between the rest-frame equivalent width of \siv\ absorbers, $W_{\rm\siv}$, as a function of the equivalent width of the matched \civ\ system, $W_{\rm\civ}$. Aligned absorbers from class 3 (purple) are distinguished from class 2 systems (orange). The predictions from {\sc Cloudy} ionization models (see text for details) are shown as solid and dashed lines, marking $\log(W_{\rm\siv}/W_{\rm\civ})=+0.5$ and $\log(W_{\rm\siv}/W_{\rm\civ})=-1.0$ respectively. The shaded region highlights the range occupied by the $\approx 2/3$ of the distribution of aligned detections, corresponding to $-0.5<\log(W_{\rm\siv}/W_{\rm\civ})<0$. Arrows show upper limits corresponding to $\log{W_{\rm\siv}}\leq\log{W_{\rm\civ}}$ for \siv\ non detections.}
\label{fig:EWratio_obs}
\end{figure*}

We complemented the comparison between \civ\ and \siv\ absorbers shown in Section \ref{sec:siv} with a more theoretical approach: indeed, one way to test the assumption that \civ\ and \siv\ absorbers trace the same gas phase is to examine whether absorbers arising at the same redshift are composed of the same number of components that are aligned in redshift. By searching for \siv\ lines detected within $|\Delta v|\leq150\,\rm km\,s^{-1}$ from a \civ\ absorber, we identify 63 matched systems (corresponding to the $\approx29\%$ of the full \civ\ sample, and $\approx58\%$ of the \siv\ systems). We model the absorption of these profiles by fitting the lines with the MC-ALF code, following the method presented in \citet{Rudie2019} to identify the number of aligned components.
Based on the alignment of each component, we classify the matched systems into three classes (see Figure \ref{fig:SiIV_Class} for examples): i) systems that are clearly misaligned, suggesting that \civ\ and \siv\ absorbers are not tracers of the same gas structure. This class contains $\approx10\%$ ($6/63$) of the matched systems; ii) systems in which the strong components are offset or only partially aligned, i.e. separated by $\Delta v\gtrsim10-20\rm\,km\,s^{-1}$ in velocity space ($\approx13\%$, $8/63$, of the matched systems belong to this class, plus the 157 \civ\ systems that do not show a \siv\ counterpart); iii) systems with the same number of components that are clearly aligned (this class contains $\approx77\%$ (49/63) of the matched systems). 
The rate of matches we observe is qualitatively in line with the recent statistical measures from \citet{DOdorico2022}. With a sample of over 600 \siv\ absorbers detected in 147 quasar sightlines at redshift $2.1\lesssim z_{\rm QSO}\lesssim6.5$, they studied the column density distribution function and found that the abundance of \civ\ and \siv\ ions trace each other across a large redshift range. However, the \siv\ number density per unit redshift path is a factor $\approx2$ lower than the \civ\ absorbers at any redshift. This difference implies a fraction of $\lesssim 50\%$  of the \civ\ systems with a detected \siv\ counterpart, as we observe.

\begin{figure} 
\centering
\includegraphics[width=\columnwidth]{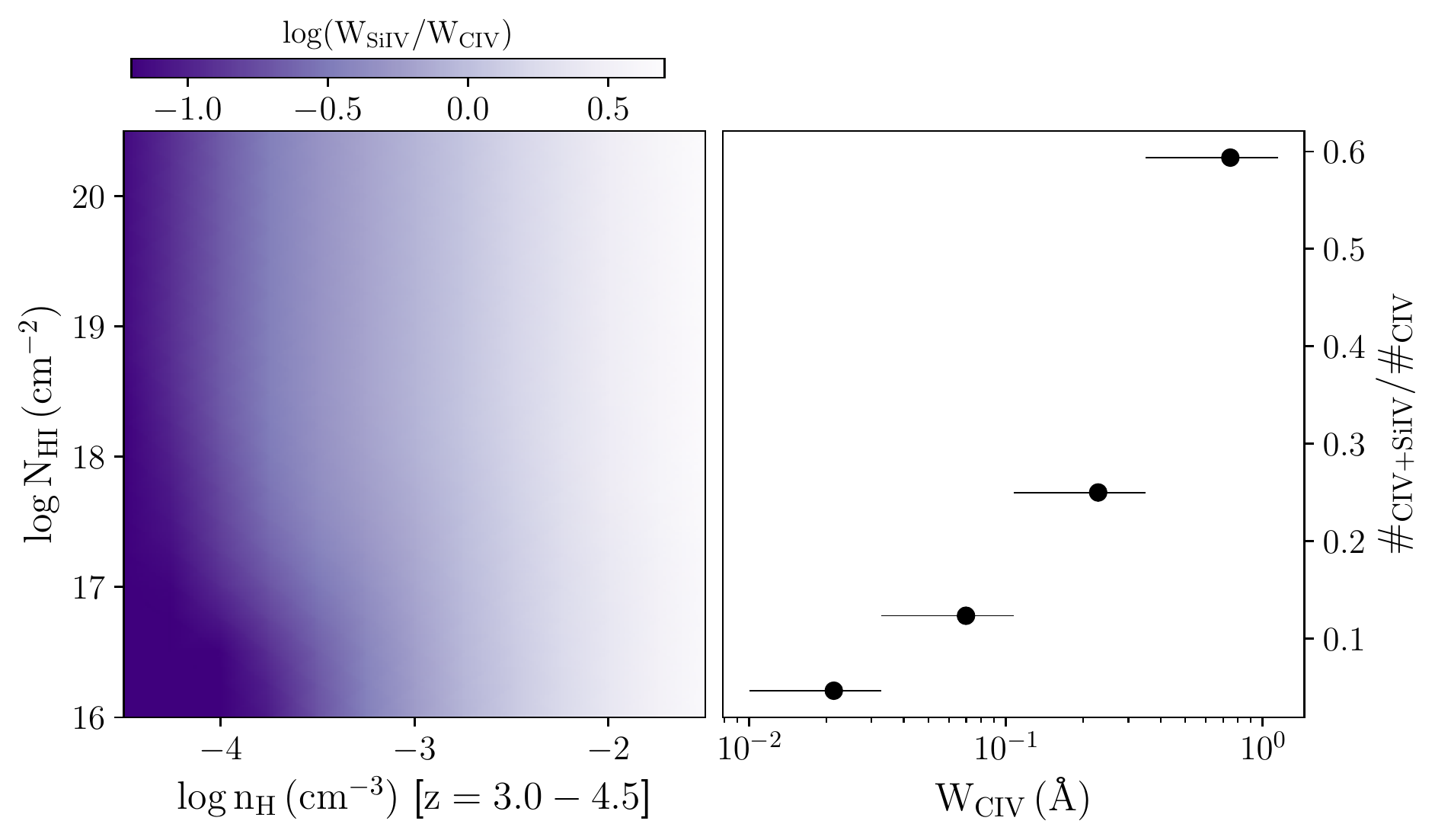}
\caption{Left panel: mean equivalent width ratio over the redshift range $z=3.0-4.5$ as predicted from a grid of ionization models in the $\log N_{\rm HI}-\log n_{\rm \ion{H}{}}$ plane. The metallicity is fixed to $\log (Z/Z_{\odot})=-2.0$. Right panel: distribution of the matching \civ-\siv\ detections, normalized to the number of \civ\ systems in each $W_{r}$ interval, as a function of the \civ\ rest frame equivalent width. Horizontal error-bars account for the width of each bin.}
\label{fig:EWratio}
\end{figure}

Having identified the aligned components in each transition, we measure the rest-frame equivalent width for class 2 and 3 systems and recover how the \siv\ equivalent width, $W_{\rm\siv}$, scales as a function of the equivalent width of the matched \civ\ absorber, $W_{\rm\civ}$. This estimate allow us to better understand the nature of the class 2 systems and test whether the partial alignment or the offset of the strong components is a sign that systems are not tracing the same gas strucutres. To complete the picture, we also compare the observed trend with the theoretical predictions from a set of ionization models using {\sc Cloudy} \citep{Ferland2013}. The results from the observations are shown in Figure \ref{fig:EWratio_obs}, where the shaded region and its limits are derived from {\sc Cloudy},
using the ``minimal grid'' from \citet{Fumagalli2016} 
Assuming the metallicity to be fixed at $\log (Z/Z_{\odot})=-2.0$, we show the mean equivalent width ratio over the redshift range $z=3.0-4.5$ as a function of the neutral hydrogen column density and the density in the left-panel of Figure~\ref{fig:EWratio}. 

As it can be seen from this grid, the equivalent width ratio is $\log(W_{\rm\siv}/W_{\rm\civ})\approx-1.0$ at densities $\log (n_{\rm \ion{H}{}}/\rm cm^{-3})\lesssim-4$ (for reference, the mean cosmic density at redshift $z\sim3$ is $\log (n_{\rm \ion{H}{}}/\rm cm^{-3})\approx-5.0$), increases to $\log(W_{\rm\siv}/W_{\rm\civ})\approx0$ at $-4\lesssim\log (n_{\rm \ion{H}{}}/\rm cm^{-3})\lesssim-2.5$ and about $\log(W_{\rm\siv}/W_{\rm\civ})\approx0.5$ at higher densities $\log (n_{\rm \ion{H}{}}/\rm cm^{-3})\gtrsim-2.5$. 
Thus, for common CGM/IGM densities of $\log (n_{\rm \ion{H}{}}/\rm cm^{-3})\lesssim-2.5$, it is reasonable to expect that only a fraction of \civ\ absorbers is detected also in \siv, which will typically appear as weaker line. Indeed, the right panel of Figure~\ref{fig:EWratio}, which shows the distribution of the \civ-\siv\ detections as a function of the \civ\ rest frame equivalent width, provides clear evidence that detection of \siv\ counterparts decline with decreasing strength of the \civ\ absorbers, significantly dropping close to the sensitivity limits of our spectra at $W_{\rm \siv} \lesssim 10^{-2}~$\AA. 

Figure~\ref{fig:EWratio_obs} provides  evidence that our observations of the aligned components are consistent with the ionization models predictions since the class 3 systems are scattered between $-1.0\lesssim\log({W_{\rm\siv}/W_{\rm\civ}})\lesssim0.5$. We observe most of the matched systems at $\log{W_{\rm\siv}}\lesssim\log{W_{\rm\civ}}$, with only few \siv\ absorbers stronger than their \civ\ counterpart. About $38\%$ (3/8) of the partially-aligned systems (class 2) follow the trend observed for the aligned systems, while the remaining sample is scattered toward a lower ratio $\log({W_{\rm\siv}/W_{\rm\civ}})\approx-1.0$ pinpointing to the presence of \civ\ absorption that is significantly stronger or with a higher number of components compared to the \siv\ counterpart. We show upper limits (gold arrows in Figure \ref{fig:EWratio_obs}) for the \civ\ absorbers without \siv\ counterpart within $\pm150\rm\,km\,s^{-1}$, assuming $\log{W_{\rm\siv}}\leq \log{W_{\rm\civ}}$. As expected from the analysis above, we observe that a large fraction of non-detections corresponds to weak \civ\ systems with $\log (W_{\rm\civ}/\text{\AA})\lesssim-1.5$, with only few strong absorbers ($\log(W_{\rm\civ}/\text{\AA})\gtrsim-0.5$) lacking associations. The last result support the hypothesis that weak \civ\ systems would be associated to aligned \siv\ absorbers below the detection limit of the survey. In the end, this analysis supports the idea that \civ\ and \siv\ absorbers are, to first approximation, two tracers of the same gas phase and are expected to lead to comparable results if used to trace an intermediate ionized phase of the CGM and IGM near LAEs.


\bsp	
\label{lastpage}
\end{document}